\newcommand{\arxiv}[2]{#2} 

\arxiv{
\documentclass[sigconf]{acmart}
}{
\documentclass[sigconf, authorversion=true, nonacm=true]{acmart}
}
\AtBeginDocument{\providecommand\BibTeX{{\normalfont B\kern-0.5em{\scshape i\kern-0.25em b}\kern-0.8em\TeX}}}

\usepackage{url,hyperref}
\usepackage{xspace}
\usepackage{svg} 
\usepackage{amsmath}
\usepackage{bbm}
\usepackage{subcaption}
\usepackage{xcolor}
\usepackage[bibliography=common]{apxproof}
\usepackage[bbgreekl]{mathbbol}
\usepackage{verbatim}
\usepackage{booktabs}
\usepackage{times}

\usepackage{caption}
\newcommand{\catalog}{\ensuremath{\Omega}}
\newcommand{\candidate}{\ensuremath{C}}
\newcommand{\reals}{\mathbb{R}}
\newcommand{\naturals}{\mathbb{N}}
\newcommand{\expect}{\mathbb{E}}
\newcommand{\prob}{\mathbf{P}}
\newcommand{\argmax}{\mathop{\arg\,\max}}
\newcommand{\argmin}{\mathop{\arg\,\min}}

\newcommand{\ulim}{\ensuremath{\bar{u}}}
\newcommand{\collection}{\mathcal{C}}
\newcommand{\obj}{\mathcal{U}}
\newcommand{\proj}{\mathtt{proj}}

\newcommand{\probset}{\mathcal{P}}

\newcommand{\constrset}{\mathcal{D}}
\newcommand{\allprobs}{\mathbb{\Pi}}
\newcommand{\clk}{\ensuremath{\mathtt{CLK}}\xspace}
\newcommand{\org}{\ensuremath{\mathtt{ORG}}\xspace}
\newcommand{\hm}{\ensuremath{\mathtt{H}}}
\newcommand{\nh}{\ensuremath{\mathtt{NH}}}

\newtheoremrep{thm}{Theorem}
\newtheoremrep{cor}{Corollary}
\newtheoremrep{lemma}{Lemma}

\newcommand{\remove}[1]{}

\newcommand{\jc}[1]{\textcolor{red}{JC: #1}}

\newenvironment{packed-itemize}
{\leftmargini=15pt\begin{itemize}
\setlength{\itemsep}{-0.3pt}}
{\end{itemize}\vspace{-0.5em}}

\allowdisplaybreaks[4]

\newcommand\blfootnote[1]{
    \begingroup
    \renewcommand\thefootnote{}\footnote{#1}
    \addtocounter{footnote}{-1}
    \endgroup
}

\title[{Harm Mitigation in Recommender Systems under User Preference Dynamics}]{Harm Mitigation in Recommender Systems under User Preference Dynamics}

\author{Jerry Chee}
\email{jerrychee@cs.cornell.edu}
\affiliation{ 
    \institution{Cornell University}  
    \city{Ithaca}
    \state{NY}
    \country{USA}
}

\author{Shankar Kalyanaraman}
\email{kshankar@meta.com}
\affiliation{ 
    \institution{Meta} 
    \city{Menlo Park}
    \state{CA}
    \country{USA}
}

\author{Sindhu Kiranmai Ernala}
\email{sindhuernala@meta.com}
\affiliation{ 
    \institution{Meta} 
    \city{Menlo Park}
    \state{CA}
    \country{USA}
}

\author{Udi Weinsberg}
\email{udi@meta.com}
\affiliation{ 
    \institution{Meta}
    \city{Menlo Park}
    \state{CA}
    \country{USA}
}

\author{Sarah Dean}
\email{sdean@cornell.edu}
\affiliation{ 
    \institution{Cornell University} 
    \city{Ithaca}
    \state{NY}
    \country{USA}
}

\author{Stratis Ioannidis}
\email{ioannidis@ece.neu.edu}
\affiliation{ 
    \institution{Northeastern University} 
    \city{Boston}
    \state{MA}
    \country{USA}
}

\renewcommand{\shortauthors}{Chee, Kalyanaraman, Ernala, Weinsberg, Dean, \& Ioannidis}

\begin{abstract}
We consider a recommender system that takes into account the interplay between recommendations, the evolution of user interests, and harmful content. We model the impact of recommendations on  user behavior, particularly the tendency to consume harmful content. We seek recommendation policies that establish a tradeoff between maximizing click-through rate (CTR) and mitigating harm. We establish conditions under which the user profile dynamics have a stationary point, and propose algorithms for finding an optimal recommendation policy at stationarity. We experiment on a semi-synthetic movie recommendation setting initialized with real data and observe that our policies outperform baselines at simultaneously maximizing CTR and mitigating harm.
\end{abstract}



\begin{document}
\maketitle

\arxiv{}{\blfootnote{This is the ArXiv version of a KDD'24 paper of the same name.}}
\section{Introduction}

The algorithm of choice  for many recommender systems in production today is the classic top-$k$ recommendation algorithm \citep{karypis2001evaluation,deshpande2004item,cremonesi2010performance}. In short, a top-$k$ recommender
uses a score function to rank items from a candidate pool and subsequently recommends the $k$ highest-scoring items. Such recommendations  also come with optimality guarantees: for example, if scores are proportional to a user's selection probabilities, top-$k$ recommendations maximize the click-through rate (CTR)~\citep{oh2019thompson,oh2021multinomial}, i.e., the probability that the user will pick at least one item from the recommended set.

 A prominent criticism against top-$k$ recommendations (and, more generally,  maximizing engagement) is centered around the concept of \emph{amplification} \citep{kalimeris2021preference}, i.e., a feedback loop that arises between recommended content and user preferences.  Adverse effects of amplification  include decreasing content diversity and amplifying biases~\citep{mansoury2020feedback,huszar2022algorithmic},  as well as increasing the spread of misinformation~\citep{zhang2022network} or extreme content  \citep{restrepo2021social, whittaker2021recommender}.  Recent studies  also explore the impact of recommendations from a user's perspective: recommendations may create pathways that steer users towards  radicalization  \citep{fabbri2022rewiring}, polarization    \citep{leonard2021nonlinear,levin2021dynamics}  but may also aid the migration towards ever more extreme  \citep{ribeiro2020auditing,restrepo2021social} or even  harmful content~\citep{lin2016association, smith2022recommender, hou2019social}.

Motivated by these concerns, we model a recommender system that takes into account the  interplay between recommendations and the evolution of user interests. In doing so, our work follows a long line of research in understanding  the interplay between recommender systems  and user behavior (see, e.g., \cite{kalimeris2021preference,lu2014optimal,dean2022preference}). Our goal is to understand (a) the impact of recommendations on pathways towards, e.g., more extreme or harmful content, and (b) how this impact should be  accounted for when making recommendations. In particular, the fundamental question that we try to answer is \emph{how should recommendations depart from top-$k$/CTR-maximizing recommendations when user preference dynamics are present, and mitigating subsequent harm is part of the objective}.

Our model gives rise to complicated, interesting phenomena related to recommendation safety and harm mitigation. For example, our results (Thms.~\ref{thm:staticoptimal} and \ref{thm:subopt}) suggest that \emph{na\"ive approaches to mitigating harm may fail catastrophically}. On one hand,  explicitly incorporating harm as a penalty in the recommender's objective may have no effect in the optimal recommendation policy, if user preference dynamics are ignored. On the other hand, designing optimal recommendations when dynamics are accounted for does not come easy: surprisingly, we prove  that data-mining workhorses such as alternating optimization   lead to arbitrarily suboptimal recommendations. 
Overall, our contributions are as follows:
\begin{itemize}
    \item We introduce a model that incorporates harm mitigation in a recommender's objective. When user preferences are static, we show  that  top-$k$ recommendations maximize CTR while \emph{simultaneously}   minimizing harm. 
    This finding suggests that \emph{to properly  mitigate harm, it is necessary to account for user dynamics} in the recommender's reasoning.
\item To that end, we incorporate the user preference dynamics of attraction~\citep{lu2014optimal,krauth2020offline,ge2020understanding,mansoury2020feedback}, which explictly model the aforementioned amplification phenomenon. Finding optimal recommendations under these  dynamics is a non-convex problem. We prove a negative result:  the most  natural algorithm, namely, alternating optimization, 
     leads to recommendation policies that are 
     in fact \emph{arbitrarily suboptimal}. 
    \item In light of this, we turn our attention to gradient-based algorithms.
We  propose a tractable method for  computing the gradients when  preference dynamics are incorporated in the objective. We also  evaluate resulting gradient-based recommendation algorithms on a semi-synthetic movie recommendation setting initialized with MovieLens data, where harmful movies are determined by IMDB parental guidelines. We show that our policies are superior at \emph{maximizing CTR and mitigating harm} over baselines, and perform up to 77\% better.
\end{itemize}
Our model, though simple, is based on  the multinomial-logit (MNL) model of choice~\citep{mcfadden1973conditional}, which is ubiquitous in recommender systems literature~\citep{danaf2019online, chaptini2005use, yang2011collaborative,kalimeris2021preference}. From a technical standpoint, our analysis requires characterizing the stationary point implied by the combination of recommendations with attraction dynamics; we accomplish this  by using the Banach  fixed point theorem. Having characterized the fixed-point, implementing gradient-based optimization algorithms in our setting  poses a significant challenge, as our objective is not expressed in closed form. Nevertheless, we propose a tractable algorithm for computing gradients 
via the implicit function theorem.

The remainder of this paper is organized as follows. We discuss related work in Section~\ref{sec:related_work}. We define our   model accounting for harm-mitigation, and determine optimal recommendations in the absence of preference dynamics  in Section~\ref{sec:model}. We present our analysis under user preference dynamics in Section~\ref{sec:main_results}, and our empirical evaluations in Section~\ref{sec:experiments}. We conclude is Section~\ref{sec:conclusions}.

\section{Related Work}\label{sec:related_work}

\noindent\textbf{Characterizing Harm.}
The  literature observing and empirically characterizing the harmful impact of algorithmic recommendations is quite extensive.  Past works  have experimentally observed the existence radicalization pathways \citep{ribeiro2020auditing}, polarization and filter bubbles \citep{ledwich2022radical,levin2021dynamics,rossi2021closed}, 
the amplification of extreme content \citep{restrepo2021social, whittaker2021recommender}, alignment with human values \citep{gormann2022dangers}, and loss of physical and mental well-being \citep{lin2016association, smith2022recommender}. Though these studies focus on empirical observations as opposed to the design of harm-mitigating recommendation policies, which is our main goal, they directly motivate our attempt to model the tension between maximizing engagement and reducing exposure to harm.

\noindent\textbf{Multiple Objectives.}
Multi-objective recommender systems  are extensively studied--see, e.g., the surveys by \citet{zheng2022survey} and \citet{jannach2022multi}. Examples include balancing engagement with diversity \citep{vargas2011rank}, fairness~\citep{xiao2017fairness}, and multi-stakeholder utility~\citep{surer2018multistakeholder}. \citet{wu2019unified} propose alternate optimization and gradient-based approaches, albeit for maximizing utility when making recommendations collectively to a group of users. While not directly considering harm reduction as an objective, \citet{suna2021user} propose methods to account for user polarization in matrix factorization-based recommender systems. 
Closer to us, \citet{singh2020building} use  RL  to balance an engagement-based reward with a notion of harm termed the ``health risk''. We depart from all these works by explicitly modeling user behavior via attraction,   as well as by incorporating the possibility that users reject the recommendation and opt instead for organically-selected content. 
Our model allows for more interpretable decision-making  (see, e.g., our discussion around Lemmas~\ref{lem:fix} and Thms.~\ref{thm:staticoptimal}--\ref{thm:contraction}) as well as a better understanding of how different parameters impact optimal recommendations and the utility/harm they incur. 

\noindent\textbf{Multinomial-Logit Model.}
For the majority of our analysis, we model user preferences via the multinomial-logit (MNL) model~\citep{mcfadden1973conditional}, itself an instance of the more general Plackett-Luce model~\citep{luce1959individual} (see  Sec.~\ref{sec:model}). Both are used extensively to model user choices: they are  workhorses in the field of econometrics, but have also found numerous applications in computer science, particularly in recommender systems \citep{danaf2019online, chaptini2005use, yang2011collaborative,kalimeris2021preference, oh2019thompson, hazrati2022recommender,  fleder2009blockbuster, jiang2014choice,koutra2017pnp}. A popular theoretical setting is contextual MNL bandits, \citep{pmlr-v65-agrawal17a,oh2019thompson,oh2021multinomial}, in which a static user profile is  learned online while top-$k$ recommendations occur; recommendation algorithms in this setting aim to minimize regret w.r.t. the CTR of an algorithm that knows the user profile. Our main departure from this literature is to model a dynamic user profile and incorporate harm. We also depart by studying recommendation policies in ``steady-state''; revisiting our results in an online/bandit setting is an interesting and challenging open problem, particularly due to the user preference dynamics \cite{maniu2020bandits,zinkevich2003online,zhang2018dynamic}, which are not considered in the aforementioned MNL bandit works.

\noindent\textbf{Preference Dynamics.}
Models of user preferences are widely used and studied in recommendation literature. 
Historically, the focus has been on static models (e.g. matrix factorization or topic models),
but an emerging line of work models the dynamics of preferences, also termed \emph{individual feedback loops}~\cite{pagan2023classification}.
Such models of preference dynamics formalize and illuminate phenomena like political polarization~\cite{leonard2021nonlinear,rossi2021closed} and echo chambers~\cite{perra2019modelling,jiang2019degenerate,kalimeris2021preference}, in which the long term influence of recommendations plays a crucial role. 
Many of these works show how feedback loops lead to unintended consequences of traditional recommendation algorithms, critiquing the status quo and proposing alternative algorithms.
For example, \citet{carroll2021estimating} and \citet{ashton2022problem} argue that recommendations which cause user preferences to shift may be viewed as inappropriate ``manipulation,'' while \citet{carroll2021estimating} and \citet{dean2022preference} propose algorithms to avoid this.
While some authors propose general purpose algorithms in Markov decision processes~\cite{carroll2021estimating,tabibian2020design},
most others investigate specific dynamics models connected to particular phenomena of interest.
\citet{curmei2022towards} advocate for a framework mapping between such preference dynamics models in computer science and known psychological effects. 
We note that the majority of the aforementioned works~\cite{leonard2021nonlinear,jiang2019degenerate,dean2022preference,tabibian2020design,kalimeris2021preference,rossi2021closed,pagan2023classification} study steady-state dynamics of their proposed models, as we do here; however, they do not directly optimize a recommendation policy towards attaining a certain objective in steady state.

Closest to us, \citet{lu2014optimal}~propose a model of user preference dynamics including attraction, aversion, and social influence. Like us, they study optimal recommendations in steady state. However, their underlying recommendation model is based on real-valued feedback in matrix factorization, as opposed to discrete user choice in the MNL model. From a technical standpoint, their dynamics are simpler, and the steady-state profiles can be computed in a closed form (rather than via a fixed point). Moreover, their recommendation design reduces to a quadratic optimization problem in their setting, which they solve via a semi-definite relaxation. Both are technically less challenging than the optimization problem we study here. 
Furthermore, their focus is on long term user satisfaction rather than avoiding harmful content.

\noindent\textbf{Integrity Checks.}
Algorithms for detecting and filtering content that is inappropriate or harmful prior to recommendations, i.e. ``integrity checks,'' have been extensively explored by academia and industry alike~\cite{akos2021how,lee2007harmful,arora2023detecting,pierri2023does,zampieri2023offenseval,gillespie2020content}. The design of such filtering mechanisms is orthogonal to the question we study. In particular, our starting point is the assumption of \emph{perfect filtering}, as the recommender \emph{never suggests harmful content}, which  only exists outside the platform: users can find it organically irrespective of recommendations. It is natural to ask if the recommender has a role in mitigating this type of indirect harm, that it cannot directly control. Surprisingly, our work answers affirmatively. If user dynamics are present, then the recommender may cause indirect harm, and its mitigation is non-trivial: na\"ive ways of modeling harm or simple algorithms like alternating optimization may fail catastrophically.

\noindent\textbf{Operationalization.}
There are many ways to conceive of and operationalize harm. In the context of  recommendations, \citet{shelby2022sociotechnical} and \citet{hosseini2023empirical}  consider representational and allocative harms. Other work considers the meaning of harm from a causal perspective, including loss of utility \citep{beckers2022causal} and counterfactual outcome frameworks \citep{richens2022counterfactual}. We limit our scope to harm arising from engagement with problematic content. Our approach relies on modelling the causal mechanisms by which a recommender system may increase the likelihood of harm, in order to reduce it.

\section{Modeling Harm-Mitigating Recommendations}\label{sec:model}

We consider a recommender making sequential suggestions to a user.  The user can accept the recommendation, by selecting an item in this set, or reject it, by making an ``organic'' selection of items that exist off-platform; this may lead to a selection of harmful content; we would like to take this  into account when recommending items.

We describe here a model of this behavior assuming that user preferences are static, and characterize optimal recommendation policies; as we will see, these differ drastically from recommendations when user preference dynamics are taken into account; we consider these in the next section.

\begin{table}[!t]
    \centering
    \begin{small}
    \begin{tabular}{c|p{0.80\linewidth}}
        $\Omega$ & universe of items; $\Omega \subseteq \reals^d, |\Omega| = n$  \\
        $H$ & subset of $\Omega$ containing harmful items, $|H|=h$ \\
        $\candidate$ & candidates $\candidate\subseteq \Omega\setminus H$ sampled from collection $\collection$.\\
        $E$ & set of items recommended, $E \subseteq \candidate$\\
        $v$ & item in $\Omega$ \\
        $u$ & vector in $\reals^d$ representing user profile \\
$\pi$ & recommendation policy\\
        $\clk$ & user accepts an item suggested by the recommender\\
        $\org$ & user rejects recommended items and chooses an item from $\Omega$\\
         $\hm$ & user selects an item from $H$\\
        $s_v$ & score denoting user's preference for $v$\\
        $s_A$ & Sum of scores $\sum_{v\in A} s_v$\\
        $c$ & scaling param to control prob. of $\clk$ (lower $\Rightarrow$ more likely) for MNL model\\
        $g$ & map $\reals_+\rightarrow [0, 1]$; for MNL, $g(s_E) = s_E / (s_E + c)$\\
        $u_0$ & ``inherent'' user preference vector\\
        $u(t), v(t)$ & current user profile and chosen item at time $t$ (resp.)\\
        $\beta$ & weight param for $u_t$'s dependence on $u_0$ (higher means more)\\
        $\alpha_{v(t)}$ & weight param for $u_t$'s dependence on $v(t)$\\
        $k$ & num. of items shown to user by the policy\\
        $\lambda$ & regularization parameter\\
    \end{tabular}
    \end{small}
    \caption{Notation Summary}
\label{tab:notation}
\end{table}

\subsection{Recommendation Policies}
We consider a universe of $n$ items $\Omega \subset\reals^d$, with $|\Omega| = n$, where each item is represented by a $d$-dimensional \emph{item profile} $v\in\catalog$ (e.g., explicit features, latent factors). A subset 
$H\subset \catalog$, with $|H|=h$,
is known to be \emph{harmful}.  In general,  set $\Omega$ includes both \emph{on-platform} items, that a recommender may display to users, as well as \emph{off-platform} items. At each timeslot~$t$, a recommender has access to a subset $\candidate_t\subseteq \catalog\setminus H$ of candidate on-platform items for possible recommendation to  users. We assume that these candidate sets are sampled in an i.i.d.~fashion from a collection $\collection$ of (possibly overlapping) subsets of $\Omega \setminus H$. Having access to this set,  the recommender  displays to a user a set $E_t\subset \candidate_t$; we refer to $E_t$ as the \emph{recommended set} or, simply, the \emph{recommendation}.

Intuitively, the recommender could be any platform that recommends content, and $\Omega$ could be all the content on the internet, including harmful content. For example, news could be recommended on a social media platform with links out to external news sites; some ``fake news'' sites will never be recommended but users could still find them through other means. Note that, by design, the recommender \emph{never suggests harmful content}, and harmful-content is  only found off-platform.

We consider (possibly) randomized \emph{recommendation policies}: given a candidate set $\candidate$, the recommendation policy is expressed as a distribution over possible sets $E\subseteq C$ to recommend to the user. Recommendations are constrained. In the \emph{bounded cardinality} setting, we assume recommended sets $E$ have at most $k\in \naturals$ items.
In the \emph{independent sampling} setting, we assume that (a) items $v\in E$ are selected independently from $C$ and (b) the \emph{expected} size of recommendations is at most $k$, i.e., $\expect[|E|\mid C]\leq k$. 
Independent sampling constrains the size of $E$  only in expectation, but policies can be described with polynomially many parameters (in $n,k$), compared with $\Theta(2^k)$ parameters for bounded cardinality (see also \arxiv{Appendix~A in \cite{arxivversion}}{Appendix~\ref{app:recpolicies}}).

\subsection{User Preferences \& Selection Behavior}
At each timeslot, the user can accept the recommendation $E$, by selecting an item in this set,
or reject it, by selecting an arbitrary item in $\catalog.$
We refer to the former event as a $\clk$ event and the latter as an $\org$ (for ``organic selection'') event. Only \org events can lead to a harmful engagement.

 We assume that user selections
are governed by the Plackett-Luce model~\citep{luce1959individual}. In particular,  for every $v\in \catalog$, there exists a non-negative score $s_v\in \reals^+$ quantifying a user's preference toward $v$. Moreover, the probability that a user selects an item among a set of alternatives is proportional to this score. Formally, for a set of items $A\subseteq \Omega$, let the total score be $s_A \equiv \textstyle\sum_{v\in A}s_v$. Plackett-Luce postulates that the probability that a user selects item $v\in A$ against alternatives $A$ is given by $s_v/s_A$. 
In our setting, given recommendation $E$, the user is also faced with an additional alternative of rejecting the recommendation and selecting an item organically. We assume that
\begin{align}
 p_{\clk\mid E} &\equiv g(s_E), &\text{and, thus}& &  p_{\org\mid E} &= 1-g(s_E),\label{eq:gfun} \end{align} where $g:\reals_+\to[0,1]$ is a function of the total score of $E$. 
 Moreover, applied to our setting, the Plackett-Luce model gives the following conditional probabilities of items $v\in E$, $v'\in \Omega $ being selected, respectively:
\begin{align}\label{eq:pl}
p_{v\mid E,\clk} &=\frac{s_v}{s_E} & \text{and}&  &p_{v'\mid E,\org} &= \frac{s_{v'}}{s_\Omega}.
\end{align}

\begin{comment}
\noindent\textbf{CTR and Probability of Harm.}
We are primarily concerned with two quantities of interest. The first is the (standard) \emph{click-through rate} (CTR), that traditional recommender systems aim to maximize: this is given by probability that the user accepts the recommendation, i.e., a \clk event occurs.   The second is the \emph{probability of harm}, i.e., the probability that the user selects harmful content; note that this can only happen organically, and is thus only indirectly controlled by the recommender. Under our modelling assumption in Eq.~\eqref{eq:gfun}, combined with the Plackett-Luce model, the probabilities of click (\clk) and harm (\hm) events are, respectively:
\begin{align}\label{eq:pclk}
    p_\clk &=\expect_{E,\candidate}[g(s_E)], &
p_\hm &=(1- p_\clk) \frac{s_H}{s_\catalog}.  \label{eq:phm}
\end{align}
where $p_{E\mid \candidate}$ is the probability that set $E\subseteq C$ is recommended, and $p_\candidate$ is the probability that $\candidate\in \collection$ is the candidate set.

\end{comment}

\subsection{Top-$k$ Recommendations}
\label{subsec:topk}
As items are linked to scores, a simple, intuitive  policy is the \emph{top-$k$ recommendations} policy \citep{karypis2001evaluation} (see also \arxiv{Appendix~A in \cite{arxivversion}}{Appendix~\ref{app:recpolicies}}). Given $C$ and budget $k$, the policy recommends the $k$ highest-scoring items $v\in \candidate$, w.r.t., scores  $s_v$. This can be implemented in $O((|C|+k)\log k)$ time by traversing the list of scores $s_v$, $v\in \candidate$, and maintaining a sorted list of the highest-scoring items.

\subsection{Optimal Recommendations under\\ Static Preferences}
Departing from traditional recommender systems, but assuming static preferences, we would like to select a  policy that \emph{maximizes CTR while simultaneously minimizing harm};
to do so, we wish to solve:
\begin{subequations}\label{eq:probstatic}\begin{align}\text{Maximize}: &\quad f_0(\pi) = p_{\clk}(\pi)-\lambda p_\hm(\pi) \label{eq:probstaticobj} \\
\text{subj.~to}:&\quad\pi\in \probset,
\end{align}\end{subequations}where 
$\lambda$ is a trade-off parameter, $\pi$ is the recommendation policy specifying $[p_{E\mid \candidate}]_{C\subseteq \Omega \setminus H, E\subseteq \candidate}$,   and
\begin{align}p_\clk &=\expect_{E,\candidate}[g(s_E)], &
p_\hm &=(1- p_\clk) \frac{s_H}{s_\catalog}, \label{eq:phm}
\end{align}
are the  probabilities of click (\clk) and harm (\hm) events, respectively. The set $\probset$ is the set of valid policies,  determined either the bounded cardinality or independent sampling constraints with capacity $k\in \naturals$ (see \arxiv{see Eqs.~(19) and (20)--(21) in Appendix B of \cite{arxivversion}}{Eqs.~\eqref{eq:Pi} and~~\eqref{eq:prodform}--\eqref{eq:R} in Appendix~\ref{app:recpolicies}}).   In short, Prob.~\eqref{eq:probstatic} aims to find a valid recommendation policy $\pi \in \probset$ that establishes an optimal tradeoff between click-through rate $p_\clk$ and probability of harm $p_\hm$, as determined by parameter $\lambda\geq 0$.

Note that, despite the fact that it never recommends harmful content, the recommender does impact the probability of harm: this is because it can reduce harm by attracting the user to its (non-harmful) recommendations. This has a rather suprising consequence: the top-$k$ recommendations policy is optimal  \emph{even if the recommender objective includes harm mitigation (i.e., for all $\lambda>0)$}: 

\begin{thm}\label{thm:staticoptimal}
Assume that function $g:\reals_+\to [0,1]$ in Eq.~\eqref{eq:gfun} is non-decreasing. Then for all $\lambda\geq 0$ the  top-$k$ recommendation policy is an optimal solution to Prob.~\eqref{eq:probstatic} under the bounded cardinality setting. If, in addition, function $g$ is  concave, then top-$k$ recommendation policy is also optimal under the independent sampling setting. \end{thm}
The proof can be found in \arxiv{Appendix B of \cite{arxivversion}}{Appendix~\ref{app:proofofthm:staticoptimal}}. It is straightforward for the bounded cardinality setting; the independent sampling case is somewhat more involved, and requires a submodularity argument. Intuitively, 
since the recommender never recommends harmful content, \emph{maximizing the click-through rate is also harm-minimizing}: recommending content likely to be clicked reduces the chance of an organic selection and
thus of selecting harmful content.
Thm.~\ref{thm:staticoptimal} is in stark contrast to what happens when the user preferences are affected by recommendations, as we will see in the following section.
Moreover, it tells  an important cautionary tale: if recommenders treat users as static, \emph{ recommendations that seemingly account for harm mitigation, by directly including it in the objective, 
may end up just maximizing click-through rate}. 
This, however, can be catastrophic in terms of the actual harm (and the corresponding objective attained), as we will again see in the next section.

\subsection{Multinomial Logit (MNL) Model}
\label{sec:mnl} 
We conclude this section by presenting
the standard multinomial logit  (MNL) model,  used extensively to  model user choices in recommender systems~\cite{hazrati2022recommender,   jiang2014choice, pmlr-v65-agrawal17a,oh2019thompson,oh2021multinomial,kalimeris2021preference}. 
A special case of the Plackett-Luce model, 
it is a natural generalization of matrix factorization from ratings to choices, and an extension of ``soft-max'', allowing for a no-choice alternative (see also \arxiv{Appendix D in~{\cite{arxivversion}}}{Appendix~\ref{app:mnl}}).  Formally, for every $v\in \catalog$, the  non-negative score $s_v\in \reals^+$ quantifying a user's preference toward $v$  is \begin{align}
    s_v \equiv e^{v^\top u}, \label{eq:score}
\end{align} i.e., it is  parameterized by both the item profile $v\in \reals^d$ as well as by a \emph{user profile} $u\in\reals^d$.
Moreover,  the conditional probability $p_{\clk\mid E}$ is given by: \begin{align}
    g(s_E)  \equiv\frac{s_E}{s_E+c}~\label{eq:clk},
\end{align}
where $c\geq 0$ is a non-negative constant. The remaining probabilities are given again by the Plackett-Luce model (Eq.~\eqref{eq:pl}), with scores determined by Eq.~\eqref{eq:score}.

It is easy to confirm that $g:\reals_+\to[0,1]$ is both non-decreasing and concave. Thm.~\ref{thm:staticoptimal} thus has the following immediate corollary:
\begin{cor}\label{cor:static} If user selections are governed by the MNL model under a static profile $u\in \reals^d$, then for all $\lambda\geq 0$  the  top-$k$ recommendation policy is an optimal solution to Prob.~\eqref{eq:probstatic} under both the bounded cardinality and independent sampling settings.
\end{cor}

\begin{figure}
\centering
\includegraphics[width=1\linewidth]{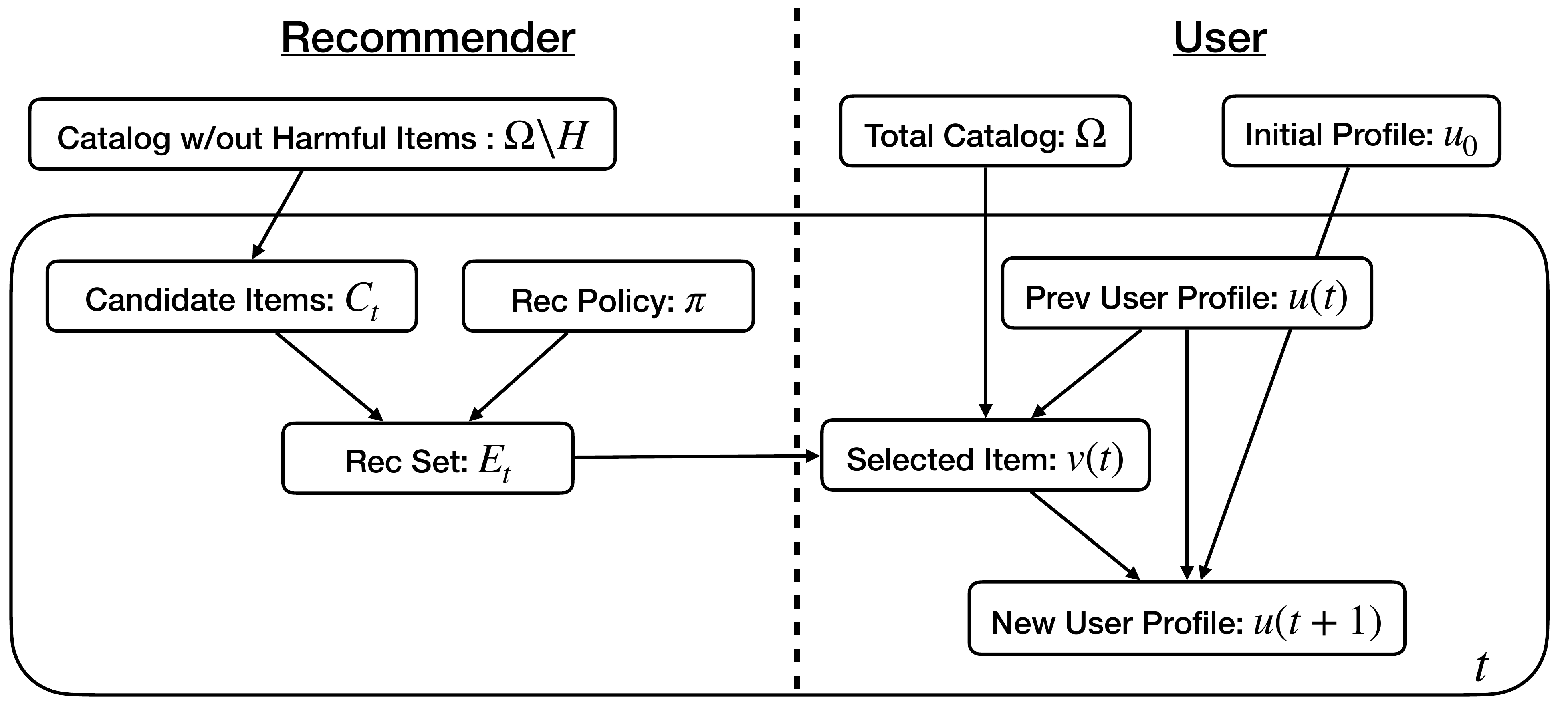}
\vspace{-2.5em}
\caption{
{
Our full model, incorporating  user dynamics in the presence of recommender system interactions.
A recommender presents a recommendation set $E_t$ to the user, who chooses to either interact with the recommended content (\clk), or organically select an item from  the entire catalog (\org), which includes harmful content.
The user profile is subsequently updated under at attraction model ~\citep{lu2014optimal,krauth2020offline,ge2020understanding,mansoury2020feedback}, leaning closer to the item {\protect{$v(t)$}} selected by the user.
}
}\label{fig:infographic}
\end{figure}

\section{Harm-Mitigating Recommendations Under Dynamic User Profiles}\label{sec:main_results}

We turn our attention to what happens 
when \emph{user preferences adapt based on the content that users consume}.
As we will see, this has drastic implications for harm-mitigating recommendations. A graphical model indicating the evolution of profiles under user dynamics is shown in Fig~\ref{fig:infographic}.

\subsection{User Profile Dynamics}
For the remainder of our analysis, we adopt MNL as the model  governing user choices. Thus, there exists a user profile $u\in\reals^d$ that, jointly with item profiles $v\in \reals^d$, determines item scores via~Eq.\eqref{eq:score}. So far (e.g., in Cor.~\ref{cor:static}), we assumed that the user profile is static; we will now depart from this assumption. 

Formally, we denote by $u(t)$ the profile of user at time $t\in \naturals$.  Let $v(t)\in \catalog$ be the item selected and consumed by the user at timeslot $t$. We model the influence of an item with an \emph{attraction} model \citep{lu2014optimal,krauth2020offline,ge2020understanding,mansoury2020feedback}. First, the user is characterised by an inherent profile $u_0\in \reals^d$, that captures their predisposition towards content. At time $t+1$, the  profile $u(t+1)$ is generated as a convex combination of (a) the inherent profile, (b) the select item $v(t)$, and (c) the current profile $u(t)$. Formally, starting from some $u(0)\in \reals^d$, we have:
\begin{align}
    u(t+1) =   \expect\left[\beta u_0 +\alpha_{v(t)}v(t)+(1-\alpha_{v(t)}-\beta)u(t)\right], \label{eq:u}
\end{align}
where the randomness is due to the selected item $v(t)$, and $\beta\in[0,1]$, $\alpha_{v}\in[0,1-\beta]$, $v\in \Omega$, control the the relative importance of the inherent profile and the item, respectively. 
This is a ``mean-field'' or ``fluid'' model \citep{borkar2009stochastic,benaim2006dynamics,benveniste2012adaptive}, in which the evolution of the user profile is governed by mean dynamics. It is also an  attraction model, as
at every timeslot, the user profile ``nudges'' towards selected object $v(t)$ with a rate $\alpha_{v(t)}\in [0,1]$: by Eq.~\eqref{eq:score}, this increases the probability that the user will select items similar to the ones it  consumed in the past. Note that $v(t)$ is a random variable whose distribution   depends on  the recommendation policy $\pi$ and the current user profile $u(t)$ (see Eq.~\eqref{eq:pv}).
In general, we allow the ``attractiveness'' to be content dependent. For simplicity of exposition, from here forward, we assume that $\alpha_v =  \alpha_\hm,$ if $v\in H,$ and  $\alpha_v = \alpha_\nh,$ if $v\notin H.$

\noindent\textbf{Stationary Profiles and Steady State Behavior.} 
Given a recommendation policy, we are interested in understanding the steady-state behavior of a user, as described by the limit $\lim_{t\to\infty} u(t)$. We thus turn our attention to the stationary points of this process, which, as it turns out, we can compute. We define the mean profile drift as:
\begin{align}
   \begin{split} \Delta u(t) &\equiv u(t+1)-u(t)\\&= \expect_\pi\left[\alpha_{v(t)}\left(v(t)-u(t)\right)\right] + \beta\left(u_0 - u(t)\right). \end{split}\label{eq:deltau}
\end{align}
A profile $\ulim\in\reals^d$ is \emph{stationary} when $\Delta u(t)=0$  for $u(t)=\ulim$: that is, if the system starts from this profile, it will remain there indefinitely. Note that, if $\lim_{t\to\infty} u(t)$ exists, it must be a stationary profile. 

\subsection{Recommendation Objective}
\label{sec:rec_obj}
We are now ready to revisit our recommendation objective. We denote by $\allprobs$ the set of all possible recommendation policies (including invalid ones, not in $\probset$).
Given policy $\pi\in \allprobs$, let $\pi\mapsto \ulim(\pi)$ be the map from $\pi$ to the stationary profile $\ulim$, as defined above (see also Eq.~\eqref{eq:fp} below). We  consider the following  optimization problem: \begin{subequations}
\label{eq:prob2}
\begin{align}
\text{Maximize}: &~~f(\pi) = p_{\clk}(\pi,\ulim(\pi))\!-\!\lambda p_\hm(\pi,\ulim(\pi)) \label{eq:prob2obj} \\
\text{subj.~to}:&~~\pi\in \probset,
\end{align}
\end{subequations}
for some regularization parameter $\lambda\geq 0$. Crucially, both the CTR and the probability of harm are measured \emph{at the stationary profile $\ulim(\pi)$}, our proxy for the steady-state limit of the user dynamics. 
Note that, even though the recommendation policy never suggests harmful content, recommendations affect user selections, which in turn affect their predisposition towards harmful content. 

Solving Prob.~\eqref{eq:prob2} is quite challenging. Beyond the fact that it is not a convex optimization problem, finding a solution is complicated by the fact that the stationary profiles depend on the recommendation policy $\pi$ via map $\ulim(\pi)$, which \emph{we cannot express  in a closed-form}. Thus, we have no direct way of computing it;  it is a priori unclear even whether such a stationary profile exists and is unique. One of our  contributions is to resolve both of these issues (see Thm.~\ref{thm:contraction}).

\noindent\textbf{Computing $\ulim(\pi)$.}
We now turn our attention to finding the stationary user profile. We fist characterize the stationary profile via a fixed-point equation.  Eq.~\eqref{eq:deltau} implies the following lemma:
\begin{lemma}\label{lem:fix} Stationary user profiles $\ulim\in \reals^d$ satisfy the following fixed-point equation:
\begin{align}
    \ulim = F(\pi,\ulim), \label{eq:fp}
\end{align}
where map $F:\allprobs\times \reals^d$ is given by: 
\begin{align}
    F(\pi,u) = \textstyle \frac{\beta u_0 + \alpha_\hm\sum_{v\in H}[v\cdot p_v(\pi,u)] + \alpha_\nh \sum_{v\notin H}[v\cdot p_v(\pi,u)] } {\beta+\alpha_\hm 
  p_\hm(\pi,u)+\alpha_\nh 
  p_\nh(\pi,u)}.\label{eq:F}
\end{align}
where  $p_\hm$,  $p_\nh$, are 
given by  Eq.~\eqref{eq:phm}, and
 $p_v$ is the probability that the user selects item  $v\in \Omega$. \end{lemma}
The proof is in \arxiv{Appendix~C in \cite{arxivversion}}{Appendix~\ref{app:proofoflem:fix}}.
Intuitively, a stationary profile consists of an interpolation between the inherent profile $u_0$ and the ``average'' harmful and non-harmful item profiles.
The weights of this average are determined by the policy. 

Next, we turn to the question of finding a stationary profile which satisfies the fixed point equation.
In particular, we establish conditions under which map $F$ in Eq.~\eqref{eq:F} is a contraction:

\begin{thm}\label{thm:contraction} Let $\|\cdot\|$ denote the Euclidean norm in $\reals^d$. If \begin{align}\|v\| <\frac{1}{6}\big(\sqrt{\textstyle\|u_0\|^2+12\frac{\alpha_\nh+\beta}{5nd\alpha_H }}-\|u_0\|\big),\label{eq:condition}\end{align}
for all $v\in \Omega$, then  $F:\allprobs\times \reals^d$ given by  Eq.~\eqref{eq:fp} is a contraction w.r.t.~$\|\cdot\|$ uniformly on all distributions $\pi\in \allprobs$. That is, for all distributions $\pi\in \allprobs$ (not necessarily in $\probset$), and all  $u,u' \in \reals^{d}$,
  $ \|F(\pi, u ) -F(\pi,u')\|\leq L \|u-u'\|, $ 
for some $L<1$ that does not depend on $\pi$.\end{thm}
We prove this in \arxiv{Appendix~E in \cite{arxivversion}}{Appendix~\ref{app:proofofthm:contraction}}.  Thm.~\ref{thm:contraction} along with the Banach fixed-point theorem~\citep{banach1922operations} imply  that  a stationary profile exists and is unique when Eq.~\eqref{eq:condition} holds. Most importantly, it can be found by the following iterative process: starting from any $\ulim^0\in \reals^d$, the iterations
\begin{align}\label{eq:fpiterate}
    \ulim^{\ell+1} = F(\pi,\ulim^\ell), \quad \ell\in\naturals, 
\end{align}
are guaranteed to converge to the unique fixed-point $\ulim^*$ of Eq.~\eqref{eq:fp}. Convergence happens exponentially fast (see \arxiv{Appendix~E in \cite{arxivversion}}{Appendix.~\ref{app:proofofthm:contraction}}).

Note that Eq.~\eqref{eq:condition} holds w.l.o.g.: multiplying every  $v\in \Omega$ with a constant $\tau<1$ and   $u_0$ with $1/\tau$ yields exactly the same scores in  Eq.~\eqref{eq:score}. Hence,  inherent user profiles and item profiles  learned from data (e.g., user-item clicks), can be rescaled so that  Eq.~\eqref{eq:condition} holds.

\subsection{Alternating Optimization}
\label{sec:alt_alg} Armed with Thm.~\ref{thm:contraction}, we turn our attention to algorithms for solving the (non-convex) Prob.~\eqref{eq:prob2} in its general form.
A simple approach is via an alternating optimization/EM-like algorithm. In particular, one could start from, e.g., the user's initial profile $u_0$, and iterate as follows, for $\ell\in \naturals$:
\begin{subequations}\label{eq:alternate}
\begin{align}
    \pi^{\ell+1} &= \textstyle\argmax_{\pi\in \probset}\left( p_\clk(\pi,u^{\ell}) - \lambda p_\hm(\pi,u^{\ell})\right)
    \label{eq:findpi},\\
        u^{\ell+1} &= \ulim(\pi^\ell). \label{eq:findu}
\end{align}
\end{subequations}
In step  \eqref{eq:findpi}, given $u^\ell$, the optimal $\pi$ is determined by the  top-$k$ policy (by Thm.~\ref{thm:staticoptimal}).
In step \eqref{eq:findu}, given  $\pi^\ell$, the map $\ulim$ is computed via the iterative algorithm in Eq.~\eqref{eq:fpiterate}. The algorithm thus alternates between finding an optimal recommendation policy and the corresponding stationary profile.   Unfortunately, this approach fails:
\begin{thm}\label{thm:subopt}
The alternating optimization steps in Algorithm~\eqref{eq:alternate} can be arbitrarily suboptimal: for every $M>0$, we can construct an instance of Prob.~\eqref{eq:prob2} with optimal solution $\pi^*$ for which the solution $\pi$ produced by Algorithm~\eqref{eq:alternate} satisfies $f(\pi^*)-f(\pi)>M$. 
\end{thm}
The proof is in \arxiv{Appendix~F in \cite{arxivversion}}{Appendix~\ref{app:proofofthm:subopt}}. On a high level, even though Alg.~\eqref{eq:alternate} produces a different solution than the top-$k$ policy with profile $u_0$, as it takes into account the steady state profile, Thm.~\ref{thm:staticoptimal} implies that \emph{this trajectory is independent of $\lambda$}.
This leads to arbitrarily suboptimal decisions when $\lambda$ is high, and minimizing the probability of harm is paramount.

\subsection{Gradient-Based Algorithms}
\label{sec:grad_alg}
The suboptimality of alternating optimization highlights the importance of understanding the evolution $\pi$ and $\ulim(\pi)$ \emph{jointly}. This can indeed be accomplished via a continuous optimization algorithm like projected gradient ascent (PGA)~\citep{bertsekas1997nonlinear}. The constraint set in both the bounded cardinality and the independent sampling settings is a  convex polytope. Applying first-order methods to solve this requires  computing the gradient of the objective~\eqref{eq:prob2obj}. However, this poses a challenge \emph{exactly because we cannot describe $\ulim(\cdot)$ in  closed form.} Nevertheless, we show that $\nabla \ulim(\cdot)$ and, subsequently $\nabla f (\cdot)$, can be computed using the implicit function theorem~\citep{folland2001advanced,blondel2022efficient}. We briefly review PGA for the bounded cardinality setting below, and provide implementation details for PGA in both constraint settings in the supplement. Nevertheless, we stress that these gradients can be used in other standard solvers; we demonstrate this in our experiments, where we use the (more powerful) SLSQP solver \citep{kraft1988software} instead.

\noindent\textbf{Projected Gradient Ascent}.
  PGA starts from a feasible policy $\pi_0\in \probset$ and proceeds iteratively via:
\begin{align}\label{eq:pga}
    \pi_{\ell+1} = \proj_{\probset} \left(\pi_\ell +\gamma_\ell \nabla_\pi   f(\pi^\ell,\ulim(\pi^\ell)) \right),~~\ell \in\naturals,
\end{align}
where 
${\proj_\probset} (\pi')= \argmin_{\pi\in \probset}\|\pi-\pi'\|_2^2 $
is the  projection to $\probset$. As  $\probset$ is a union of simplices (see \arxiv{Appendix A in \cite{arxivversion}}{Appendix~\ref{app:recpolicies}}), $\proj_\probset$  has  efficient implementations (see, e.g., \cite{michelot1986finite}).

\noindent\textbf{Computing the Gradient for the Bounded Cardinality Setting.}
 Let  $m\equiv\sum_{\candidate\in \collection}|\constrset_\candidate|= \sum_{\candidate\in \collection}{|\candidate| \choose k}$ be the number of parameters in $\pi\in \probset$. By the chain rule, we have:
\begin{align}\label{eq:nabf}
    \nabla_\pi f(\pi,\ulim(\pi)) \!= \!   \nabla_\pi \obj(\pi,\! \ulim)\! + \! 
       \left[\nabla_{\pi} \ulim(\pi)\right]^\top\!\! \nabla_{u} \obj(\pi,\!u),\!\!
\end{align}
where  $
\obj(\pi,u) \equiv p_{\clk}(\pi,u)-\lambda p_\hm(\pi,u),$ and $\nabla_\pi \obj(\pi, \ulim)\in \reals^{m}$, $\nabla_{u} \obj(\pi,\ulim)\in \reals^d$ are the gradients of $\obj$ w.r.t.~its first and second arguments, respectively, evaluated at inputs $\pi$ and $\ulim = \ulim(\pi)$. As we have closed-form, differentiable expressions for $p_\hm$ and $p_\nh$, these two gradients can be computed via standard means, while $\ulim = \ulim(\pi)$ can be computed via Eq.~\eqref{eq:fpiterate}.

We thus turn our attention to computing  $\nabla_{\pi} \ulim(\pi)\in \reals^{d\times m}$, i.e., the Jacobian of map $\ulim(\cdot)$ w.r.t.~$\pi$, which we cannot express in closed form. 
Observe that $\pi\in [0,1]^m$, while $\ulim\in \reals^d$. Eq.~\eqref{eq:fp} implies that $
    F(\pi,\ulim) - \ulim= 0,
$
where $F$ is given by \eqref{eq:F}.
This is a system of $d$ non-linear equations, involving both $\pi$ and $\ulim$ as unknowns. 
Hence, by the implicit function theorem \citep{folland2001advanced}, the Jacobian $\nabla_\pi \ulim(\pi)$, can be computed via:
\begin{align}
    \nabla_\pi \ulim(\pi) = -\left( \nabla_{u} F(\pi,\ulim)- I_{d\times d}\right)^{-1}\cdot \nabla_\pi F(\pi, \ulim) ,\label{eq:ulimjac}
\end{align}
where $\nabla_\pi F(\pi, \ulim)\in \reals^{d\times m}$, $ \nabla_{u} F(\pi,\ulim)\in \reals^{d\times d}$ are the Jacobians of map $F$ w.r.t.~its two arguments, evaluated again at $\pi$ and $\ulim=\ulim(\pi)$. As $F$ is a closed-form, differentiable function, constituent matrices in the r.h.s. can  be computed  by standard techniques, and $\nabla_\pi \ulim(\pi)$ can be obtained by solving a linear system (to avoid matrix inversion).  Additional details, along with ways of computing the gradient for the independent sampling setting, can be found in \arxiv{Appendix~G in~\cite{arxivversion}}{Appendix~\ref{app:gradientcomp}}.

\noindent\textbf{Computational Complexity and Tractability.}
For the bounded cardinality setting, gradient computations via Eqs.~\eqref{eq:nabf}--\eqref{eq:ulimjac} involve (a) linear algebra operations (matrix multiplications, solving linear systems) and (b) computing the stationary point $\ulim(\pi)$. The former can be done in polynomial time in constituent matrix dimensions  $m$ and $d$. The latter can be computed via the iterations in Eq.~\eqref{eq:fpiterate}:  each iteration is again polynomial in $m$ and $d$, and convergence is exponentially fast (see also the discussion below Thm.~\ref{thm:contraction}). Having computed the gradient, using Michelot's algorithm~\citep{michelot1986finite}, the projection in Eq.~\eqref{eq:pga} is $O(m\log m)$, while the remaining operations in PGA steps \eqref{eq:pga} are linear in $m$. Finally, all of the above statements extend to the independent sampling setting, with $m$ replaced by $m'\equiv \sum_{\candidate\in\collection}|\candidate|$ (see \arxiv{Appendices~A and~G in~\cite{arxivversion}}{Appendices~\ref{app:recpolicies} and~\ref{app:gradientcomp}}).

\begin{table}[!t]
\centering
\resizebox{\columnwidth}{!}{
\begin{tabular}{|c|ccc|}
\hline
\hline
& \multicolumn{3}{c|}{\underline{Action}} \\
\midrule
 Policy & 
Objective $f$~($\uparrow$) & $p_{\clk}$~($\uparrow$) & $p_\hm$~($\downarrow$)  \\
\midrule 
\emph{Grad} &   \textbf{-4.694 ($\pm$1.330)} & \textbf{ 0.810 ($\pm$0.037}) &  \textbf{0.055 ($\pm$0.013)} \\
\emph{Alt}  &   -6.268 ($\pm$2.117) &  0.774 ($\pm$0.060) &  0.070 ($\pm$0.021)  \\
\emph{U0}   &   -6.279 ($\pm$2.417) &  0.778 ($\pm$0.066) &  0.071 ($\pm$0.024) \\
\emph{Unif} &  -15.681 ($\pm$0.555) &  0.487 ($\pm$0.011) &  0.162 ($\pm$0.005)  \\
\midrule 
\midrule
& \multicolumn{3}{c|}{\underline{Adventure}} \\
\midrule
Policy & 
Objective $f$~($\uparrow$) & $p_{\clk}$~($\uparrow$) & $p_\hm$~($\downarrow$)  \\
\midrule 
\emph{Grad} &  \textbf{-3.445 ($\pm$1.162)} & \textbf{0.676 ($\pm$0.091)} &  \textbf{0.041 ($\pm$0.011)} \\
\emph{Alt}  &  -3.793 ($\pm$1.304) &  0.666 ($\pm$0.088) &  0.045 ($\pm$0.012) \\
\emph{U0}   &  -4.160 ($\pm$1.738) &  0.645 ($\pm$0.119) &  0.048 ($\pm$0.016)  \\
\emph{Unif} &  -9.243 ($\pm$0.222) &  0.303 ($\pm$0.009) &  0.095 ($\pm$0.002)  \\
\midrule
\midrule
& \multicolumn{3}{c|}{\underline{Comedy}} \\
\midrule
Policy & 
Objective $f$~($\uparrow$) & $p_{\clk}$~($\uparrow$) & $p_\hm$~($\downarrow$)  \\
\midrule 
\emph{Grad} &   \textbf{-3.816} ($\pm$0.673) &  \textbf{0.717 ($\pm$0.049)} &  \textbf{0.045 ($\pm$0.006)}  \\
\emph{Alt}  &   -4.725 ($\pm$1.536) &  0.684 ($\pm$0.084) &  0.054 ($\pm$0.015)  \\
\emph{U0}   &   -5.556 ($\pm$2.032) &  0.645 ($\pm$0.111) &  0.062 ($\pm$0.019) \\
\emph{Unif} &  -11.318 ($\pm$0.520) &  0.306 ($\pm$0.013) &  0.116 ($\pm$0.005)  \\
\midrule
\midrule
& \multicolumn{3}{c|}{\underline{Fantasy}} \\
\midrule
Policy & 
Objective $f$~($\uparrow$) & $p_{\clk}$~($\uparrow$) & $p_\hm$~($\downarrow$)  \\
\midrule %aistats24/figs_mf/Fantasy_mf_4_beta015/table.tex
\emph{Grad} &  \textbf{-2.754 ($\pm$0.862)} &  \textbf{0.649 ($\pm$0.081)} &  \textbf{0.034 ($\pm$0.008)}  \\ 
\emph{Alt}  &  -2.760 ($\pm$0.849) &  0.662 ($\pm$0.073) &  \textbf{0.034 ($\pm$0.008) } \\
\emph{U0}   &  -3.425 ($\pm$1.416) &  0.603 ($\pm$0.122) &  0.040 ($\pm$0.013) \\
\emph{Unif} &  -7.957 ($\pm$0.368) &  0.223 ($\pm$0.009) &  0.082 ($\pm$0.004)  \\
\midrule
\midrule
& \multicolumn{3}{c|}{\underline{Sci-Fi}} \\
\midrule
Policy & 
Objective $f$~($\uparrow$) & $p_{\clk}$~($\uparrow$) & $p_\hm$~($\downarrow$)  \\
\midrule
\emph{Grad} &   \textbf{-2.925 ($\pm$1.319)} &  \textbf{0.812 ($\pm$0.056)} &  \textbf{0.037 ($\pm$0.013)}  \\
\emph{Alt}  &   -4.772 ($\pm$2.339) &  0.742 ($\pm$0.094) &  0.055 ($\pm$0.022)  \\
\emph{U0}   &   -4.618 ($\pm$2.108) &  0.749 ($\pm$0.082) &  0.054 ($\pm$0.020) \\
\emph{Unif} &  -12.485 ($\pm$0.473) &  0.385 ($\pm$0.012) &  0.129 ($\pm$0.005) \\
\midrule
\midrule
\end{tabular}
}
\caption{
Key metrics ($f$, $p_\clk$, and $p_\hm$) for all policies under the 5 genre datasets, for  $\alpha_\hm=0.25$, $a_\nh=0.5$, $\beta=0.15$,  $\lambda=100$,  and a dataset-specific $c\in[1,20]$, as reported in \arxiv{Appendix~H in~\cite{arxivversion}}{Appendix~\ref{sec:supp_exp}}, for the bounded cardinality setting with $k=1$. We report means and standard deviations across 100 users. Across all genre datasets, our gradient-based policy achieves superior recommendation objective. Note that, even though we report standard deviations, the performance  improvement also occurs on a per-user basis (see also the PDFs of objective differences, which we report in Fig.~\ref{fig:pdfdiffobj_mini} and \arxiv{Appendix~H in~\cite{arxivversion}}{Appendix~\ref{sec:supp_exp}}). 
Interestingly, across all genres, \emph{Grad} attains both a lower $p_\hm$ and a higher $p_\clk$ than competitors.
}
\label{tab:big_table}
\end{table}

\begin{figure}
\centering
% \begin{minipage}[t]{0.55\linewidth}
\begin{minipage}[t]{0.49\linewidth}
\centering
\includegraphics[width=1.05\linewidth]{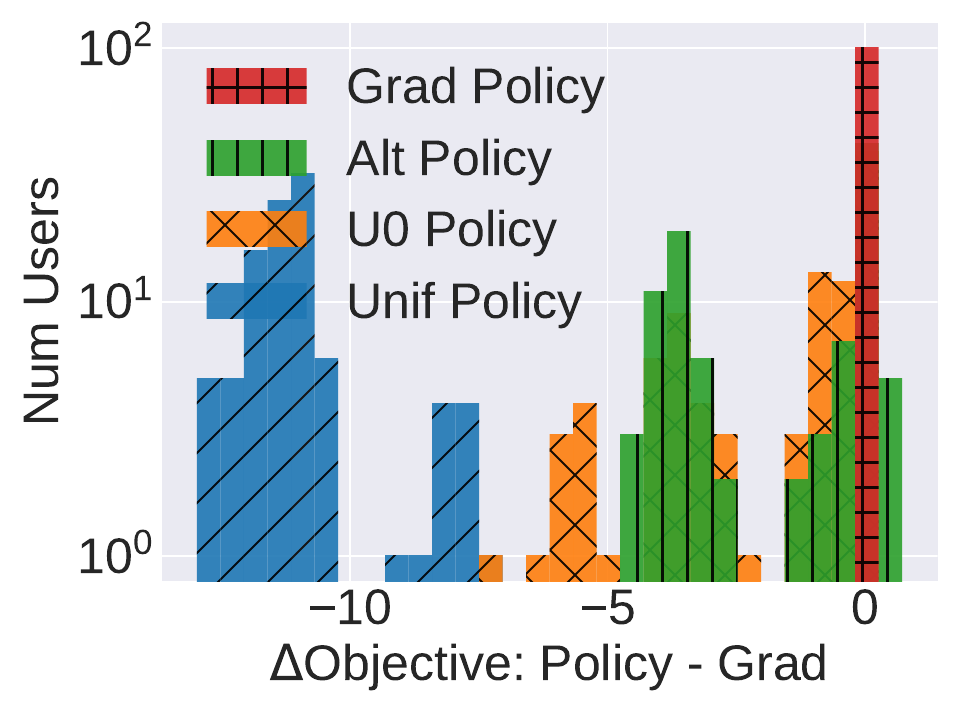}
\vspace{-2.5em}
\caption*{(a) Action}
\end{minipage}
% \begin{minipage}[t]{0.55\linewidth}
\begin{minipage}[t]{0.49\linewidth}
\centering
\includegraphics[width=1.05\linewidth]{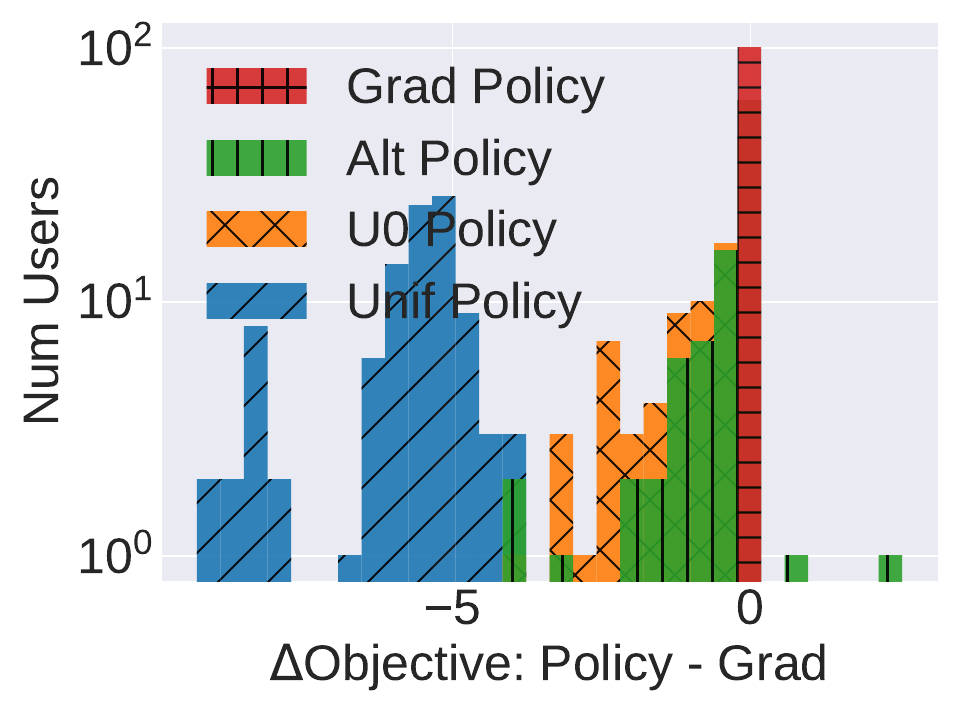
}
\vspace{-2.5em}
\caption*{(b) Adventure}
\end{minipage}
\vspace{-1em}
\caption{
PDF of the per-user difference between the objective attained by different policies minus the objective by attained by \emph{Grad}, with experimental settings as in Table~\ref{tab:big_table}. 
\emph{Grad} dominates other policies in terms of the objective on a per user basis.
Additional genres, and impact on $p_\clk$, $p_\hm$ are shown in the appendix.
} 
\label{fig:pdfdiffobj_mini}
\end{figure}
\begin{figure}
\centering
\begin{minipage}[t]{0.49\linewidth}
\vspace{-1em}
\centering
\includegraphics[width=1.05\linewidth]{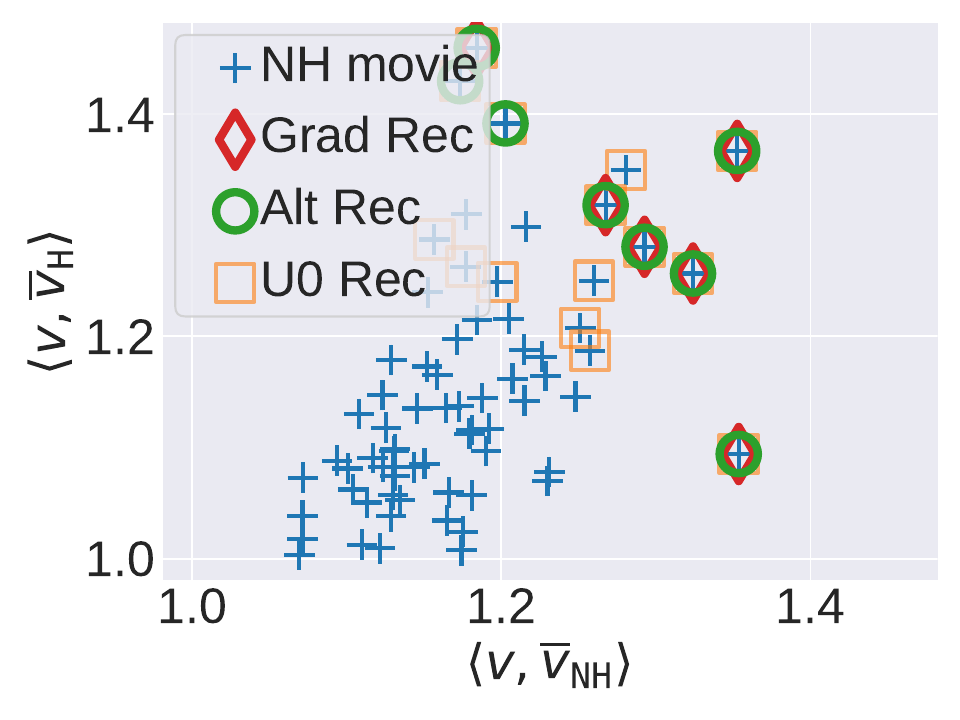}
\vspace{-2.5em}
\caption*{(a) Action}
\end{minipage}
\begin{minipage}[t]{0.49\linewidth}
\vspace{-1em}
\centering
\includegraphics[width=1.05\linewidth]{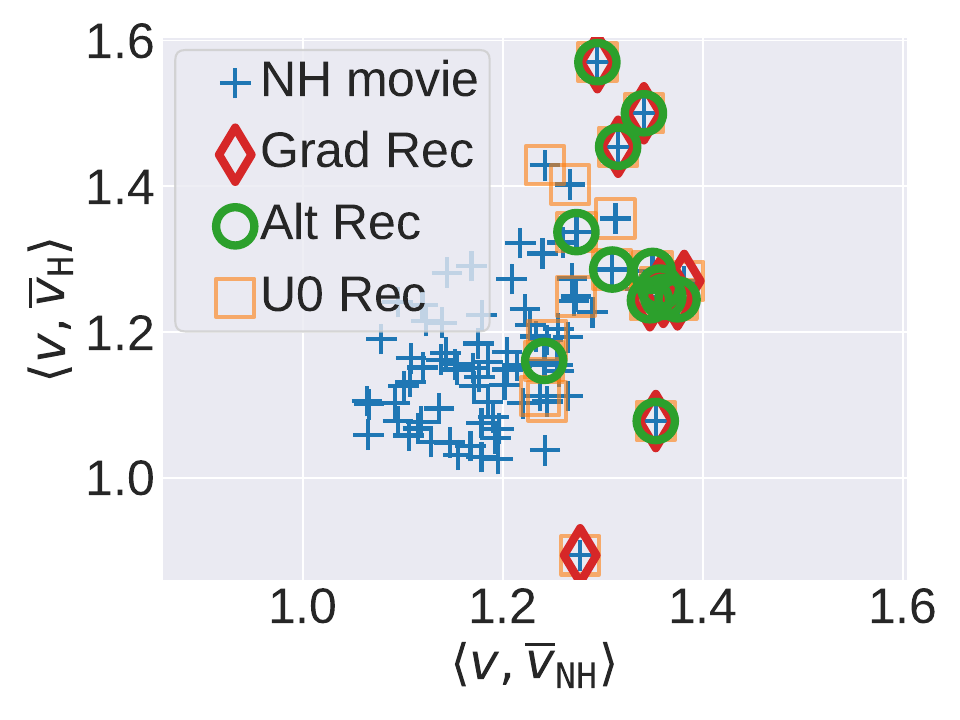}
\vspace{-2.5em}
\caption*{(b) Adventure}
\end{minipage}
\vspace{-1em}
\caption{
Visualization of the movies each policy recommends,  with respect to their inner product with the mean non-harmful and harmful vector, under the setting reported in Table~\ref{tab:big_table}, for a specific user in the Action and Adventure genres. All non-harmful movies are embedded via a $+$ sign in this plane; the support of each policy is indicated by additional symbols (we omit \emph{Unif} as its support is everything). 
We observe that the gradient-based policy is more concentrated towards the right, i.e., on  movies which have high inner product with the average non-harmful vector. Remaining genres are shown in \arxiv{Appendix~H in~\cite{arxivversion}}{Appendix~\ref{sec:supp_exp}}.
}
\label{fig:ip_select_small}
\end{figure}

We  stress   that bounded cardinality setting operations
are efficient only if $m$ is polynomial in $n$; this is the case only when $k$ is fixed (e.g., when $k=3$), and the recommender only recommends a few items to the user.  In contrast, as $m'=\sum_{\candidate\in\collection}|\candidate|$, the independent sampling setting can be polynomial in $n$ \emph{for all $k$}: for example, if $\collection$ is a partition of $\Omega\setminus H$, then $m'=n-h$, which does not depend on $k$.

\section{Experiments}\label{sec:experiments}

\subsection{Experimental Setup}
\noindent\textbf{Dataset}.
We use the  MovieLens25m dataset to extract movie and item profiles and the IMDB parental guideline ratings~\citep{Hay2023,IMDBParent} to determine our definition of harmful movies. The IMDB parental guideline ratings comprise 5 categories: (1) Sex \& Nudity, (2) Violence \& Gore, (3) Profanity, (4) Alcohol, Drugs, \& Smoking, and (5) Frightening \& Intense Scenes.
For these experiments, we consider a movie harmful if it is ``Severe''---the worst parental rating---in any of these categories.
We join the MovieLens25m dataset  with the IMDB data set via the IMDB movie identifiers, and then 
create 5 sub-datasets by filtering containing movies from 5 categories: ``Action'', ``Adventure'', ``Comedy'', ``Fantasy'', and ``Sci-Fi''.
We restrict each dataset to the 100 movies with the most ratings and 100 users, with user and movie profiles of dimension $d=12$ constructed via matrix factorization. 
Learned  profiles play the role of $u_0$ and $v$ in our experiments.
Full details on the this pre-processing pipeline, including hyperparameters, can be found in \arxiv{Appendix~H in \cite{arxivversion}}{Appendix~\ref{sec:supp_exp}}. 
We make our code publicly available.\footnote{\url{https://github.com/jerry-chee/HarmMitigation}} 

\begin{figure*}
\includegraphics[width=0.95\textwidth]{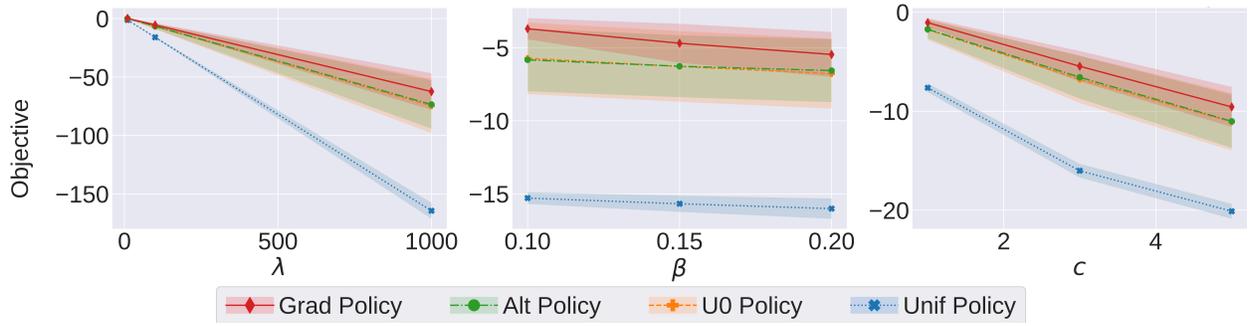}
\caption{Effect of modifying $\lambda$, $\beta$, and $c$ on the objective attained by different policies for the Action genre. Additional genres, and impact on $p_\clk$, $p_\hm$, are shown in \arxiv{Appendix~H in~\cite{arxivversion}}{Appendix~\ref{sec:supp_exp}}. Increasing any parameter decreases the objective attained by every policy.  We observe that increasing $\lambda$ naturally increases the performance gap of the \emph{Grad} policy. Increasing $\beta$ has the opposite effect, as it limits the ability of all policies to impact a user's profile. Parameter $c$ also increases the improvement of \emph{Grad} over other policies as, the larger $c$ is, the less likely the recommendation is to be accepted, and the more important it becomes to succesfully minimize harm. }\label{fig:mod_obj_small}
\end{figure*}

\noindent\textbf{Model Parameters.}\label{sec:modmet}
We use MF user profiles as $u_0$ (inherent profiles), and item profiles $v$ as extracted from the above matrix factorization process. We use these values into our model to compute the steady-state behavior of each of different policies, but also simulate the evolution of the user profiles under these policies.  Unless otherwise noted, we use $\alpha_\hm=0.25$, $a_\nh=0.5$, $\beta=0.15$, and $\lambda=100$ (though we also explore the impact of these parameters).
{Unless otherwise noted, we use $k=1$ under the bounded cardinality setting for most experiments, exploring the impact of $k$ on recommendations separately under the independent sampling setting.}
We calibrate $c$  in function $g$ (Eq.~\eqref{eq:g}) in the range $[1,20]$ to balance the steady-state mean $p_\clk$ and $p_\hm$~( Eq.\eqref{eq:phm}) under a benchmark recommendation policy (the uniform recommendation policy, described below). 
The full set of values of $c$ per dataset, as well as the corresponding justification for these choices, is given in \arxiv{Appendix~H in~\cite{arxivversion}}{Appendix~\ref{sec:supp_exp}}. 

As we use profiles learned from data, rather than user ratings directly, our experiments are semi-synthetic. 
Note that going beyond semi-synthetic experiments poses several challenges which we share with the broader recommender system literature (e.g., \citet{oh2019thompson,oh2021multinomial}):   assessing the performance of different policies would require knowing how a user would react to alternatives they have not seen, which we cannot extract from offline datasets.

\noindent\textbf{Algorithms.}\label{sec:algorithms}
We implement the following algorithms:  \emph{Gradient-Based Algorithm} (\emph{Grad}) is  the SLSQP implementation of SciPy~\citep{2020SciPy-NMeth}, for both the bounded cardinality and the independent sampling setting, using our gradient computation approach as described in Sec.~\ref{sec:main_results} and \arxiv{Appendix~G in~\cite{arxivversion}}{Appendix.~\ref{app:gradientcomp}}; \emph{Alternating Optimization }\textbf{(Alt)} is 
is the (suboptimal) alternating optimization algorithm in Eq.~\eqref{eq:alternate};  \emph{Static Profile Optimization} (\emph{U0}) is the optimal policy under static profile $u_0$, as determined by Thm.~\ref{thm:staticoptimal}; finally, \emph{Uniform Recommendations} (\emph{Unif}) is the policy that selects the recommended set u.a.r. Additional implementation details, including convergence tolerance criteria, are in \arxiv{Appendix~H in~\cite{arxivversion}}{Appendix~\ref{sec:supp_exp}}. We stress here that, as Prob.~\eqref{eq:prob2} is non-convex, it is not a priori clear that \emph{Grad} reaches a globally optimal policy.

\noindent\textbf{Metrics.}
We report the recommendation objective~(Eq.~\eqref{eq:prob2}), as well as its constituent $p_\clk$ and $p_\hm$.
We also measure $\| \lim_{t\to\infty} u(t) - \ulim\|$~(Eq.~\eqref{eq:u},\eqref{eq:fp}), the distance of the user profile dynamics as time evolves under a given policy and the corresponding fixed point.

\subsection{Results}

\noindent\textbf{Recommendation Policy Comparison.}
Table~\ref{tab:big_table} reports our key metrics over the 5 movie genre data sets, averaged across the 100 users in the dataset.
Across all genre datasets, \emph{Grad} attains superior recommendation objective values, while the improvement over uniform is as much as 77\% (for Sci-Fi).
Surprisingly,
\emph{our policy does not need to sacrifice $p_\clk$ to attain low $p_\hm$} in comparison to other policies; it is superior in both terms. 
This is  because it recommends an item (a) users become more likely to click, in steady state, and (b) is distinct enough from harmful content, so that $p_\hm$ remains low. 

We note that, even though we report standard deviations, \emph{Grad} dominates other policies in terms of the objective also on a per user basis: this is evident {in Fig.~\ref{fig:pdfdiffobj_mini},}  where we report also the PDFs of the difference of objectives from the one attained by \emph{Grad} (additional genres are in the appendix). In Fig.~\ref{fig:ip_select_small}, we visualize the which movies each policy recommends for  two users in the Action and Adventure genres. We display each vector $v$ with respect to their inner product with the mean non-harmful and harmful vector. 
We observe that the gradient-based policy is more concentrated towards the right, i.e., on  movies which have high inner product with the average non-harmful vector. Similar observations hold for remaining genres (see \arxiv{Fig.~7 in Appendix~H in~\cite{arxivversion}}{Fig.~\ref{fig:ip_all} in Appendix~\ref{sec:supp_exp}}).

\noindent\textbf{Understanding the Effect of Model Parameters.}
We plot the effect of modifying $\lambda$, $\beta$, and $c$ on the objective attained by different policies for the Action genre in Fig.~\ref{fig:mod_obj_small}; additional genres, as well as the effect these parameters have on $p_\clk$ and $p_\hm$, are shown in \arxiv{Figs.~11--13
in Appendix H in \cite{arxivversion}}{Figs.~\ref{fig:mod_obj_all}--\ref{fig:mod_phm_all} in Appendix~\ref{sec:supp_exp}}. Increasing any parameter decreases the objective attained by every policy, which is intuitive.  We also observe that increasing $\lambda$ naturally increases the performance gap of the \emph{Grad} policy. Increasing $\beta$ has the opposite effect, as it limits the ability of all policies to impact a user's profile. Parameter $c$ also increases the improvement of \emph{Grad} over other policies as, the larger $c$ is, the less likely the recommendation is to be accepted, and the more important it becomes to minimize harm. 

\begin{figure}
\centering
\begin{minipage}[t]{0.55\linewidth}
\centering
\includegraphics[width=1.05\linewidth]{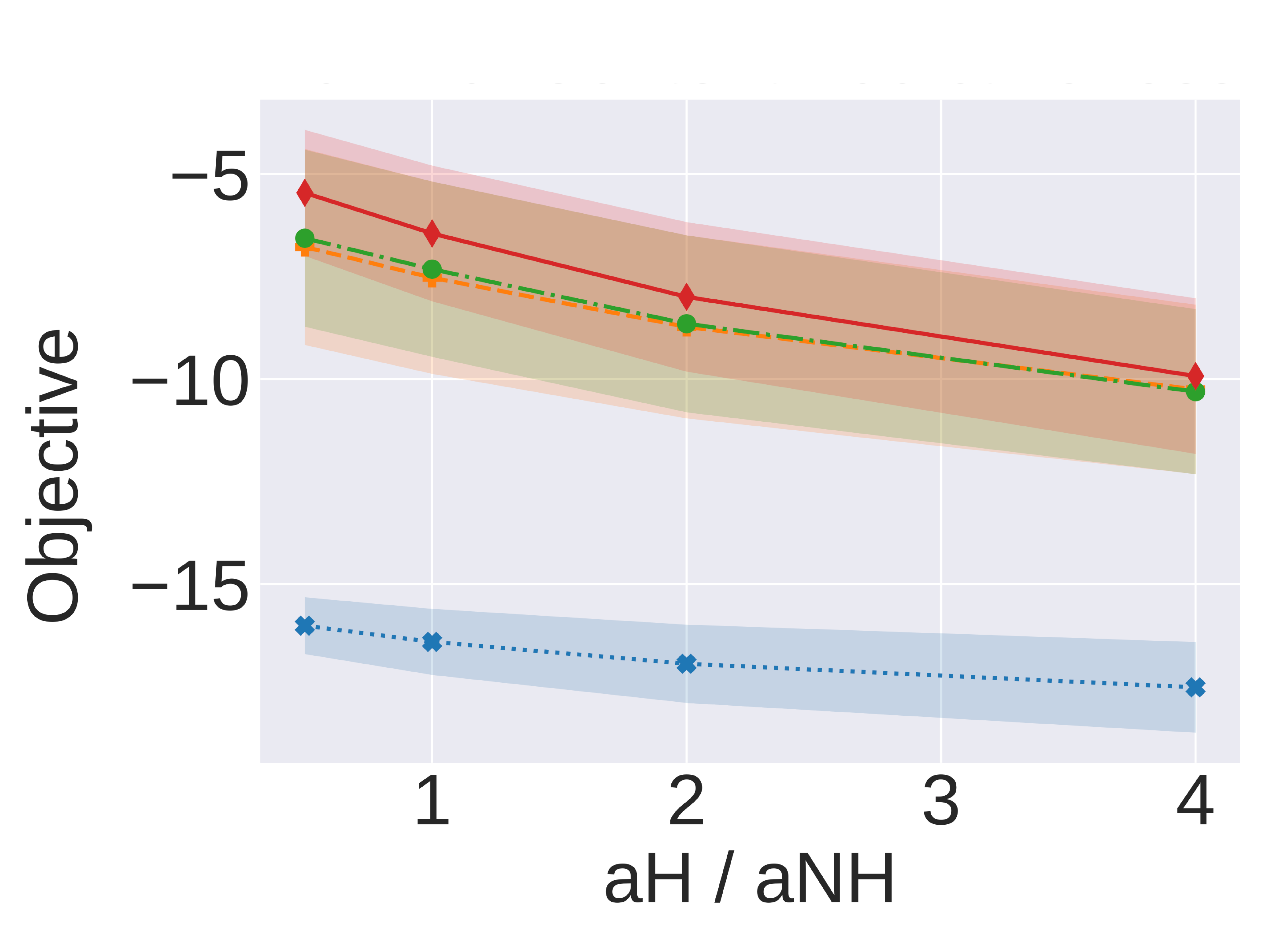}
\vspace{-2.5em}
\caption*{(a) Action}
\end{minipage}
\begin{minipage}[t]{0.55\linewidth}
\centering
\includegraphics[width=1.05\linewidth]{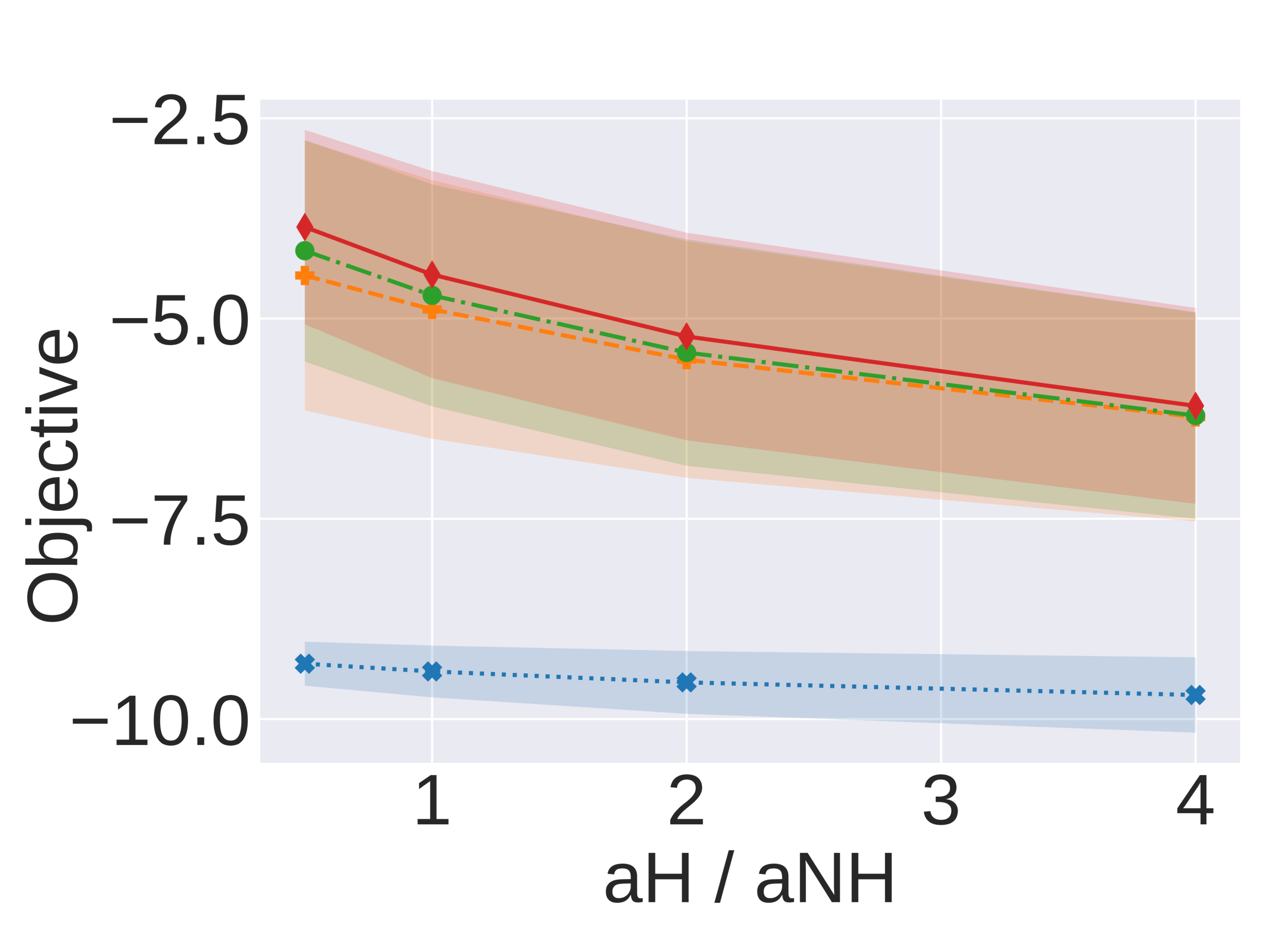}
\vspace{-2.5em}
\caption*{(b) Adventure}
\end{minipage}
\vspace{-1em}
\caption{
Effect of modifying ratio $\alpha_\hm/\alpha_\nh$ on the objective attained by different policies for the Action and Adventure genres. Additional genres, and impact on $p_\clk$, $p_\hm$, are shown in \arxiv{Appendix~H in~\cite{arxivversion}}{Appendix~\ref{sec:supp_exp}}. Increasing the ratio decreases the objective attained by every policy as well as the gap from \emph{Grad}, as policies have less leeway in minimizing harm. 
}
\vspace{-0.5em}
\label{fig:ahanh_small}
\end{figure}

In Fig.~\ref{fig:ahanh_small}, we also plot the impact of the  ratio $\alpha_\hm/\alpha_\nh$ on the objective attained by different policies for the Action and Adventure genres. Increasing the ratio decreases the objective attained by every policy as well as the gap from \emph{Grad}, as policies have less leeway in minimizing harm. Similar observations hold for other genres (see \arxiv{Figs.~14--16
in Appendix H in \cite{arxivversion}}{Figs.~\ref{fig:ahanh_all}--\ref{fig:ahanh_phm_all} in Appendix~\ref{sec:supp_exp}}).

\begin{table}[]
\centering
\begin{tabular}{c|ccc}
 & \multicolumn{3}{c}{$\|\lim u(t) - \ulim\|$} \\
\midrule
Policy & Action & Adventure & Comedy \\
\midrule
Grad & 6.8e-4 ($\pm$4.2e-4) & 7.3e-4 ($\pm$4.2e-4) & 7.1e-4 ($\pm$4.7e-4) \\
Alt  & 3.6e-4 ($\pm$2.0e-4) & 6.1e-4 ($\pm$4.1e-4) & 7.3e-4 ($\pm$4.9e-4) \\
U0   & 3.5e-4 ($\pm$1.9e-4) & 4.2e-4 ($\pm$3.3e-4) & 5.3e-4 ($\pm$4.5e-4) \\
Unif & 3.7e-4 ($\pm$1.1e-4) & 3.1e-4 ($\pm$1.1e-4) & 3.4e-4 ($\pm$1.4e-4) \\
\midrule
\midrule
Policy & Fantasy & Sci-Fi & \\
\midrule
Grad & 4.7e-5 ($\pm$7.0e-5) & 4.2e-4 ($\pm$3.1e-4) & \\
Alt  & 4.7e-5 ($\pm$7.0e-5) & 2.2e-4 ($\pm$1.4e-4) & \\
U0   & 1.7e-4 ($\pm$1.3e-4) & 2.5e-4 ($\pm$1.6e-4) & \\
Unif & 4.3e-4 ($\pm$1.6e-4) & 3.6e-4 ($\pm$1.2e-4) & \\
\midrule
\midrule
\end{tabular}
\caption{
The user dynamics converge to the fixed-point for all policies and all genres; $\|\lim u(t) - \ulim\|$ is small and on par with the convergence threshold.
We report the mean$\pm$std.
}
\vspace{-1em}
\label{tab:big_table2}
\end{table}

\noindent\textbf{Convergence.} For all datasets, policies, and users in the experiments reported in Table~\ref{tab:big_table}, we also compute the user profile dynamics in Eq.~\eqref{eq:u}. In all experiments, they converge to the corresponding stationary user profile in Lemma~\ref{lem:fix}, within a $10^{-3}$ tolerance; see Table~\ref{tab:big_table2} for per genre tolerances.

\begin{figure}
\begin{minipage}[t]{0.65\linewidth}
\centering
\includegraphics[width=1.0\linewidth]{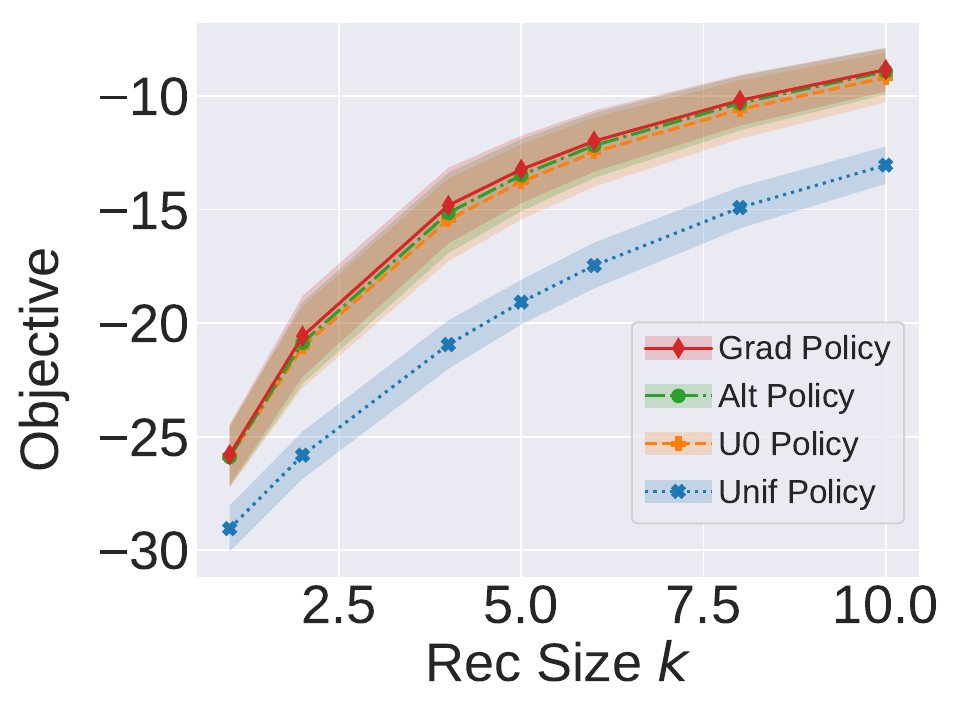}
\vspace{-2.5em}
\caption*{(a) Fixed $c \approx$ Varying $p_{\clk}$}
\end{minipage}
\begin{minipage}[t]{0.65\linewidth}
\centering
\includegraphics[width=1.0\linewidth]{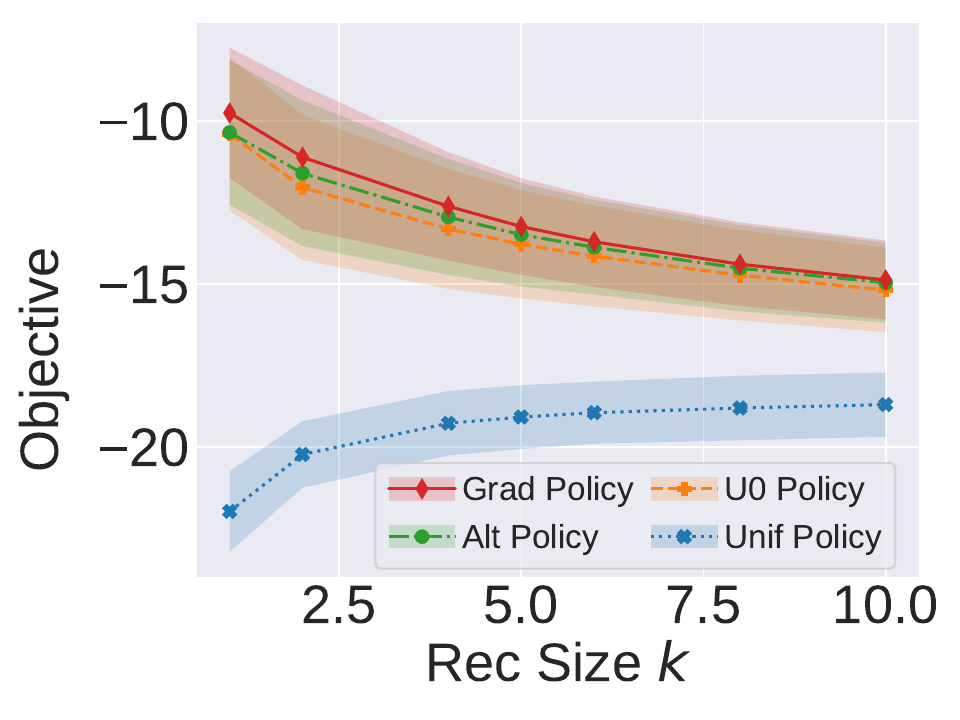}
\vspace{-2.5em}
\caption*{(b) Fixed $k/c \approx$ Fixed $p_{\clk}$}
\end{minipage}
\vspace{-1em}
\caption{
Effect of increasing the recommendation set size $k$ on the objective attained by different policies for the Action genre.
Impact on $p_\clk$ and $p_\hm$ are show in \arxiv{Appendix~H in~\cite{arxivversion}}{Appendix~\ref{sec:supp_exp}}
(a) We first increase the number of recommended items $k$. 
From the MNL model in Eq.~\eqref{eq:clk} we see that $S_E$ can increase with larger $k$, effectively increasing $p_\clk$-and thus the objective-for all policies. 
This trend is observed when plotting $p_\clk$ (\arxiv{Appendix~H in~\cite{arxivversion}}{Appendix~\ref{sec:supp_exp}}). 
(b): we keep the ratio $k/c$ fixed in order to counteract the effect of rising $p_\clk$ upon increasing $k$. 
The gradient-computed policy is superior over a variety of $k$.
}\label{fig:mod_k_obj}
\end{figure}

\noindent\textbf{Increasing Recommendation Size under the Independent Sampling Setting.}
Next, we study the scalability of our algorithms by  increasing the recommendation size $k$, under the independent sampling setting.
In Fig.~\ref{fig:mod_k_obj} we plot the impact of $k$ on the objective for the Action genre, under two different scenarios: in one, we vary $c$ along with $k$, while in the other we keep it fixed.
Intuitively, as $k$ increases, $p_\clk$ converges to one, and $p_\hm$ collapses to zero. Thus, to better  understand the performance of different policies as $k$ increases we make the optimization ``harder'' as we increase $k$. To accomplish this,   
recall from Sec.~\ref{sec:model}  the total score of a recommendes set $E \subseteq \Omega$ is $S_E \equiv \sum_{v \in E} s_v$, and that the conditional probability $p_{\clk|E} = g(s_E) \equiv s_E / (s_E + c)$. Assuming $S_E$ grows proportionally with $k$, setting $c$ to also grow proportionally with $k$ keeps $p_{\clk|E}$ relatively constant, allowing us to better understand how increasing $k$ effects harm.
In Fig.~\ref{fig:mod_k_obj}(a), when $c$ is fixed, we see that all policies improve (as expected) w.r.t. the overal objective, with \emph{Grad}  dominating other policies.   In Fig.~\ref{fig:mod_k_obj}(b), when $c$ is proportional to $k$, the problem becomes harder and the objective decreases, with \emph{Grad} still being dominant; this is confirmed when observing the impact on $p_\clk$ and $p_\hm$, reported in \arxiv{Appendix~H in~\cite{arxivversion}}{Appendix~\ref{sec:supp_exp}}.

\section{Conclusions}\label{sec:conclusions}

Overall, our goal is to gain a better understanding of real-world phenomena through a simplified but insightful model.
We study a model in which the likelihood of consuming harmful content, though never directly the result of a recommendation, can nonetheless be influenced by a recommendation policy through its impact on user preferences.
The notion of harm that we study is simple: content is known to be either harmful or non-harmful, and recommenders will never recommend harmful content.
Despite this, considerable complexity arises from the user behavior model which is influenced on both short (via clicks) and long (via interests) timescales by the choices of the recommender.
We develop algorithms for designing recommenders which simultaneously maximize a CTR objective while minimizing the likelihood of harm and verify their performance on a setting initialized with real movie rating data.
We present compelling evidence that a user pathway to harm should be accounted for; this is missing in both literature and practice.
We rigorously substantiate that solely maximizing the click-through-rate may come at the cost of shifting user preferences towards harm.

Our focus on the long term impacts of recommendations leads to conceptions of harm that are both mathematically challenging and conceptually rich.
Many opportunities for future work remain in the setting that we study, for example: under what conditions are user profiles guaranteed to converge to a fixed point?
Is it possible to make guarantees about tractable algorithms for optimizing our non-convex objective?
Many open questions arise from relaxing our assumptions, like:
what if users interests evolve according to something other than attraction (e.g. repulsion~\citep{lu2014optimal}, biased assimilation~\citep{dean2022preference}).
What if user and item profiles are not directly observed, and must be learned from data?
Finally, we hope that our perspective, centered on the dynamics of users, can contribute to the rich literature on harms caused by algorithmic systems and inspire effective methods for mitigation.

\begin{acks}
This work was partly funded by NSF CCF-2312774, NSF OAC-2311521, and NSF DGE-1922551.
\end{acks}
\newpage

\bibliographystyle{abbrvnat}\bibliography{references}
\newpage
\onecolumn
\arxiv{}{
\appendix

\section{Recommendation Policies}\label{app:recpolicies}

We denote by $p_\candidate=\prob(\candidate_t=\candidate)\in[0,1],$ 
the probability that the candidate set $\candidate_t$ is $\candidate \in \collection$  at time $t$.
We denote by $p_{E\mid \candidate}\in[0,1],$ for $E\subseteq \Omega\setminus H,$ 
the probability that set $E$ is recommended to the user, given that the candidate set is $\candidate$.   
 We refer to the conditional probabilities $\pi=[p_{E\mid \candidate}]_{C\subseteq \Omega \setminus H, E\subseteq \candidate}$ as the \emph{recommendation policy}, and assume that, conditioned on $\candidate$, $E$ is sampled from $\pi$ independently at each timeslot. We denote by $\allprobs$ the set of all possible recommendation policies.  We denote by $\probset\subset \allprobs$  the set of \emph{valid}  policies, that satisfy desired constraints; 
 we  describe several examples below. 

\noindent\textbf{Bounded Cardinality.} In this setting, only sets $E$ of cardinality $|E|\leq k$ may be recommended, for some  $k\in \naturals$. Formally, given $\candidate\in\collection$, let
\begin{align}
\constrset_\candidate=\{E\subseteq \candidate: |E|\leq k\},\label{eq:k-constr}\end{align}
be  the set of  subsets of cardinality $k$. Then, a valid recommendation policy is a probability distribution over these sets, i.e.:
 \begin{align}
 \label{eq:Pi}
 \begin{split}\probset &=\big\{\pi\in\allprobs: \textstyle\sum_{E\in \constrset_\candidate} p_{E\mid\candidate} =1, \text{ for all}~\candidate\in \collection\big\}.\end{split}\end{align}
\noindent\textbf{Independent Sampling.}
In this setting, we assume that, given the candidate set $\candidate\in\collection$, the recommended items are selected independently with a fixed probability. Formally, the probability that the recommended set is $E$ is given by: 
\begin{align}
    p_{E\mid \candidate} =\textstyle \prod_{v\in E}\rho_{v\mid C}\prod_{v'\notin E} (1-\rho_{v'\mid C}), \quad\text{for all}~E\subseteq \candidate, \label{eq:prodform}
\end{align}
i.e., the events $\{v\in E\}$ are independent Bernoulli random variables parameterized by   probabilities $\rho_{v\mid C}=\prob(v\in E\mid C)$, $v\in \candidate$. We also require that, conditioned on $C$, the expected size of the recommended set $E$ is at most $k\in \naturals$, i.e.:
\begin{align}\expect_E[|E|\mid C]&\leq k, ~\text{or, equivalenty},~\textstyle\sum_{v\in \candidate}\rho_{v\mid \candidate} \leq k,~\text{for all}~\candidate\in\collection.
\label{eq:R}\end{align}
The set $\probset$ comprises all distributions of form~\eqref{eq:prodform} for which $\rho$ satisfies Eq.~\eqref{eq:R}.

Bounded cardinality policies  ensure that the cardinality of recommendations $|E|$ is at most $k$, but describing them requires $m\equiv \sum_{\candidate\in\collection}{|\candidate|\choose k}$ parameters; this is practical only for small $k$.  Independent sampling policies require $m'\equiv \sum_{\candidate\in\collection}|\candidate|$ parameters. This polynomial in the number of in $n$ if, e.g., $\collection$ is a partition of $\Omega\setminus H$. However, independent sampling constrains cardinality  only in expectation.

\noindent\textbf{Top-$k$ Recommendations.} A simple, intuitive deterministic policy is the \emph{top-$k$ recommendations} policy \citep{karypis2001evaluation,deshpande2004item,cremonesi2010performance}. Given $C$ and budget $k$, the policy recommends the $k$ top-scoring items $v\in C$, w.r.t.~some scores  $s_v\in \reals_+$ quantifying a user's preference toward $v$,

Formally, given $\candidate$ and $k$, the top-$k$ recommendation is given by:  \begin{align}E^*_\candidate =\textstyle\argmax_{E\in \constrset_\candidate} s_E,\label{eq:topkrec2}\end{align}
be the highest-scoring subset of $C$ of size at most $k$. As $s_E$ is a modular function, the greedy algorithm that starts from an empty set and iteratively traverses $\candidate$ and adds the highest-scoring items in the solution is in fact optimal, and produces $E^*_\candidate$ in polynomial time.

We note that this deterministic policy is included in $\probset$, when the latter is given either bounded cardinality and independent sampling settings. In the bounded cardinality setting, the top-k policy is captured by the Dirac distribution on $E^*$, i.e., $\pi_{E^*}=1$ and 0 everywhere else. In the independent sampling case, it is captured by policy in which $\rho_{v\mid \candidate}=1$ for all $v\in E^*$, and $\rho_{v\mid \candidate}=0$ for all $v\not\in E^*$. Conversely,  an optimization over policies $\pi\in \probset$, with $\probset$ defined by Eqs.~\eqref{eq:Pi} or~~\eqref{eq:prodform}--\eqref{eq:R}, the top-$k$ policy is included as a valid policy.

\section{Proof of Thm.~\ref{thm:staticoptimal}} \label{app:proofofthm:staticoptimal}
Observe that by Eq.~\eqref{eq:phm}
the objective is given by
$$f_0(\pi) = \expect_{E,\candidate}\left[g(s_E)\right]\left(
1+\frac{\lambda s_H}{s_\catalog}\right) -\lambda \frac{s_H}{s_\catalog}.$$
As the remaining quantities do not depend on $\pi$,  to maximize $f_0$, it suffices to maximize \begin{align}  f_0'(\pi)\equiv \expect_{E,\candidate}\left[g(s_E)\right]=\textstyle\sum_{C}\sum_{E\subseteq C} g(s_E) p_{E\mid \candidate}p_\candidate,\end{align}
which is exactly $p_\clk$.  In other words, for any $\lambda\geq 0$, \emph{the optimal policy $\pi\in \probset$ is the one that minimizes the click-through rate, as captured by $p_\clk$.}

Consider first the case where $\probset$ is given  by Eq.~\eqref{eq:Pi} (i.e., the bounded cardinality setting).
Given a set $\candidate\in \collection$, let
\begin{align}
f'_{0,C}(\pi_C) \equiv \expect_{E}[g(s_E)\mid C] =  \textstyle \sum_{E\subseteq C} g(s_E) p_{E\mid \candidate}
\end{align}
be the click probability conditioned on the candidate set being $C$, where $\pi_C=[p_{E|\candidate}]_{E\subseteq E}$ is the recommendation policy conditioned on the candidate set being $\candidate$. Observe that, as $p_\candidate\geq 0$ for all $\candidate\in \collection$, maximizing $f_0'(\pi)$ over $\pi\in\probset$ is \emph{separable}: the optimal solution can be computed by solving $|\collection|$ independent problems of the form:
\begin{subequations}\begin{align}\text{Maximize}: &\quad f'_{0,C}(\pi_C) = \textstyle \sum_{E\subseteq C} g(s_E) p_{E\mid \candidate} \\
\text{subj.~to}:&\quad \textstyle\sum_{E\in \constrset_\candidate} p_{E\mid C}=1. \label{eq:sepconstr2}
\end{align}\end{subequations}Each of these problems is a linear program. 
Hence, by the fundamental theorem of linear programming, there exists an optimal solution that is an extremum of these constraints in Eq.~\eqref{eq:sepconstr}, which is a Dirac distribution on a set $E\subseteq \candidate$.  The monotonicity of $g(\cdot)$, as a function of $E$ implies that the largest value is attained at the $E\in \constrset_\candidate$ with the largest $s_E$, namely $E^*=\textstyle\argmax_{E\in \constrset} s_E$, and the theorem follows.

Consider next the case where $\probset$ is given  by Eqs.~\eqref{eq:prodform}--\eqref{eq:R} (i.e., the independent sampling setting). Assume now that, in addition to non-decreasing, $g$ is also concave.
Observe that, as a result, (i) as a function of set $E$, function $s_E$ is a modular function with non-negative weights $s_v$, $v\in E$, and (ii) $g(s_E)$ is a monotone (non-decreasing) submodular function of set $E$. 

For the same reasons as above, the optimization is again separable over parameters $\rho_C= [\rho_{v\mid C}]_{v\in C}$ conditioned on $C$, and reduces to $|\collection|$ independent problems of the form:
\begin{subequations}\begin{align}\text{Maximize}: &\quad f'_{0,C}(\rho_C) =  \sum_{E\subseteq C} g(s_E) \ \prod_{v\in E}\rho_{v\mid C}\prod_{v'\notin E} (1-\rho_{v'\mid C}) \\
\text{subj.~to}:&\quad \sum_{v\in \candidate} \rho_{v\mid C}\leq k. \label{eq:sepconstr}
\end{align}\end{subequations}Then, $f'_{0,C}$ is the so-called  multilinear relaxation \citep{calinescu2011maximizing} of submodular function $g(s_E)$. As a result,  is $\epsilon$-convex in any pair of its coordinates, and for any fractional solution $\rho\in R$ the exists an integral $\rho' \in R$ s.t. $f'_{0,C}(\rho'_C)\geq f'_{0,C}(\rho)$\citep{ageev2004pipage}; moreover, $\rho'_C$ can be constructed via pipage rounding~\citep{ageev2004pipage}. Hence, there exists an optimal solution $\rho_C$ that is integral. As its value $f'_{0,C}(\rho_C)$ cannot exceed $f'_{0,C}(s_{E^*_C})$, and $E^*_C$ corresponds to a feasible  $\rho_C$ (that has $\rho_v^*=1$ for every $v\in E^*_C$), $\rho^*$ is optimal.   \qed

\section{Proof of Lemma~\ref{lem:fix}}\label{app:proofoflem:fix}
Observe that, setting  $u=u(t)$, by Eq.~\eqref{eq:deltau} we have
\begin{align*}\Delta u &=  \expect\left[\alpha_{v(t)}\left(v(t)-u\right)\right] + \beta\left(u_0 - u\right)\\
&=\left(\alpha_\hm\sum_{v\in H}p_v(v-u)+\alpha_\nh\sum_{v\notin H}p_v(v-u)\right) + \beta\left(u_0 - u\right) 
\end{align*}
The lemma follows by setting $\Delta u =0$ and solving w.r.t.~$u$. The formula for $p_v$, the probability that the user selects item  $v\in \Omega$, follows from Eqs.\eqref{eq:gfun}--\eqref{eq:pl}. In particular:
\begin{align}\label{eq:pv}
p_v = \expect_{E,\candidate}[p_{v\mid E}]=\textstyle\sum_{\candidate \in \collection}\sum_{E\subseteq \candidate} p_{v\mid E}p_{E\mid \candidate}p_{\candidate},
\end{align}
where \begin{align}\label{eq:pve}
    p_{v\mid E} &= \textstyle \begin{cases}
    \frac{s_v}{s_E}g(s_E)+\frac{s_v}{s_\catalog}(1-g(s_E)), & \text{for }v\in E,\\
    \frac{s_v}{s_\catalog}(1-g(s_E)),& \text{for }v\notin E.
    \end{cases}\end{align}
\qed

\section{Multinomial Logit (MNL) Model.}\label{app:mnl} 
We give here a more detailed description of the standard multinomial logit  (MNL). MNL special case of the Plackett-Luce model, 
and it is popular in modeling user choices in recommender, search engine, and ad display settings \citep{danaf2019online, chaptini2005use, yang2011collaborative}. This is precisely because it has several useful and natural properties. First, it is a natural generalization of matrix factorization \citep{koren2009matrix} from ratings to choices, as scores depend on the inner product $v^\top u$  (i.e., how well item and user profiles are ``aligned''). It is also a natural extension of ``soft-max'', allowing for a no-choice alternative. We elaborate further in these connections below. 

Formally, under MNL, for every $v\in \catalog$, the  non-negative score $s_v\in \reals^+$ quantifying a user's preference toward $v$  is \begin{align}
    s_v \equiv e^{v^\top u}, \label{eq:score2}
\end{align} i.e., it is  parameterized by both the item profile $v\in \reals^d$ as well as by a \emph{user profile} $u\in\reals^d$.
Moreover,  the conditional probability $p_{\clk\mid E}$ is given: by \begin{align}
    g(S_E)  \equiv\frac{s_E}{s_E+c}~\label{eq:clk2}.
\end{align}
where $c\geq 0$ is a non-negative constant. The remaining probabilities are given again by the Plackett-Luce model (Eq.~\eqref{eq:pl}), with scores determined by \eqref{eq:score}. 

In particular, conditioned on recommendation $E$, the probability that the user selects (i.e., clicks) on item $v\in E$ is given by:
\begin{align}
    p_{v\land \clk\mid E} =\frac{s_v}{s_E+c}, \quad\text{for }v\in E, \label{eq:pclkvsel}
\end{align}
where $c\geq 0$ is a non-negative constant.
In other words, the probability that $v\in E$ is selected is proportional to $s_v\in \reals^+$ and is, in effect, a ``soft-max'' over inner products $v^\top u$, $v\in E$, with $c$ controling the no-choice alternative.

\section{Proof of Thm.~\ref{thm:contraction}}\label{app:proofofthm:contraction}
To show that map $F:\allprobs\times \reals^d$ given by  Eq.~\eqref{eq:fp} is a contraction w.r.t.~$\|\cdot\|$ uniformly on all distributions $\pi\in \allprobs$, we need to show that, for all distributions $\pi\in \allprobs$ (not necessarily in $\probset$), and all  $u,u' \in \reals^{d}$,
   $ \|F(\pi, u ) -F(\pi,u')\|\leq L \|u-u'\|, $ 
for some $L<1$ that does not depend on $\pi$.

Let $\delta = \max_{v\in \Omega}\|v\|. $
Dropping the dependence on $\pi$ for simplicity, and focusing on $F$ as a function of $u$, observe that 
\begin{align}F = G(p(u))\label{eq:gp}\end{align}
where $p:\reals^d \to [0,1]^n$ is the map that produces the probabilities $p_v$, given by Eq.~\eqref{eq:pv}, and 
$G:[0,1]^n\to \reals^d$ is the map:
\begin{align}
    G(p) =  \frac{\beta u_0+\alpha_\hm\sum_{v\in H}p_vv + \alpha_\nh \sum_{v\notin H}p_vv } {\beta+\alpha_\hm\sum_{v\in H}p_v+\alpha_\nh\sum_{v\notin H}p_v} = \frac{\beta u_0+\sum_{v\in \Omega} \alpha_vp_v\cdot v }{\beta+\sum_{v\in \Omega} a_vp_v}.\label{eq:g}
\end{align}
We will show that both $G$ and $p$ are contractions with and, hence, so is $F$. The Jacobian $\nabla G$ of $G$ contains the following elements:
\begin{align}
    \frac{\partial G_i(p)}{\partial p_v} &=\frac{\partial}{\partial p_v}\left( \frac{\beta u_0+\sum_{v'\in \catalog} \alpha_{v'}p_{v'} v_i' }{\beta+\sum_{v'\in \catalog} \alpha_{v'}p_{v'}}\right)\nonumber\\
    & =  \frac{ \alpha_{v} v_i \cdot (\beta+\sum_{v'\in \catalog} \alpha_{v'}p_{v'}) - (\beta u_{0i}+\sum_{v'\in \catalog} \alpha_{v'}p_{v'} v_i') \alpha_{v} }{(\beta+\sum_{v'\in \catalog} \alpha_{v'}p_{v'})^2}\nonumber\\
    & = \frac{ \alpha_{v} \sum_{v'\in \catalog}(v_i-v_i') \alpha_{v'}p_{v'} + \alpha_v \beta (v_i - u_{0i}) }{(\beta+\sum_{v'\in \catalog} \alpha_{v'}p_{v'})^2} \label{eq:dg}
\end{align}
We have
\begin{align*}
\left|\frac{\alpha_v}{\beta+\sum_{v'\in \catalog} \alpha_{v'}p_{v'}}\right|\leq \frac{\alpha_\hm}{\alpha_\nh+\beta} ,
\end{align*}
while
\begin{align*}
\left|\frac{\sum_{v'\in \catalog}(v_i-v_i') \alpha_{v'}p_{v'}}{\beta+\sum_{v'\in \catalog} \alpha_{v'}p_{v'}}\right| \leq  \frac{\sum_{v'\in \catalog}\left|v_i-v_i'\right| \alpha_{v'}p_{v'}}{\beta+\sum_{v'\in \catalog} \alpha_{v'}p_{v'}}\leq 2\delta ,
\end{align*}
and
\begin{align*}
\left|\frac{  \beta (v_i - u_{0i}) }{(\beta+\sum_{v'\in \catalog} \alpha_{v'}p_{v'})}\right| \leq \delta +\|u_0\|
\end{align*}
so $\left|\frac{\partial G_i(p)}{\partial p_v}\right|\leq \frac{(3\delta+\|u_0\|) \alpha_\hm}{\alpha_\nh+\beta} $.
We thus conclude that 
\begin{align}
\|G(p)-G(p')\|\leq \frac{(3\delta+\|u_0\|)\alpha_H\sqrt{nd} }{\alpha_\nh+\beta} \|p-p'\|, \quad \text{for all}~p,p'\in [0,1]^n.
\end{align}

To see that the map $p$ is also Lipschitz, observe that, by \eqref{eq:pv}
{
\begin{align}
p_v = \expect_C[\expect_\pi[p_{v\mid E}]]
\end{align}
}
where
\begin{align}
    p_{v\mid E} &= \begin{cases}
    \frac{s_v}{s_E}g(s_E)+\frac{s_v}{s_\catalog}(1-g(s_E)), & \text{for }v\in E,\\
    \frac{s_v}{s_\catalog}(1-g(s_E)),& \text{for }v\notin E.
    \end{cases}\\
    & \stackrel{}{=}\begin{cases}\frac{s_v}{\sum_{v'\in E}s_{v'} +c} +  \frac{c s_v}{(\sum_{v'\in E}s_{v'}+c)(\sum_{v'\in \catalog} s_{v'})},& \text{if}~v\in E,\\  \frac{c s_v}{(\sum_{v'\in E}s_{v'}+c)(\sum_{v'\in \catalog} s_{v'})}, &\text{o.w.}\end{cases}\label{eq:pvfull}
\end{align}

From Eq.~\eqref{eq:score}, we have that for all $v\in \catalog$ and all $1\leq i\leq d$:
$$\frac{\partial s_v}{\partial u_i} = v_i s_v. $$
Hence, 
\begin{align}\label{eq:dsv1}
   \left| \frac{\partial }{\partial u_i}\left(\frac{s_v}{\sum_{v'\in E}s_{v'} +c}\right)\right| &= \left|\frac{v_is_v\cdot (\sum_{v'\in E}s_{v'} +c) - s_v\cdot( \sum_{v'\in E}v'_is_{v'})}{(\sum_{v'\in E}s_{v'} +c)^2}\right| \\
    & =\left|\frac{s_v\cdot (\sum_{v'\in E}(v_i-v_i')s_{v'} +cv_i) }{(\sum_{v'\in E}s_{v'} +c)^2}\right|\leq 2\delta,
\end{align}
while 
\begin{align}\label{eq:dsv2}
   \left| \frac{\partial }{\partial u_i}\left(\frac{c}{\sum_{v'\in \catalog}s_{v'}}\right)\right| &= \left|\frac{-c\cdot( \sum_{v'\in \catalog}v'_is_{v'})}{(\sum_{v'\in \catalog}s_{v'})^2}\right| \leq \delta,
\end{align}
and, hence
\begin{align}\label{eq:dsv3}
     \left| \frac{\partial }{\partial u_i}\left(\frac{s_v}{\sum_{v'\in E}s_{v'}+c} \cdot  \frac{c}{\sum_{v'\in \catalog} s_{v'}}\right)\right| &\leq \left|\frac{c}{\sum_{v'\in \catalog} s_{v'}}\cdot \frac{\partial }{\partial u_i}\left(\frac{s_v}{\sum_{v'\in E}s_{v'} +c}\right) \right|\\
     &+\left| \frac{s_v}{\sum_{v'\in E}s_{v'} +c}\cdot\frac{\partial }{\partial u_i}\left(\frac{c}{\sum_{v'\in \catalog} s_{v'}}\right) \right|\\
     &\leq 3\delta.
\end{align}
Hence, 
\begin{align*}
   \left| \frac{\partial p_v}{\partial u_i} \right|\leq 5\delta
\end{align*}
and
\begin{align}
    \|p(u)-p(u')\|\leq 5\delta\sqrt{nd}\cdot\|u-u'\|,\quad \text{for all}~u,u'\in\reals^d.
\end{align}
We thus have that, for all $u,u'\in\reals^d$,
\begin{align}
\|F(u)-F(u')\|\leq L \cdot\|u-u'\|,
\end{align}
where $L=(3\delta^2 +\|u_0\|\delta) \frac{5nd\alpha_H }{\alpha_\nh+\beta}<1$ if
$\delta <\frac{1}{6}\left(\sqrt{\|u_0\|^2+12\frac{\alpha_\nh+\beta}{5nd\alpha_H }}-\|u_0\|\right)$ \qed

Thm.~\ref{thm:contraction} along with the Banach fixed-point theorem~\citep{banach1922operations} imply  that  a stationary profile exists and is unique. Most importantly, it can be found by the following iterative process. Starting from any $\ulim^0\in \reals^d$, the iterations
\begin{align}\label{eq:fpiterate2}
    \ulim^{\ell+1} = F(\pi,\ulim^\ell), \quad \ell\in\naturals, 
\end{align}
are guaranteed to converge to the unique fixed-point $\ulim^*$ of Eq.~\eqref{eq:fp}.

The Banach fixed-point theorem also implies that convergence happens exponentially fast. To see this, observe that, for $L<1$ the Lipschitz parameter of $F$, \begin{align*}
\|\ulim^* - \ulim^\ell\|&=\|F(\pi,\ulim^*)-F(\pi,u^{k-1})\|
\leq L\|\ulim^*-\ulim^{k-1}\| =L \|F(\pi,\ulim^*) - F(\pi, \ulim^{k-2})\|\\
& \leq L^2 \|| \ulim^*- \ulim^{k-2}\| =\ldots
\leq L^{\ell}\|\ulim^*-\ulim^0 \|,\end{align*}
for all $\ell\in\naturals$.

\section{Proof of Thm.~\ref{thm:subopt}}\label{app:proofofthm:subopt}
Consider a setting where candidate set is always the entire catalog, i.e., $\candidate=\Omega$.
 Let $d=3$, $k=1$, and $B_1>B_2\gg 1$ a large positive number.  Consider a user that starts from $u_0 = [B_1,0,0]$.  Assume that $\Omega$ contains a harmful item with $v_\hm=[0,B_2,-B_2]$, and two non-harmful  items in $\Omega\setminus H$: $v_0=[1,B_2,-B2]$ and $v_1=[0.99,-B_2,B_2]$. Set $\alpha_{\nh}=\alpha_\hm$ and $\beta = 0$. Note that we can select $B_1,B_2$ so that the conditions of Thm~\ref{thm:contraction} hold. Observe that, under the user's initial profile, $$s_{v_0}=e^{B_1},~s_{v_1}=e^{0.99B_1},~\text{and}~s_{v_\hm}=1.$$ Set $c$ to be small enough so that $g(s_{v_0})>g(s_{v_1})=0.8$, i.e., so that irrespective of the recommendation, the recommended item has a high probability of being selected. 
 
 By Thm.~\ref{thm:staticoptimal}, in its first iteration, Alg.~\eqref{eq:alternate} will recommend $v_0$, as between the two options ($v_0$,$v_1$), it has the highest score.  Under this recommendation policy, function $F$ will be an interpolation between the three items, with the weight of $v_0$ being larger than $0.8$. This means that the first application of $F$ to $u_0$ will produce a vector that will contain a high value (at least $0.6B_2$) on the second coordinate, and a negative value (at most $-0.6B_2$) on the third coordinate. Repeated applications of $F$ will make the contribution of $v_0$ and $v_\hm$ higher, simultaneously suppressing the contribution of $v_1$, as the latter will have a score  $e^{-\Theta(B_2^2)}$. As a result, the fixed point $\ulim$ will contain a larger contribution from $v_0$ and $v_\hm$ and, hence, the second coordinate will be positive and of order $\Theta(B_2)$, while the third coordinate will be negative and of order $-\Theta(B_2)$. Thus, in the next iteration of the alternating optimization algorithm, $v_0$ will again be selected as an optimal recommendation, and the algorithm will terminate (as $\ulim$/$\pi$) will again be the same). At this operating point, however, as $\ulim$ has a non-zero, positive second coordinate, the probability of harm will be non-zero. In fact, it will be arbitrarily close to 1-$p_{v_0}$, as the probability of selecting of $v_1$ will be negligible. Moreover, $p_{v_0}$ will be bounded away from 1, as the probability of organic selection of $v_0$ and $v_\hm$ will be comparable for large $B_2$ (approximately $\approx 0.5$).
 
 A very different behavior will occur however if $v_1$ is recommended. By our choice of $c$, this will again be selected with probability $>0.8$, and repeated applications of $F$ will lead to a $\ulim$ that has a $\theta_{B_2}$ For the same reasons, $\ulim$ will have a high positive  value ($\Theta(B_1)$) on the last coordinate and a high negative value (-$\Theta(B_2)$) on the second coordinate. As a result, the probability (organic) selection of both $v_0$ and $v_1$ will be arbitrarily small under the stationary profile constructed by this recommendation policy.
 
 The (approximate) symmetry between recommending $v_0$ and $v_1$ and the $\ulim$s they lead to imply that these two policies will have a very similar $p_\clk$; in contrast, they will have very different $p_\hm$: this is precisely because with the first the probability of selecting $v_\hm$ is non-negligible and comparable to the one selecting $v_0$, whereas with the second $p_\hm$ is negligible. Hence, by appropriately setting $\lambda$, we can make the gap between the utilities constructed by two polices arbitrarily large. \qed

\section{Gradient Computations}
\label{app:gradientcomp}

\subsection{Bounded Cardinality Setting}\label{app:pgasimp}

\noindent\textbf{Gradient of $\ulim(\cdot)$ Operator.}
We begin with the computation of the gradient of the $\ulim(\cdot)$ operator, given by:
\begin{align}
    \nabla_\pi \ulim(\pi) = -\left( \nabla_{\ulim} F(\pi,\ulim)- I_{d\times d}\right)^{-1}\cdot \nabla_\pi F(\pi, \ulim) \label{eq:ulimjac2}
\end{align}
The terms in the r.h.s. are as follows:

\begin{itemize}
    \item $I_{d\times d}$ is the $d$-dimensional identity matrix.
    \item 
$\ulim = \ulim(\pi)\in \reals^d$, computed via Eq.~\eqref{eq:fpiterate}.
\item $\nabla_{\ulim} F(\pi,\ulim)\in \reals^{d\times d}$ is the Jacobian of $F$ w.r.t.~its second argument (i.e.,~$\ulim$). The coordinates of this matrix are given by:
\begin{align}\label{eq:Fugrad}
    \frac{\partial F_i}{\partial \ulim_j}\stackrel{\eqref{eq:gp}}{=} \sum_{v}\frac{\partial G_i(p)}{\partial p_v}\cdot \frac{\partial p_v}{\partial\ulim_j}, \quad \text{for}~i,j\in\{1,\ldots,d\},
\end{align}
where
\begin{align}\label{eq:dG_i}
    \frac{\partial G_i(p)}{\partial p_v}
    \stackrel{\eqref{eq:dg}}{=}
    \frac{ 
        \alpha_{v} \sum_{v'\in \catalog}(v_i-v_i') \alpha_{v'}p_{v'} 
        + \alpha_v \beta (v_i - u_0)
    }{
        (\beta+\sum_{v'\in \catalog} \alpha_{v'}p_{v'})^2
    },
\end{align}
while {\begin{align}\frac{\partial p_v}{\partial\ulim_j}=\expect_C\left[\expect_{\pi}\left[\frac{\partial p_{v|E}}{\partial\ulim_j}\right] \right]\end{align}}
where
\begin{align}
 \frac{\partial p_{v|E}}{\partial\ulim_j}\stackrel{\eqref{eq:pvfull}}{=}  \begin{cases} \frac{\partial}{\partial \ulim_j}\left(\frac{s_v}{\sum_{v'\in E}s_{v'} +c}\right) + \frac{\partial}{\partial \ulim_j}\left( \frac{c s_v}{(\sum_{v'\in E}s_{v'}+c)(\sum_{v'\in \catalog} s_{v'})}\right),& \text{if}~v\in E,\\ \frac{\partial}{\partial \ulim_j}\left( \frac{c s_v}{(\sum_{v'\in E}s_{v'}+c)(\sum_{v'\in \catalog} s_{v'})}\right), &\text{o.w.}\end{cases}
\end{align}

The latter are given by:
\begin{align}
&\frac{\partial}{\partial \ulim_j}\left(\frac{s_v}{\sum_{v'\in E}s_{v'} +c}\right) \stackrel{\eqref{eq:dsv1}}{=} \frac{s_v\cdot (\sum_{v'\in E}(v_j-v_j')s_{v'} +cv_j) }{(\sum_{v'\in E}s_{v'} +c)^2},\\
&\frac{\partial}{\partial \ulim_j} \left(\frac{s_v}{\sum_{v'\in E}s_{v'} +c}\cdot\frac{c}{\sum_{v'\in \catalog} s_{v'}} \right) \stackrel{\eqref{eq:dsv2},\eqref{eq:dsv3}}{=} \nonumber\\
&\quad=
\frac{c}{
    \sum_{v'\in \catalog} s_{v'}
}
\cdot
\frac{
    s_v (\sum_{v'\in E}(v_j-v_j')s_{v'} +cv_j)
}{
    (\sum_{v'\in E}s_{v'} +c)^2
}\nonumber\\
&\quad~~~~-\frac{s_v}{
    \sum_{v'\in E}s_{v'} +c
}\cdot 
\frac{
    c\cdot( \sum_{v'\in \catalog}v'_js_{v'})
}{
    (\sum_{v'\in \catalog}s_{v'})^2
}.\label{eq:fugradlast}
\end{align}
\item  $\nabla_{\pi} F(\pi,\ulim)\in \reals^{d\times m}$ is the Jacobian of $F$ w.r.t.~its first argument, (i.e.,~$\pi$). The coordinates of this matrix are given by:\begin{align}\frac{\partial F_i}{\partial p_{E\mid C}}\stackrel{\eqref{eq:gp}}{=} \sum_{v}\frac{\partial G_i(p)}{\partial p_v}\cdot \frac{\partial p_v}{\partial p_{E\mid C}},\quad \text{for}~i\in\{1,\ldots,d\},E\in D_C, C\in \collection,
    \label{eq:fpigrad1}\end{align}where $ \frac{\partial G_i(p)}{\partial p_v}$ is again given by Eq.~\eqref{eq:dG_i} and 
\begin{align}
    \frac{\partial p_v}{\partial p_{E\mid C}} = p_C\cdot p_{v\mid E}, 
    \label{eq:fpigrad2}
    \end{align} 
\end{itemize}
for $p_{v\mid E}$ as in Eq.~\eqref{eq:pve}.

\noindent\textbf{Gradients of $p_\clk$, $p_\hm$.}
Then, the gradient  $\nabla_\pi p_\clk(\pi,\ulim(\pi))\in \reals^m$ will be:
\begin{align}
    \nabla_\pi p_\clk(\pi,\ulim(\pi)) =    \nabla_\pi p_\clk(\pi, \ulim) +  
       \left[\nabla_{\pi} \ulim(\pi)\right]^\top \nabla_{\ulim} p_\clk(\pi,\ulim) ,
\end{align}
where the terms of the r.h.s.~are as follows:
\begin{itemize}
  \item $\ulim = \ulim(\pi)\in \reals^d$, computed via Eq.~\eqref{eq:fpiterate}.
   \item  $\nabla_\pi \ulim(\pi)\in \reals^{d\times m}$ is the Jacobian of $\ulim(\cdot)$, computed via Eq.~\eqref{eq:ulimjac2}.
   \item $\nabla_{\pi} p_\clk(\pi,\ulim)\in \reals^m$ is the gradient of 
   {
   \begin{align}
           p_\clk(\pi,\ulim) &\stackrel{}{=}\expect_C \left[ \expect_\pi[ g(s_E(\ulim))]\right],
   \end{align}}
   w.r.t.~to its first argument (i.e., $\pi$). Its coordinates are given by:
   {
   \begin{align}
       \frac{\partial p_\clk}{\partial p_{E\mid C}} = p_C \cdot g(s_E(\ulim)),\quad\text{for}~E\in D_C.
   \end{align}
   }
\item     $\nabla_{\ulim} p_\clk(\pi,\ulim)\in \reals^d$ is the gradient of 
          $ p_\clk(\cdot,\cdot)$ 
   w.r.t.~to its second argument (i.e.,~$\ulim$). Its coordinates are given by:
   {
   \begin{align}
       \frac{\partial p_\clk}{\partial \ulim_i} = \expect_C\left[\expect_\pi\left[ g'(s_E(\ulim))\frac{\partial s_E(\ulim)}{\partial \ulim_i}\right]\right]\quad\text{for}~E\in D_C
   \end{align}}
   where $g'(s)=\frac{c}{(s+c)^2}$ and $\frac{\partial s_E(\ulim)}{\partial \ulim_i}=\sum_{v\in E}v_is_v$.
\end{itemize}
Finally, the gradient of 
\begin{align}
    p_\hm(\pi,\ulim(\pi)) = \frac{s_H(\ulim(\pi))}{s_\catalog(\ulim(\pi))}(1-p_\clk(\pi,\ulim(\pi)))
\end{align}
w.r.t.~$\pi$ can be computed similarly, using 
\begin{align}
    \nabla_\pi p_\hm(\pi,\ulim(\pi)) =    \nabla_\pi p_\hm(\pi, \ulim) +  
       \left[\nabla_{\pi} \ulim(\pi)\right]^\top \nabla_{\ulim} p_\hm(\pi,\ulim) ,
\end{align}
along with the fact that
\begin{align}
    \nabla_\pi p_\hm(\pi,\ulim) = -\frac{s_H(\ulim)}{s_\catalog(\ulim)}\cdot \nabla_\pi p_\clk(\pi,\ulim)
\end{align}
and 
\begin{align}
    \nabla_{\ulim} p_\hm(\pi,\ulim) =  \nabla_{\ulim}\left[\frac{s_H(\ulim)}{s_\catalog(\ulim)}\right]\cdot (1-p_{\clk}(\pi,\ulim)) -\frac{s_H(\ulim)}{s_\catalog(\ulim)}\cdot\nabla_{\ulim} p_\clk(\pi,\ulim)
\end{align}
where 
\begin{align}
    \frac{\partial}{\partial \ulim_j }\left[\frac{s_H(\ulim)}{s_\catalog(\ulim)}\right] = \sum_{v\in H}   \frac{s_v\cdot (\sum_{v'\in \catalog}(v_j-v_j')s_{v'} +cv_j) }{(\sum_{v'\in \catalog}s_{v'} +c)^2},\quad\text{for}~i\in \{1,\ldots,d\}.
\end{align}

\noindent\textbf{Projected Gradient Ascent.}
Given the gradient, the PGA steps  would be:
\begin{align}
    \pi_{\ell+1} = \mathcal{P}_{\Pi} \left(\pi_\ell +\gamma_\ell \nabla_\pi f(\pi_k) \right)
\end{align}
where 
\begin{align}
{\mathcal{P}_\Pi} (\pi')= \argmin_{\pi\in \Pi}\|\pi-\pi\|_2^2 
\end{align}
is the orthogonal projection to $\Pi$, given by Eq.~\eqref{eq:Pi}. This projection is a quadratic optimization  problem that can be solved through standard techniques, but because $\Pi$ is is the canonical simplex, it actually admits a strongly poly-time algorithm (see, e.g., \cite{michelot1986finite}).

\subsection{ Independent Sampling Setting}\label{app:indep}

Recall that, under the independent sampling setting, computing the optimal $\pi$ thus reduces to computing the { $m'\equiv \sum_{\candidate\in \collection} |C|$ parameters in $\rho=[\rho_{v\mid C}]_{v\in \candidate, \candidate\in\collection}$}. This comes with a  computational advantage compared to the bounded cardinality setting; {for example, if $\collection$ is a partition of $\catalog\setminus H$,  the number of parameters becomes linear (namely, $n'=n-h$)}.

{Given a set $\candidate\in \collection$, let $\rho_\candidate=[\rho_{v\mid \candidate}]_{v\in \candidate}\in \reals^{|\candidate|}$ be the vector of probabilities condition on the candidate set being $\candidate$. Then, for any set function $z:2^\candidate\to\reals$, we can approximate $\expect_{\rho_\candidate}[z(E)]$ via
\begin{align}\widehat{\expect_{\rho_\candidate}}[z(E)] = \textstyle\frac{1}{N}\sum_{\ell=1}^N z(E_\ell),\label{eq:estimator} \end{align}
where $E_\ell$, $\ell=1,\ldots,N$ are i.i.d., sampled from the appropriate product form distribution (see Eq.~\eqref{eq:prodform}). Most importantly, we can also estimate  the gradient $\nabla_{\rho_\candidate} \expect_{\rho_\candidate}[z(E)]\in \reals^{|\candidate|}$ via
\begin{align}\textstyle \widehat{\frac{\partial}{\partial \rho_{v\mid C}}} \left(\expect_{\rho_C}[z(E)] \right) =  \widehat{\expect_{\rho_C}}[z(E\cup\{v\} -z(E\setminus\{v\})].\label{eq:estimatorgrad}\end{align}
This is classic (see, e.g., \cite{calinescu2011maximizing}) and has a long history of use in  submodular   optimization \cite{calinescu2011maximizing,hassani2017gradient}. The function $Z(\rho)=\expect_{\rho_C}[z(E)]$ is known as the \emph{multi-linear relaxation} of $z$, and  Chernoff bounds can be used to characterize the quality of the above approximations; moreover, provided $z$ is bounded (which is indeed the case for all functions we compute here) a polynomial number of samples $N$ is required for the above estimators to get within a given accuracy from the true expectation (see, e.g., Sec.~3 in~\citet{calinescu2011maximizing}). }

{These sampling techniques can be applied directly to our independent sampling setting, again combined with an appropriate combination of chain rules the an appropriate evocation of the implicit function theorem.}

\noindent\textbf{Fixed-point Operator.}
Given $\rho$, the fixed point map who's repetition gives $\ulim$ is given by: 
\begin{align}F(\rho,u) = G(p(\rho,u))\end{align}
where {$p:[0,1]^{\sum_{C\in \collection}|C|}\times \reals^d \to [0,1]^n$} is the map that produces the probabilities $p_v$, given by 
Eq.~\eqref{eq:pv}, 
and 
$G:[0,1]^n\to \reals^d$ is the map given by Eq.~\eqref{eq:g}. Hence, $G$ remains the same as in the bounded cardinality case, and the map $p_v$ can be computed by{
\begin{align}
    p_v = \expect_C [\expect_{\rho_C}[p_{v\mid E}]]\label{eq:pvrho2}
\end{align}}
with $p_{v\mid E}$ given by Eq.~\eqref{eq:pve}, by replacing the expectation with the estimator given by Eq.~\eqref{eq:estimator}.

\noindent\textbf{Projected Gradient Ascent.}
Projected gradient ascent is defined by
\begin{align}
    \rho_{k+1} = \mathcal{P}_{R} \left(\rho_k +\gamma_k \nabla_\rho f(\rho_k) \right)
\end{align}
where 
\begin{align}
{\mathcal{P}_\Pi} (\rho')= \argmin_{\rho\in R}\|\rho'-\rho\|_2^2 
\end{align}
is the orthogonal projection to set $R$, given by Eq.~\eqref{eq:R}. This is again a projection to a rescaled simplex, and can be computed by, e.g., Michelot's algorithm \citep{michelot1986finite}.

\noindent\textbf{Gradient of the $\ulim(\cdot)$ Operator.}
 The gradient of the $\ulim(\cdot)$ operator, is again given by:
\begin{align}
    \nabla_\rho \ulim(\rho) = -\left( \nabla_{\ulim} F(\rho,\ulim)- I_{d\times d}\right)^{-1}\cdot \nabla_\rho F(\rho, \ulim) \label{eq:ulimjacindep}
\end{align}
Here
\begin{itemize}
    \item $\nabla_{\ulim} F(\pi,\ulim)\in \reals^{d\times d}$ is computed exactly as in Eq.~\eqref{eq:Fugrad}--\eqref{eq:fugradlast}, the only difference being that the expectation:
     \begin{align}\frac{\partial p_v}{\partial\ulim_j}=\expect_C\left[\expect_{\rho_C}\left[\frac{\partial p_{v|E}}{\partial\ulim_j}\right]\right] \end{align}
     is computed via the sampling estimator in \eqref{eq:estimator}.
     \item  $\nabla_{\pi} F(\pi,\ulim)\in \reals^{d\times n}$ is computed as in Eq.~\eqref{eq:fpigrad1}--\eqref{eq:fpigrad2}, where
     \begin{align}\frac{\partial p_v}{\partial \rho_{v|mid C}} = \frac{\partial }{\partial \rho_{v\mid C}}\left(\expect[p_{v\mid E}]\right)\end{align}
     is computed via the gradient sampling estimator in Eq.~\eqref{eq:estimatorgrad}.
\end{itemize}

\section{Additional Experimental Details and Results}
\label{sec:supp_exp}

\subsection{Experimental Setup}
We join the MovieLens25m ratings, movies, and links csv files by {\tt movieId}, and then join with the IMDB dataset~\citep{IMDBParent} by the same column.
We filter on the desired genre, and then select the top 100 movies with the highest number of ratings, and the subsequent top 1000 users by the number of these movies that they have rated.
A matrix factorization model is trained using this user-movie ratings matrix with dimension 10 as well as additional user and item level biases.
We use stochastic gradient descent over the mean squared error, with learning rate of 0.01, regularization parameter of 0.01, and 100 iterations.
This process results in learned 12 dimensional user and item embeddings.
The RMSE of these learned embeddings are as follows: Action: 1.39, Adventure: 1.49, Comedy: 1.54, Fantasy: 1.86, Sci-Fi: 1.59. 
For the simulations, we  select a set of 100 users per genre u.a.r., with a fixed random seed of 42.
Experiments were conducted on a CPU cluster with slurm management.
To compute the embeddings, 100G of memory and 10 CPUs were requested. The matrix factorization implementation we used is a modified from an open source repository.\footnote{ \url{https://github.com/albertauyeung/matrix-factorization-in-python}}

\noindent\textbf{$p_\hm$ Calibration with $c$.}
The parameter $c$ in the multinomial logit model~\eqref{eq:clk} influences the effectiveness of our policies in garnering clicks and therefore guiding user dynamics.
For any given $s_E$, as $c \to \infty$, $g \to 0$.
By calibrating $c$, we can influence $p_\hm$ so that our simulations are in a regime where harm matters.
For each genre we take a random sample of 10 users, and compute the resulting $p_\clk$, $p_\hm$ under the uniform and alternating policies, at steady state user dynamics.
Table~\ref{tab:calib_pH_c} shows the results for each genre, and the chosen value of $c$, for $k=1$ under the bounded cardinality setting.
Our heuristic is to choose $c$ for each genre maximizing $p_\hm$ such that $p_\clk > 0.5$ under the Alt policy.

\begin{table}[]
\centering
\begin{tabular}{ccc|}
\multicolumn{3}{c}{\underline{Action, $c=3$}} \\
$c$ & Alt ($p_\clk$, $p_\hm$) & Unif ($p_\clk$, $p_\hm$) \\
\midrule
1 & 0.89, 0.03 & 0.72, 0.09 \\
2 & 0.79, 0.07 & 0.57, 0.14 \\
3 & 0.70, 0.10 & 0.47, 0.17 \\
4 & 0.62, 0.12 & 0.40, 0.20 \\
5 & 0.55, 0.15 & 0.35, 0.21
\end{tabular}
\begin{tabular}{ccc|}
\multicolumn{3}{c}{\underline{Adventure, $c=7$}} \\
$c$ & Alt ($p_\clk$, $p_\hm$) & Unif ($p_\clk$, $p_\hm$) \\
\midrule
5 & 0.67, 0.05 & 0.37, 0.09 \\
6 & 0.62, 0.05 & 0.33, 0.09 \\
7 & 0.57, 0.06 & 0.30, 0.10 \\
8 & 0.52, 0.07 & 0.27, 0.10 \\
& & 
\end{tabular}
\begin{tabular}{ccc}
\multicolumn{3}{c}{\underline{Comedy, $c=7$}} \\
$c$ & Alt ($p_\clk$, $p_\hm$) & Unif ($p_\clk$, $p_\hm$) \\
\midrule
3 & 0.80, 0.04 & 0.49, 0.09 \\
4 & 0.74, 0.05 & 0.42, 0.09 \\
5 & 0.68, 0.06 & 0.37, 0.11 \\
6 & 0.63, 0.07 & 0.32, 0.12 \\
7 & 0.58, 0.08 & 0.29, 0.13
\end{tabular}
\begin{tabular}{ccc|}
\multicolumn{3}{c}{\underline{Fantasy, $c=13$}} \\
$c$ & Alt ($p_\clk$, $p_\hm$) & Unif ($p_\clk$, $p_\hm$) \\
\midrule
5 & 0.82, 0.02 & 0.42, 0.06 \\
10& 0.65, 0.04 & 0.27, 0.08 \\
12& 0.59, 0.04 & 0.24, 0.08 \\
13& 0.56, 0.05 & 0.22, 0.08 \\
15& 0.51, 0.05 & 0.20, 0.09
\end{tabular}
\begin{tabular}{ccc}
\multicolumn{3}{c}{\underline{Sci-Fi, $c=5$}} \\
$c$ & Alt ($p_\clk$, $p_\hm$) & Unif ($p_\clk$, $p_\hm$) \\
\midrule
3 & 0.78, 0.05 & 0.49, 0.11 \\
4 & 0.72, 0.06 & 0.42, 0.13 \\
5 & 0.66, 0.08 & 0.37, 0.14 \\
6 & 0.61, 0.09 & 0.33, 0.15 \\
7 & 0.56, 0.10 & 0.30, 0.15
\end{tabular}
\caption{
Calibrating per-genre parameter $c$ to ensure a balance of sufficiently high $p_\clk$ and $p_\hm$.
We choose $c$ for each genre maximizing $p_\hm$ such that $p_\clk > 0.5$ under the Alt policy.
}
\label{tab:calib_pH_c}
\end{table}

\noindent\textbf{Algorithm Implementation Details.} Fixed-point operations (Eq.~\eqref{eq:fpiterate}) executed for at most $\tau=10$ iterations, and always converged within a $10^{-3}$ tolerance, irrespective of whether conditions of Thm.~\ref{thm:contraction} were satisfied.
Unless otherwise noted, all recommendation policies operate in the Bounded Cardinality setting (Sec.~\ref{sec:rec_obj}). 

Additional algorithm implementation details per algorithm are as  follows:
\begin{itemize}
\item \emph{Gradient-Based Algorithm (Grad).}
We implement a gradient-based algorithm  solving Prob.~\eqref{eq:prob2} using the SLSQP implementation of SciPy~\citep{2020SciPy-NMeth}. 
For the bounded cardinality setting, we encode constraints such that the sum of the policy vector is 1, and each entry of the policy vector is non-negative. 
For the independent sampling setting, we encode constraints such that the sum of the policy vector is $k$, and each entry lies within the interval $[0,1]$.
We set SLSQP precision tolerance to  {\tt ftol}$=10^{-4}$. Because this is a non-convex optimization problem, we start the SLSQP procedure at various starting points and choose the resulting policy which achieves the best objective.
For starting points we use: U0 policy, Unif policy, Unif ``interior'' policy (dividing the vector by 100 to reach an interior point of the constraints), 3 random initalizations such that the sum of the policy vector is 1, and 3 random ``interior'' initalizations taking the previous random vector and dividing by 100.
\item \emph{Alternating Optimization (Alt).}
We also implement alternating optimization~(Sec.~\ref{sec:alt_alg}) for 10 steps; in all our experiments, user profiles converged with tolerance less than $2.5\times 10^{-3}$, in $\ell_2$  distance.

\item \emph{Static Profile Optimization (U0).}
As an additional baseline, we consider  the  policy  maximizing the objective under a static  user profile,  i.e., Eq.~\eqref{eq:probstatic} with  user profile $u_0$.

\item \emph{Uniform Recommendations (Unif).}
We also implement recommending uniformly at random as a baseline. For the bounded cardinality setting, that is $\pi_E = 1/|\constrset|,$ for $E\in \constrset$, and $\rho_v=k/n'$, for $v\in \Omega\setminus H$, for the independent sampling setting. 
\end{itemize}

\noindent\textbf{Profile Evolution.} We also implement the time evolution of a user profile under a given recommendation policy, always starting from $u_0$. We iterate over  Eq.~$\eqref{eq:u}$ until $\|u(t+1
)- u(t)\|_\infty < 10^{-3}$.

\noindent\textbf{Programming and Experimental Environment.} All  code was written in Python 3.10, using numpy 1.21.5, pandas 1.5.2, and scipy 1.9.3.
Experiments were conducted on a CPU cluster with slurm management.  
Each genre and experimental configuration was parallelized across users by running multiple slurm jobs.
To learn the policies, 1G of memory and 5 CPUs were requested for each parallelized job.
\subsection{Additional Experimental Results}

\begin{figure*}
\centering
\begin{minipage}[t]{0.3\linewidth}
\centering
\includegraphics[width=1\linewidth]{figures/Action_mf_4_beta015/ml_ip_select_nolog.pdf}
\vspace{-2em}
\caption*{Action}
\end{minipage}
\begin{minipage}[t]{0.3\linewidth}
\centering
\includegraphics[width=1\linewidth]{figures/Adventure_mf_4_beta015/ml_ip_select_nolog.pdf}
\vspace{-2em}
\caption*{Adventure}
\end{minipage}
\begin{minipage}[t]{0.3\linewidth}
\centering
\includegraphics[width=1\linewidth]{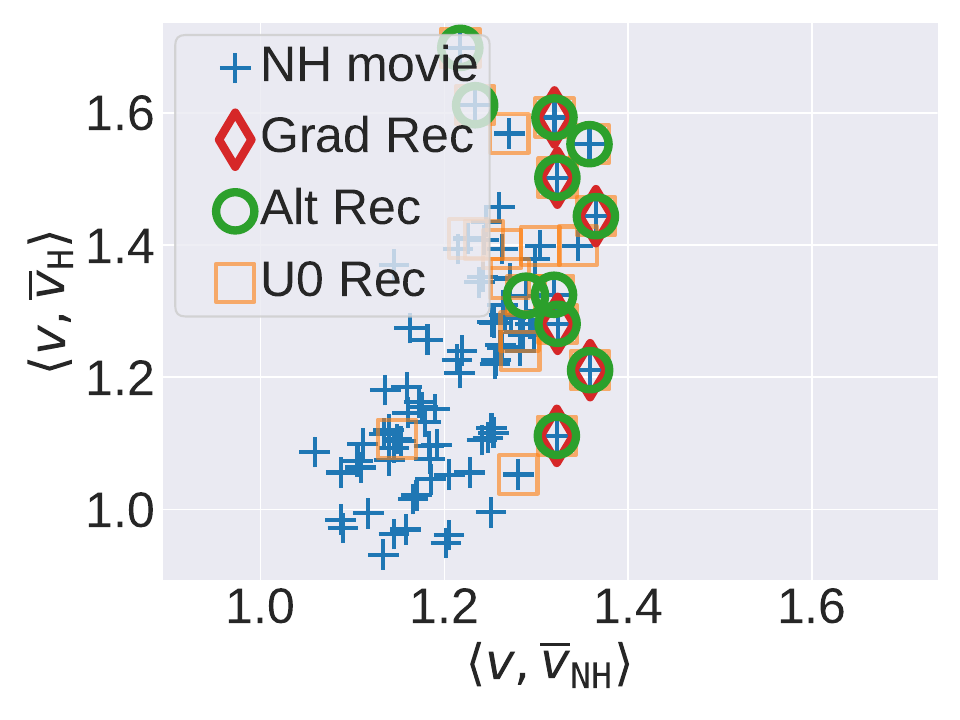}
\vspace{-2em}
\caption*{Comedy}
\end{minipage}
\begin{minipage}[t]{0.3\linewidth}
\centering
\includegraphics[width=1\linewidth]{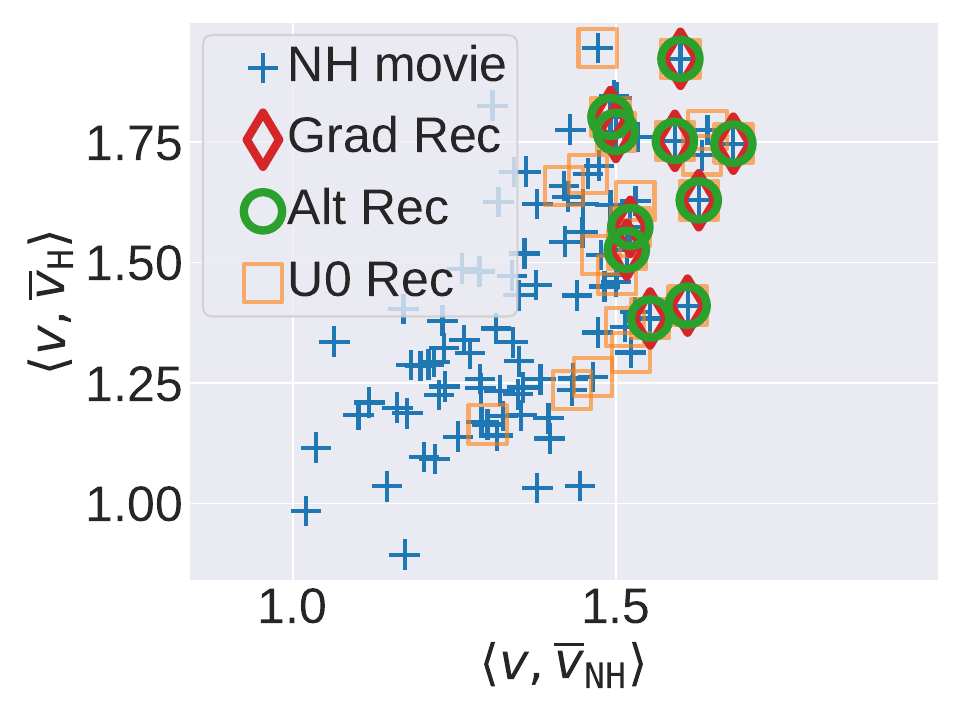}
\vspace{-2em}
\caption*{Fantasy}
\end{minipage}
\begin{minipage}[t]{0.3\linewidth}
\centering
\includegraphics[width=1\linewidth]{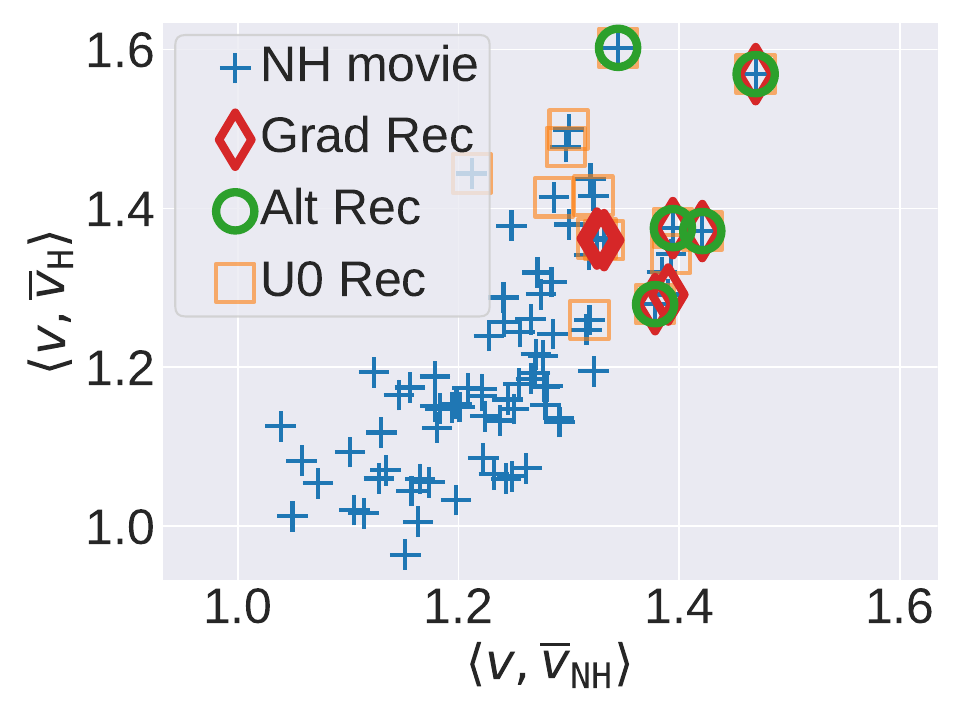}
\vspace{-2em}
\caption*{Sci-Fi}
\end{minipage}
\caption{
Visualization of the movies each policy recommends,  with respect to their inner product with the mean non-harmful and harmful vector, under the setting reported in Table~\ref{tab:big_table}, for a specific user. All non-harmful movies are embedded via a $+$ sign in this plane; the support of each policy is indicated by additional symbols (we omit \emph{Unif} as its support is everything). 
We observe that the gradient-based policy is more concentrated towards the right, i.e., on  movies which have high inner product with the average non-harmful vector.
}
\label{fig:ip_all}
\end{figure*}

\begin{figure*}
\centering
\begin{minipage}[t]{0.3\linewidth}
\centering
\includegraphics[width=1\linewidth]{figures/Action_mf_4_beta015/hist_obj_deltgrad.pdf}
\vspace{-2em}
\caption*{Action}
\end{minipage}
\begin{minipage}[t]{0.3\linewidth}
\centering
\includegraphics[width=1\linewidth]{figures/Adventure_mf_4_beta015/hist_obj_deltgrad.pdf
}
\vspace{-2em}
\caption*{Adventure}
\end{minipage}
\begin{minipage}[t]{0.3\linewidth}
\centering
\includegraphics[width=1\linewidth]{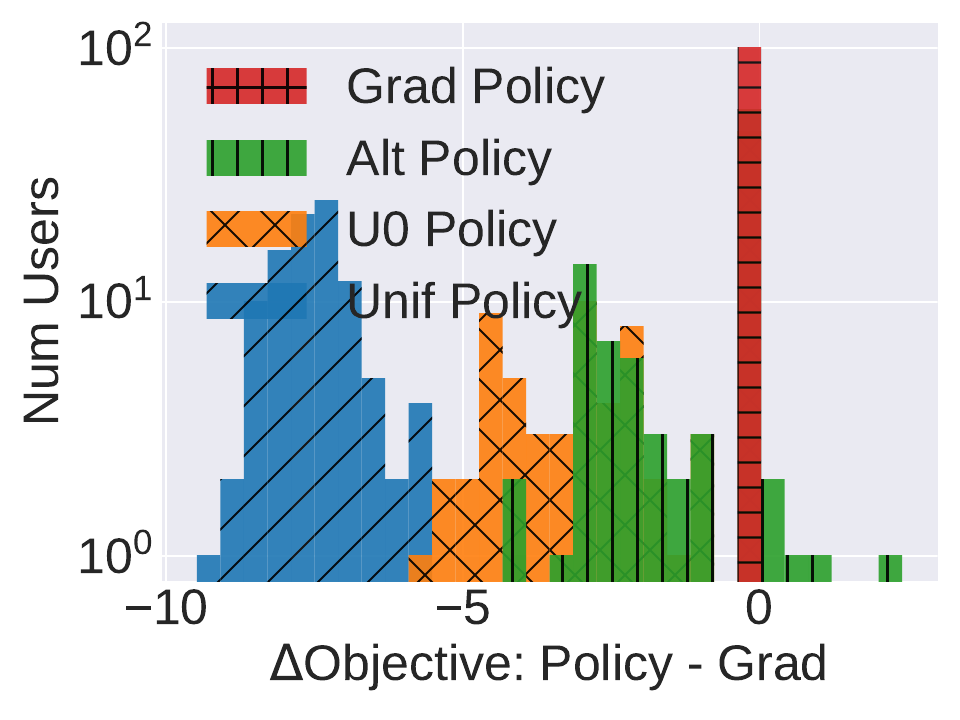}
\vspace{-2em}
\caption*{Comedy}
\end{minipage}
\begin{minipage}[t]{0.3\linewidth}
\centering
\includegraphics[width=1\linewidth]{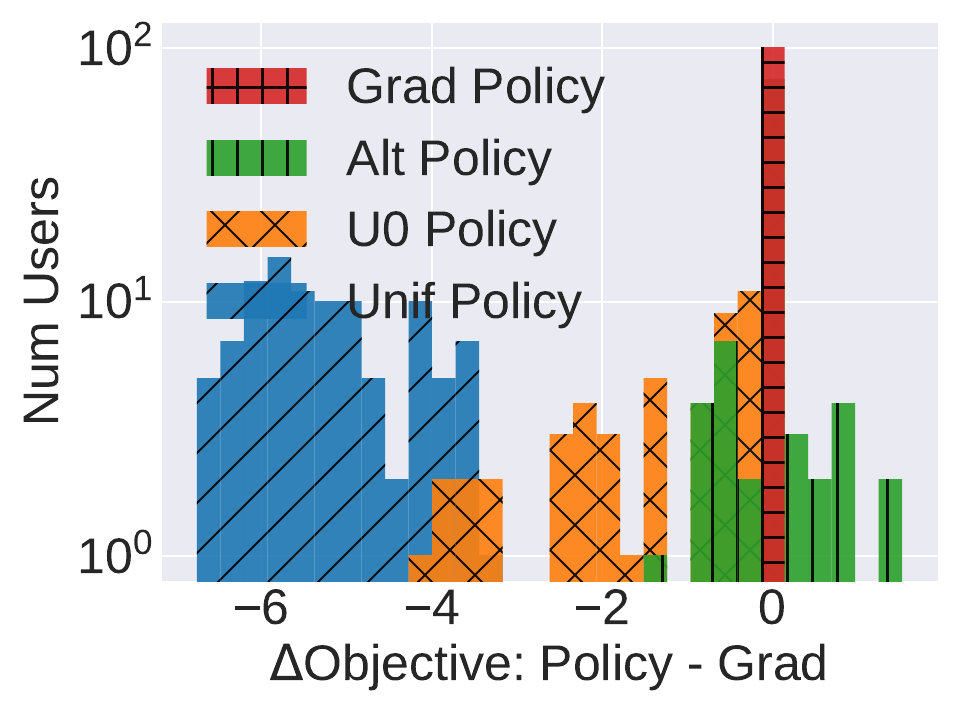}
\vspace{-2em}
\caption*{Fantasy}
\end{minipage}
\begin{minipage}[t]{0.3\linewidth}
\centering
\includegraphics[width=1\linewidth]{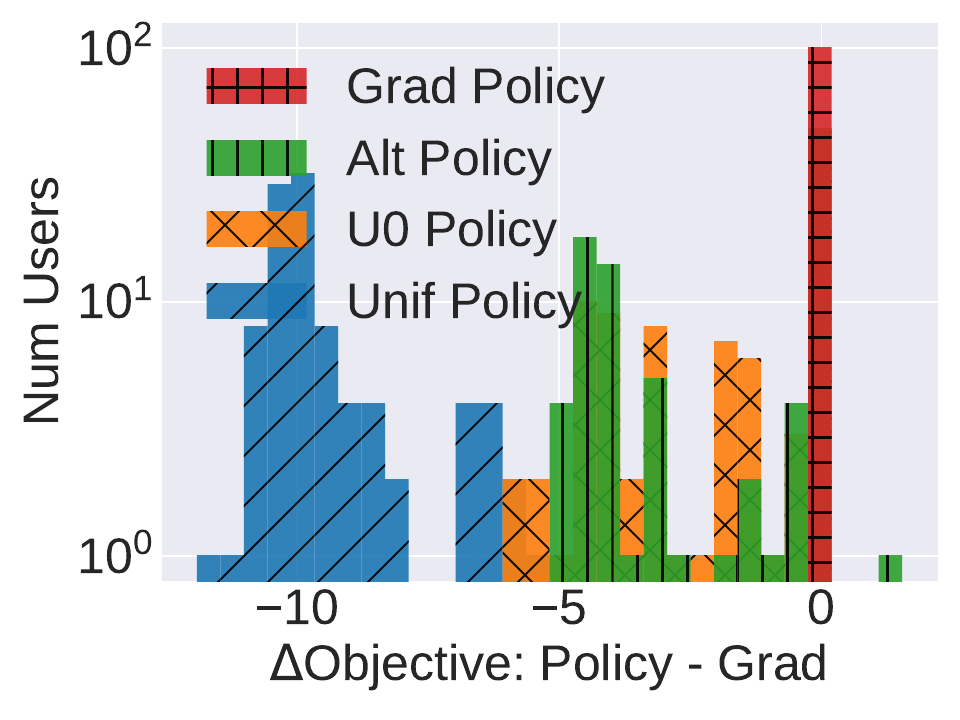}
\vspace{-2em}
\caption*{Sci-Fi}
\end{minipage}
\caption{
PDF of the difference between the objective attained by different policies minus the objective by attained by \emph{Grad}, for the same experimental settings as in Table~\ref{tab:big_table}. 
For the overwhelming majority of users, the objective obtained by the \emph{Grad} policy is superior to any other policy.
} 
\label{fig:pdfdiffobj}
\end{figure*}

\begin{figure*}
\centering
\begin{minipage}[t]{0.3\linewidth}
\centering
\includegraphics[width=1\linewidth]{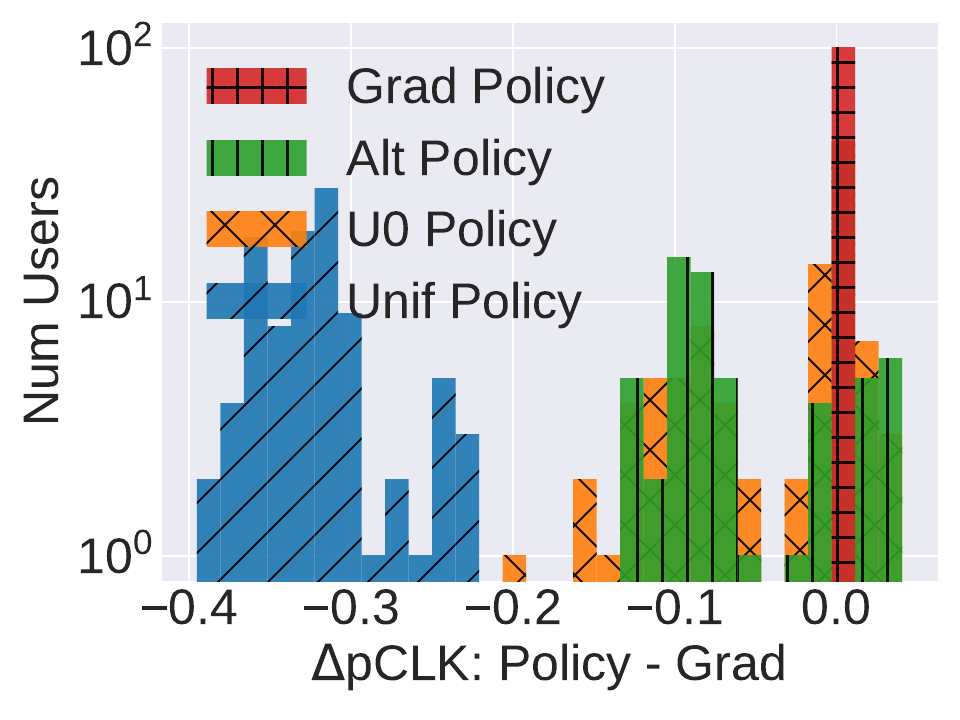}
\vspace{-2em}
\caption*{Action}
\end{minipage}
\begin{minipage}[t]{0.3\linewidth}
\centering
\includegraphics[width=1\linewidth]{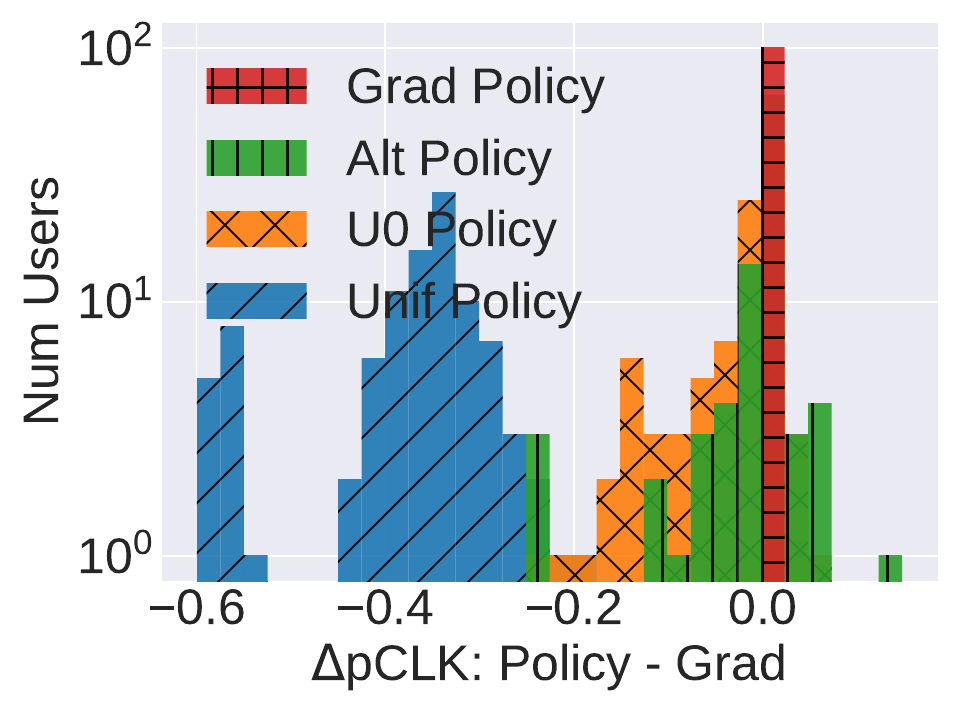
}
\vspace{-2em}
\caption*{Adventure}
\end{minipage}
\begin{minipage}[t]{0.3\linewidth}
\centering
\includegraphics[width=1\linewidth]{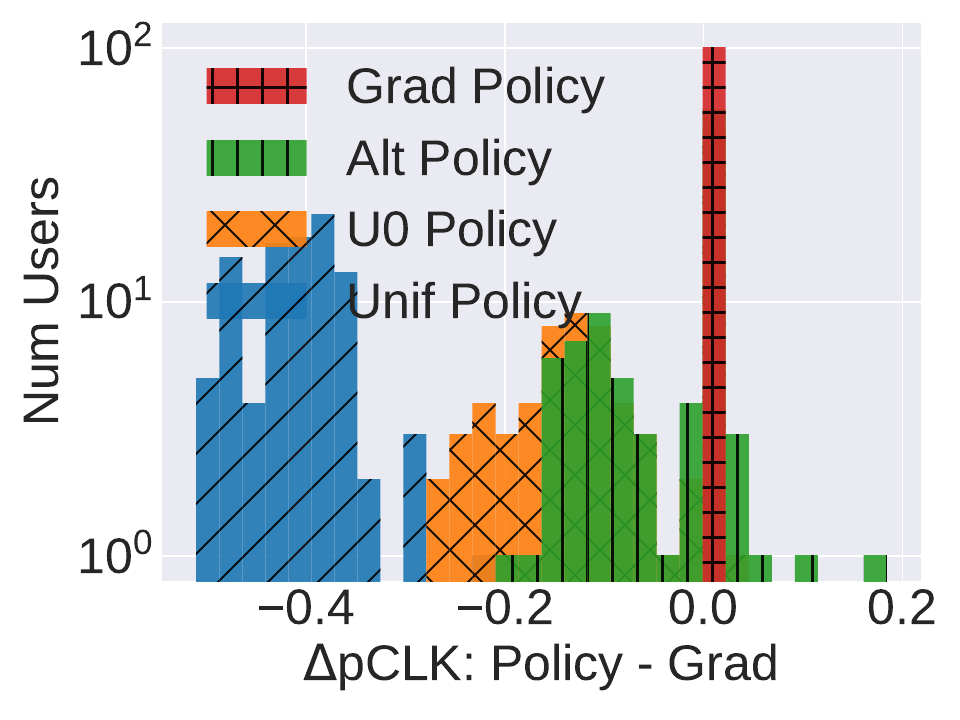}
\vspace{-2em}
\caption*{Comedy}
\end{minipage}
\begin{minipage}[t]{0.3\linewidth}
\centering
\includegraphics[width=1\linewidth]{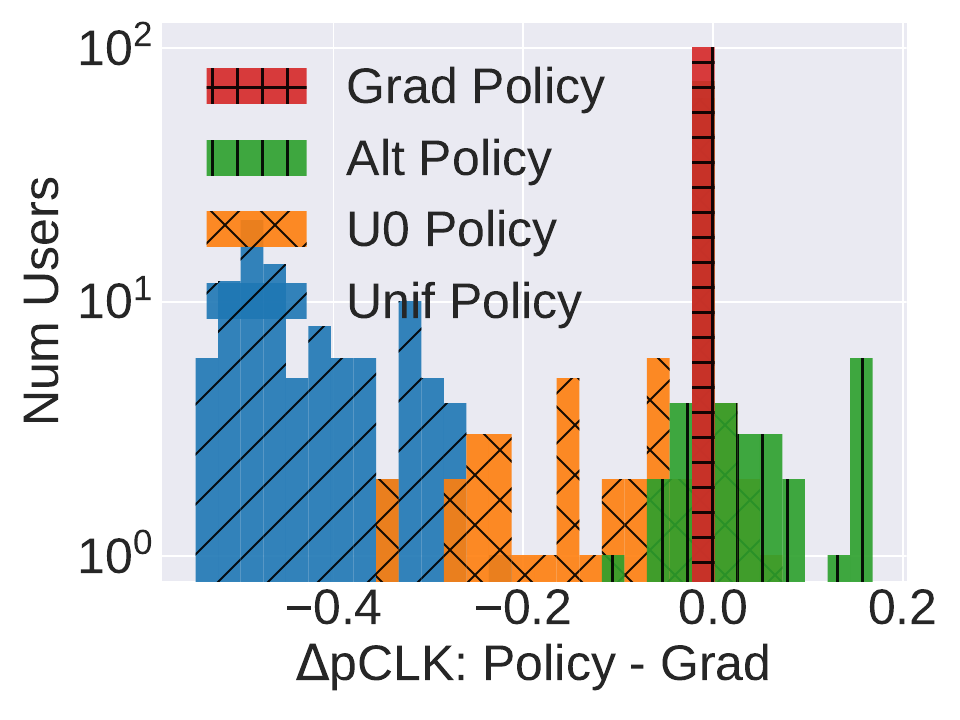}
\vspace{-2em}
\caption*{Fantasy}
\end{minipage}
\begin{minipage}[t]{0.3\linewidth}
\centering
\includegraphics[width=1\linewidth]{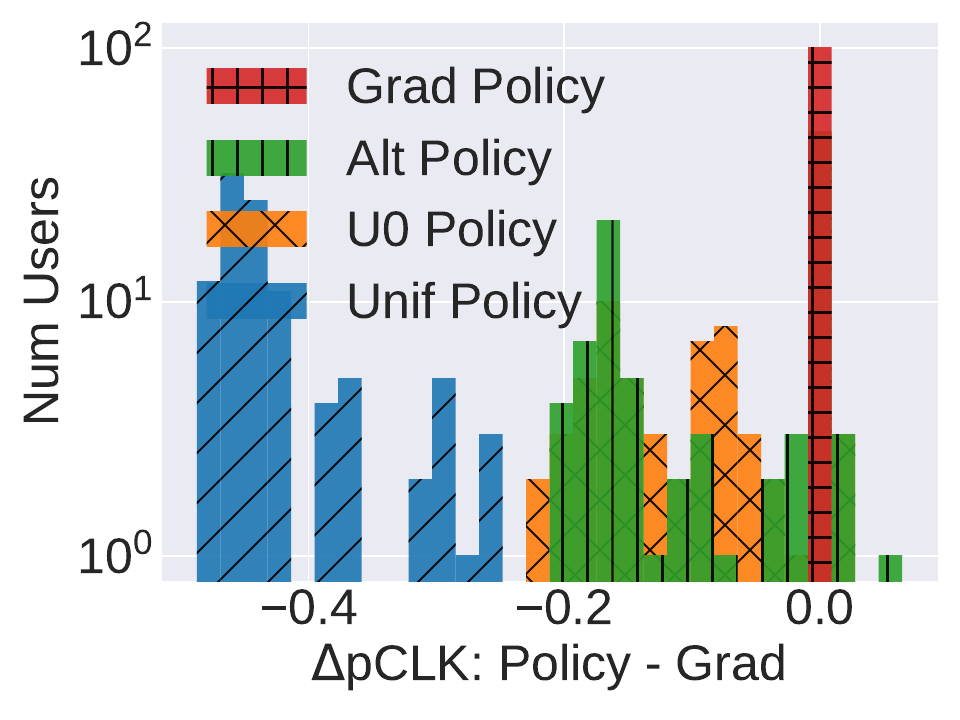}
\vspace{-2em}
\caption*{Sci-Fi}
\end{minipage}
\caption{
PDF of the difference between the $p_\clk$ attained by different policies minus the objective by attained by \emph{Grad}, for the same experimental settings as in Table~\ref{tab:big_table}. 
For the overwhelming majority of users, the $p_\clk$ values obtained by the \emph{Grad} policy is superior to any other policy.
}
\label{fig:pdfdiffpclk}
\end{figure*}

\begin{figure*}
\centering
\begin{minipage}[t]{0.3\linewidth}
\centering
\includegraphics[width=1\linewidth]{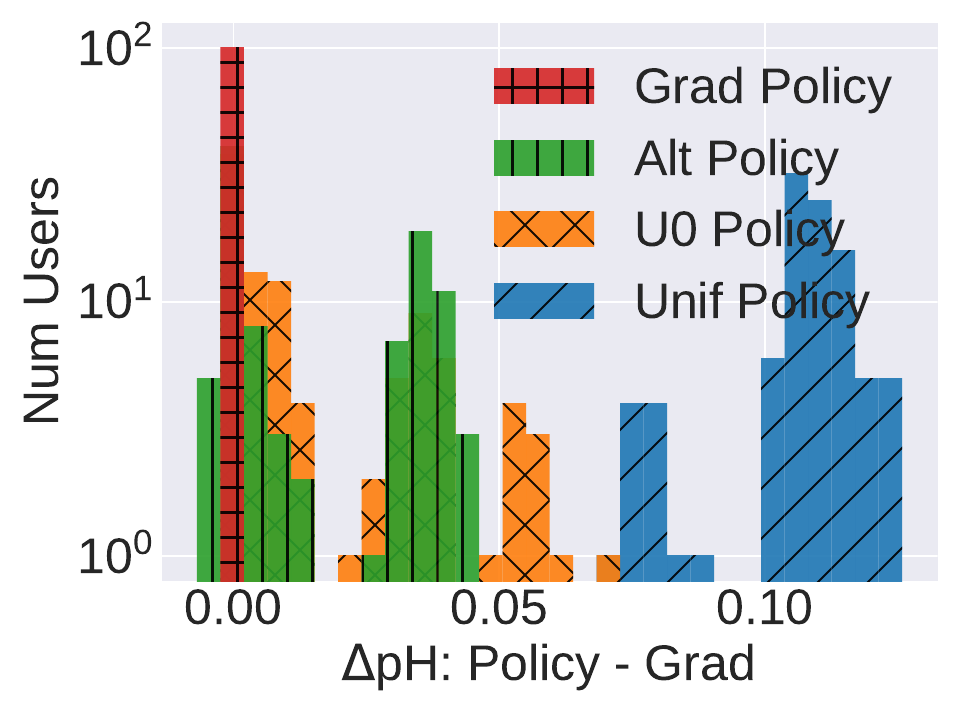}
\vspace{-2em}
\caption*{Action}
\end{minipage}
\begin{minipage}[t]{0.3\linewidth}
\centering
\includegraphics[width=1\linewidth]{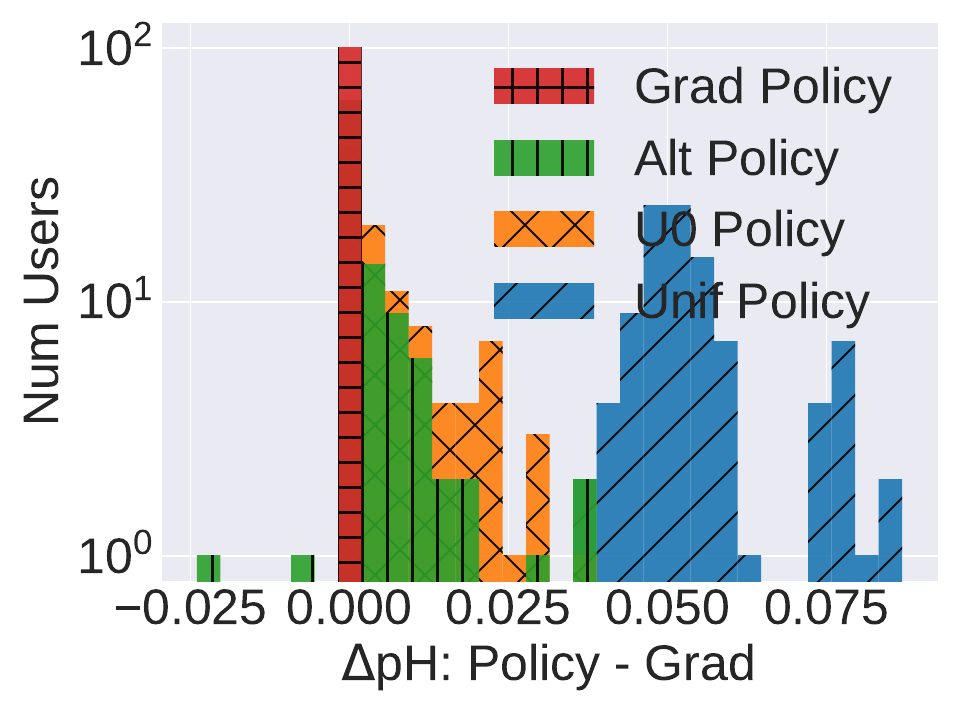
}
\vspace{-2em}
\caption*{Adventure}
\end{minipage}
\begin{minipage}[t]{0.3\linewidth}
\centering
\includegraphics[width=1\linewidth]{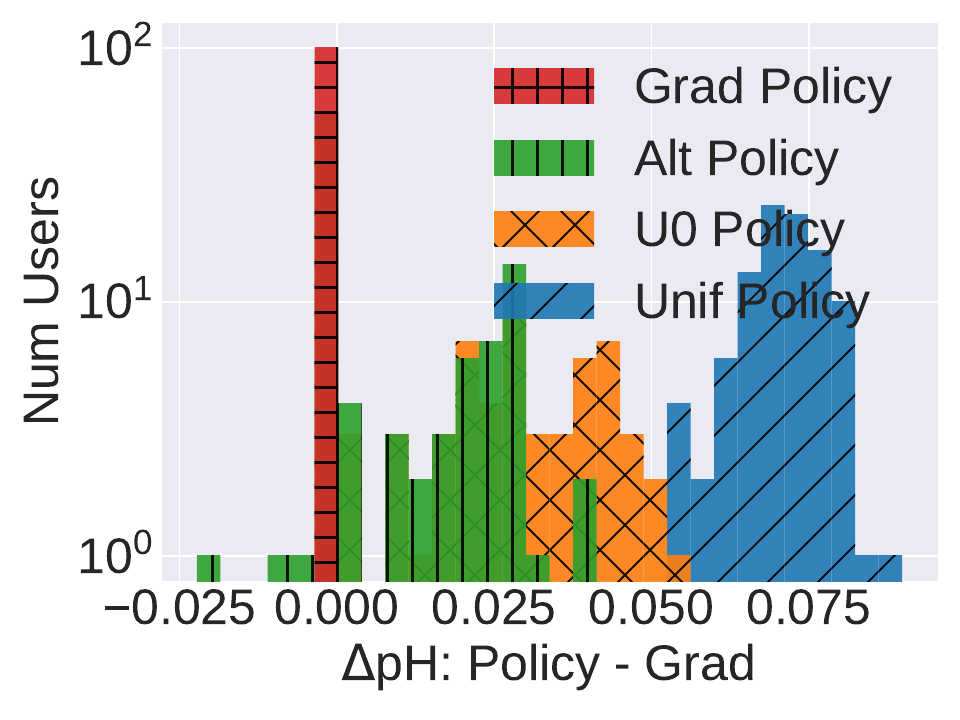}
\vspace{-2em}
\caption*{Comedy}
\end{minipage}
\begin{minipage}[t]{0.3\linewidth}
\centering
\includegraphics[width=1\linewidth]{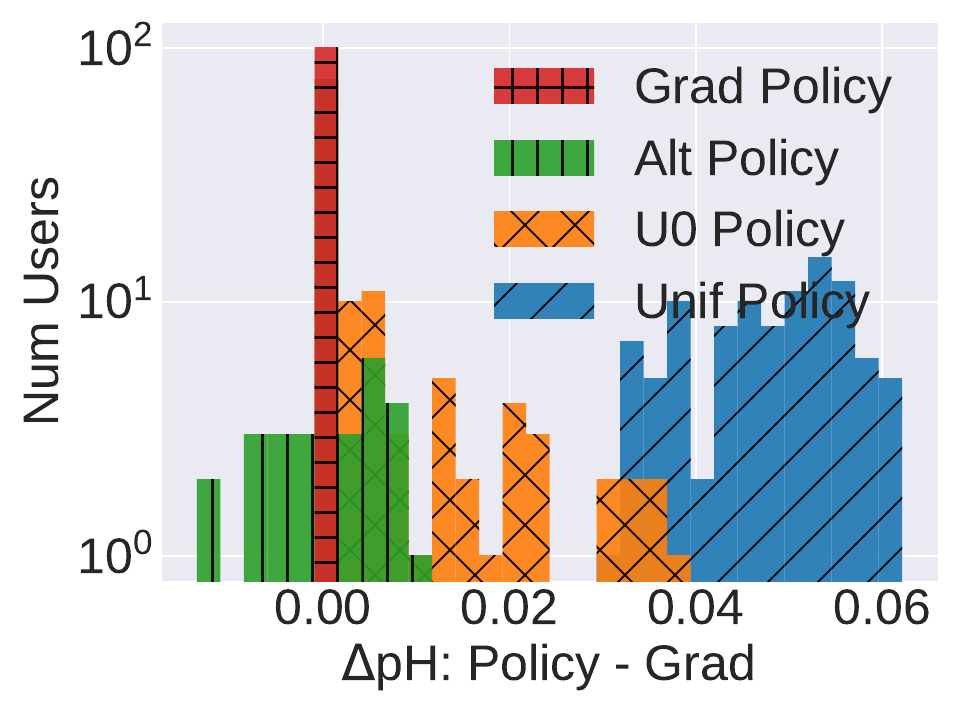}
\vspace{-2em}
\caption*{Fantasy}
\end{minipage}
\begin{minipage}[t]{0.3\linewidth}
\centering
\includegraphics[width=1\linewidth]{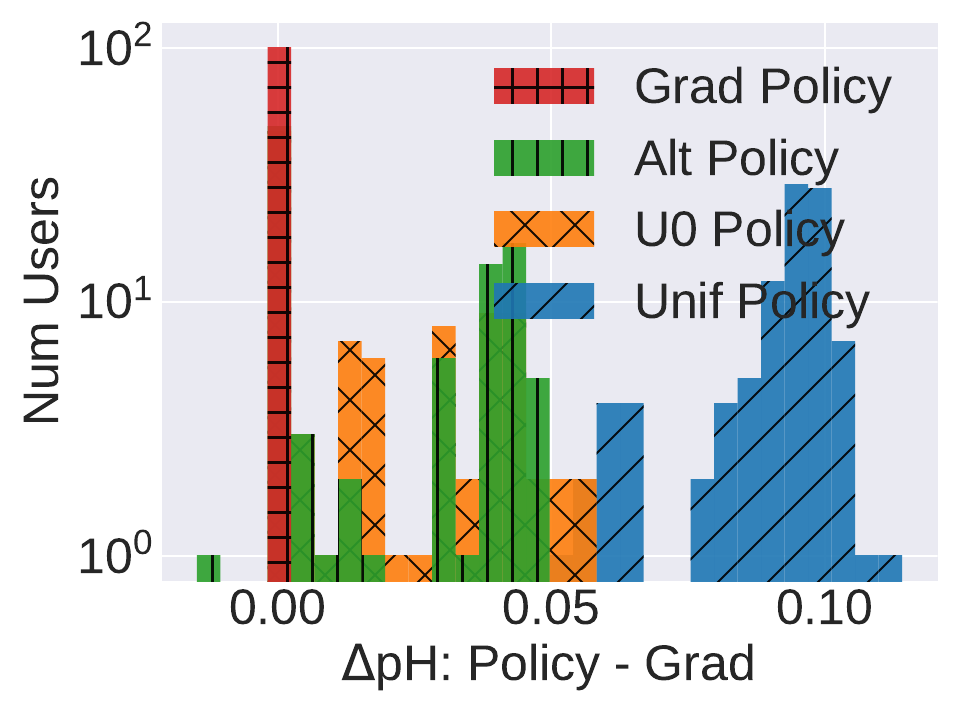}
\vspace{-2em}
\caption*{Sci-Fi}
\end{minipage}
\caption{
PDF of the difference between the $p_\hm$ attained by different policies minus the objective by attained by \emph{Grad}, for the same experimental settings as in Table~\ref{tab:big_table}. 
For the overwhelming majority of users, the $p_\hm$ values obtained by the \emph{Grad} policy is superior to any other policy.
}
\label{fig:pdfdiffph}
\end{figure*} 

\noindent\textbf{Per-User Policy Performance Comparison.} Fig.~\ref{fig:pdfdiffobj} shows  the PDF of the difference between the objective attained by different policies minus the objective by attained by \emph{Grad}, for the same experimental settings as in Table~\ref{tab:big_table}. For the overwelming majority of users, the objective attained by other policies is below the ones attained by \emph{Grad}.  
These insights are also reflected in the plots for $p_\clk$ and $p_\hm$, Figures~\ref{fig:pdfdiffpclk} and~\ref{fig:pdfdiffph}, respectively.

\begin{figure*}
\begin{center}

\includegraphics[width=0.85\textwidth]{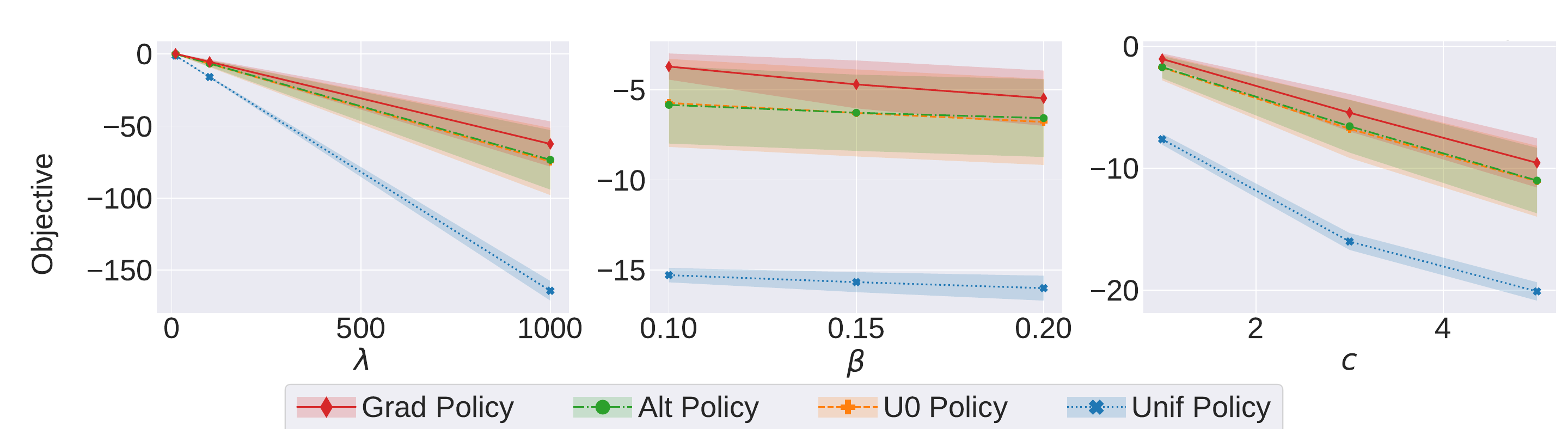}\\
(a) Action\\
\includegraphics[width=0.85\textwidth]{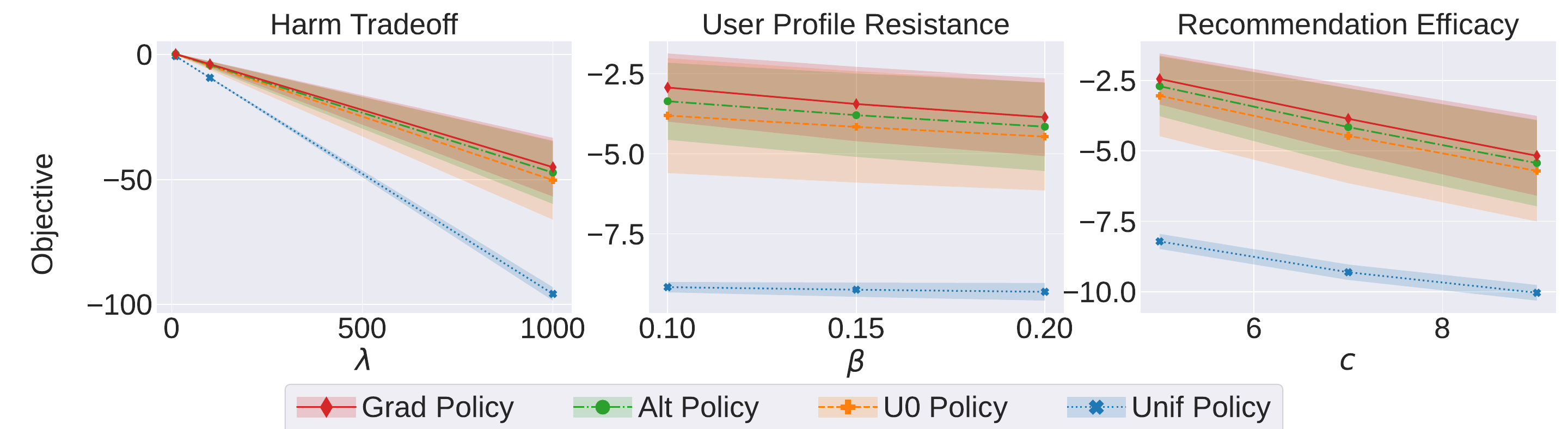}\\
(b) Adventure\\
\includegraphics[width=0.85\textwidth]{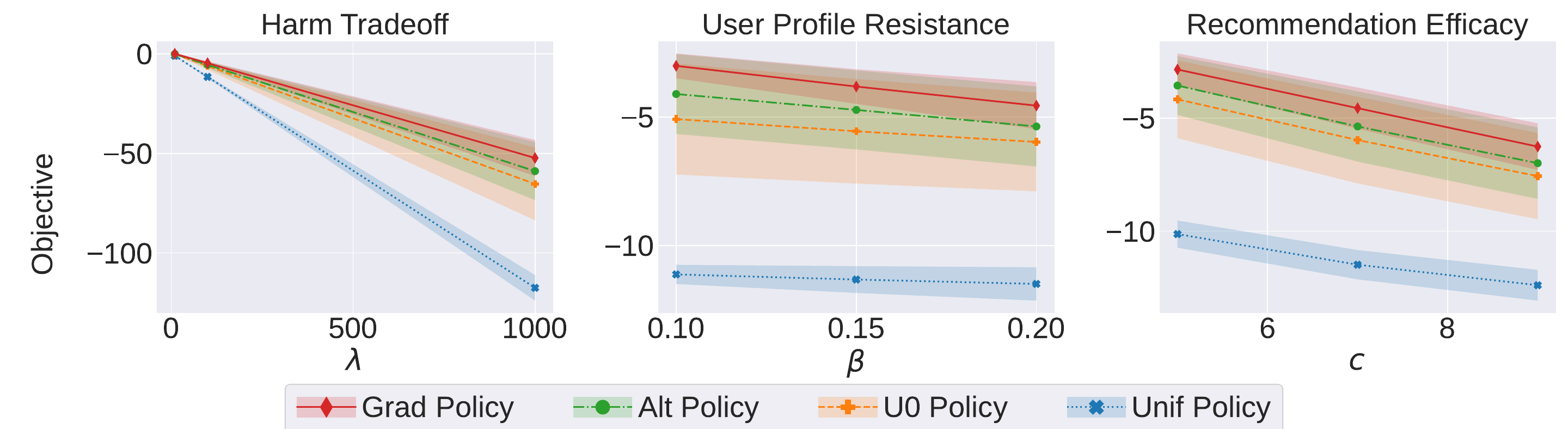}\\
(c) Comedy \\
\includegraphics[width=0.85\textwidth]{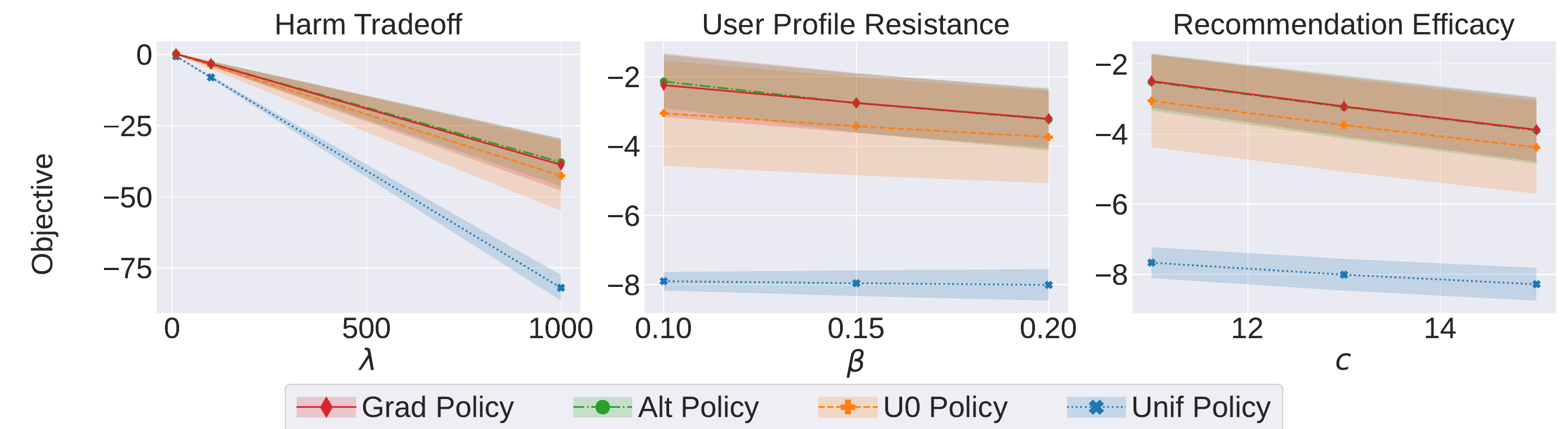}\\
(d) Fantasy \\
\includegraphics[width=0.85\textwidth]{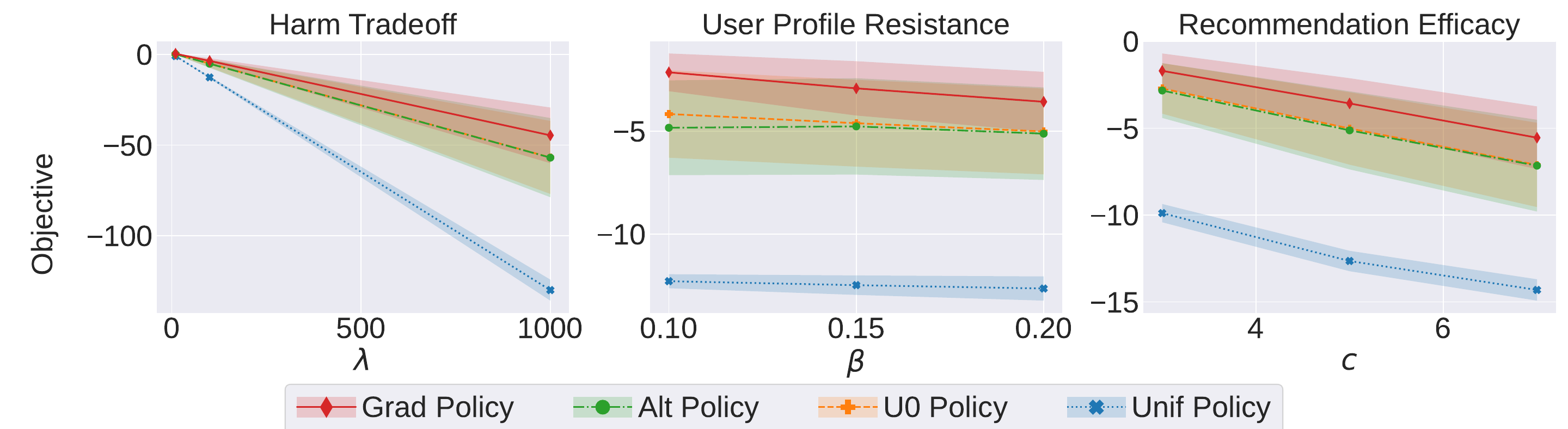}\\
(e) Sci-Fi
\end{center}
\caption{Effect of modifying $\lambda$, $\beta$, and $c$ on the objective attained by different policies for different genres. Increasing any parameter decrease the objective attained by every policy.  We observe that increasing $\lambda$ naturally increases the performance gap of the \emph{Grad} policy. Increasing $\beta$ has the opposite effect, as it limits the ability of the policy to impact a user's profile. Parameter $c$ also increases the improvement of \emph{Grad} over other policies as, the larger $c$ is, the less likely the recommendation is to be accepted, and the more important it becomes to minimize harm. 
} \label{fig:mod_obj_all}
\end{figure*}

\begin{figure*}
\begin{center}

\includegraphics[width=0.85\textwidth]{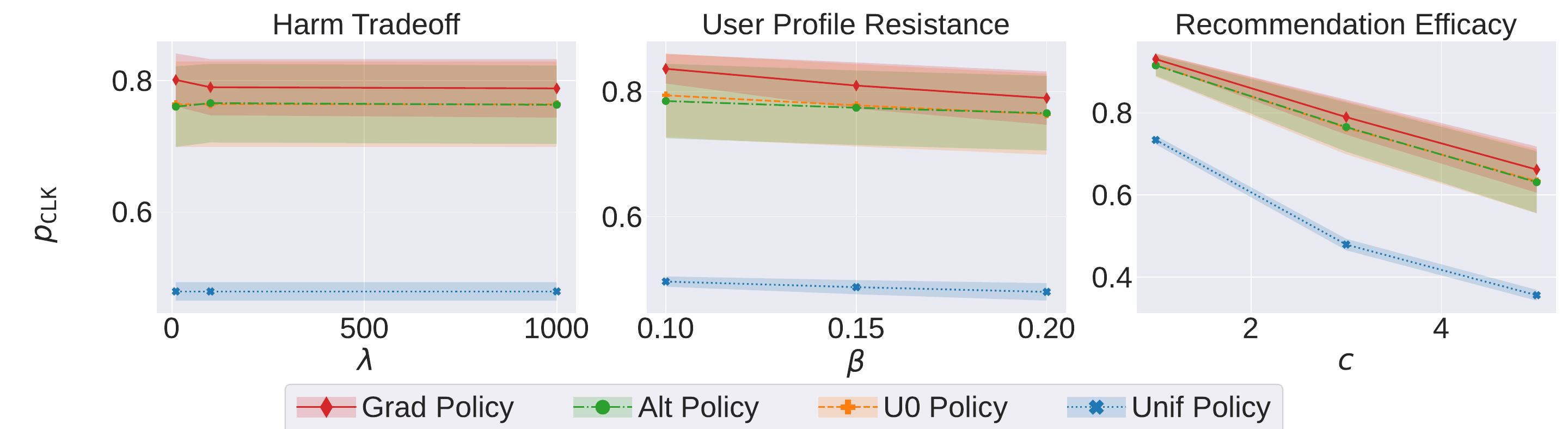}\\
(a) Action\\

\includegraphics[width=0.85\textwidth]{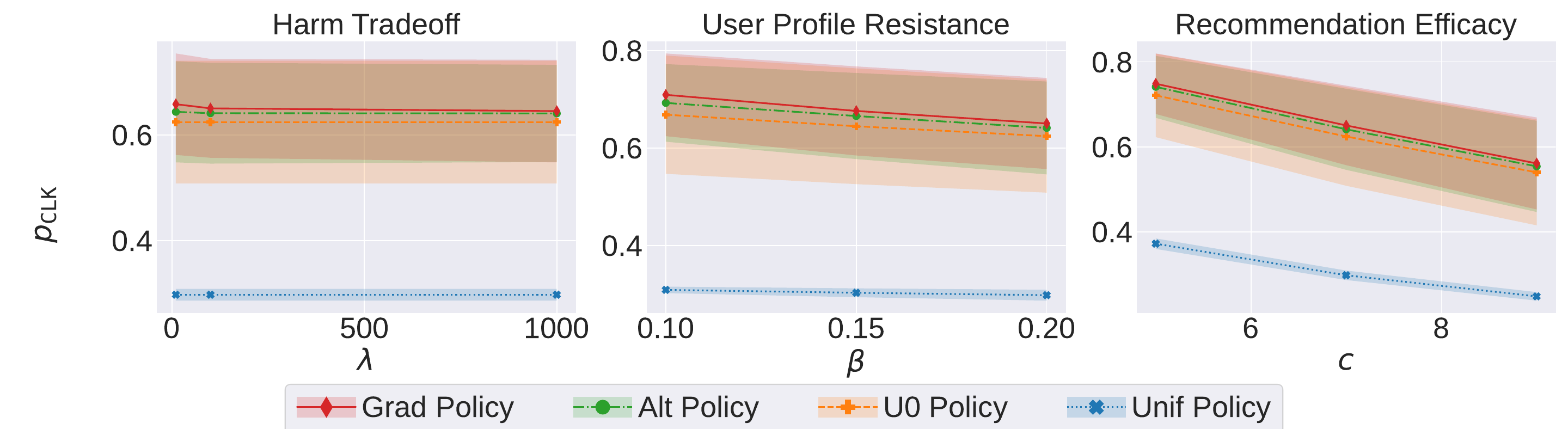}\\
(b) Adventure\\

\includegraphics[width=0.85\textwidth]{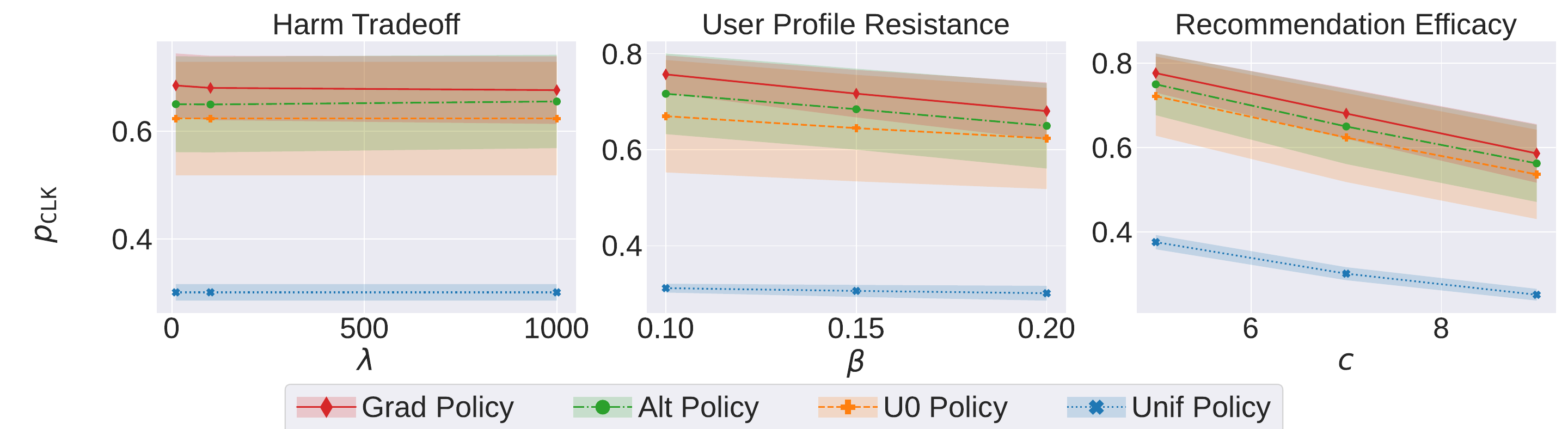}\\
(c) Comedy \\

\includegraphics[width=0.85\textwidth]{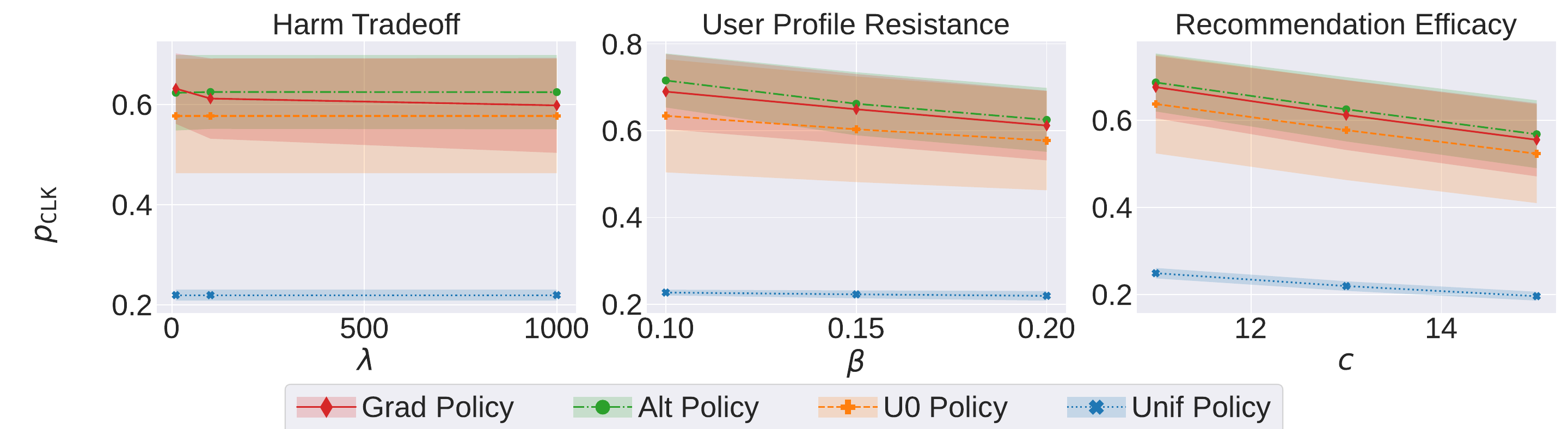}\\
(d) Fantasy \\

\includegraphics[width=0.85\textwidth]{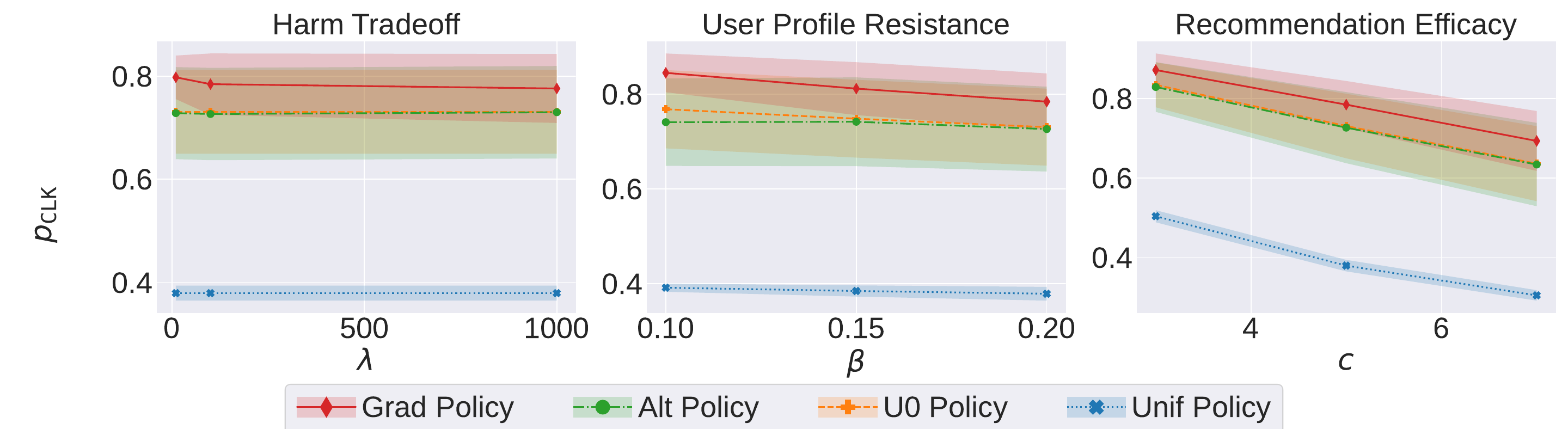}\\
(e) Sci-Fi
\end{center}
\caption{Effect of modifying $\lambda$, $\beta$, and $c$ on $p_\clk$ attained by different policies for different genres. Increasing any parameter decrease the objective attained by every policy.  We observe that increasing $\lambda$ naturally increases the performance gap of the \emph{Grad} policy. Increasing $\beta$ has the opposite effect, as it limits the ability of the policy to impact a user's profile. Parameter $c$ also increases the improvement of \emph{Grad} over other policies as, the larger $c$ is, the less likely the recommendation is to be accepted, and the more important it becomes to minimize harm. 
}\label{fig:mod_pclk_all}
\end{figure*}

\begin{figure*}
\begin{center}

\includegraphics[width=0.85\textwidth]{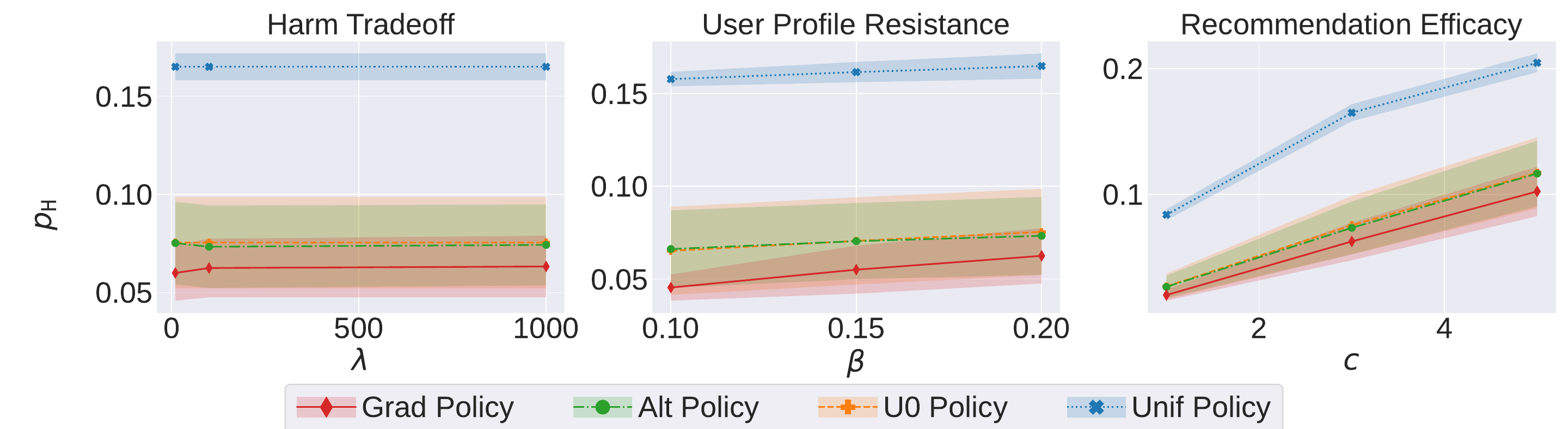}\\
(a) Action\\
\includegraphics[width=0.85\textwidth]{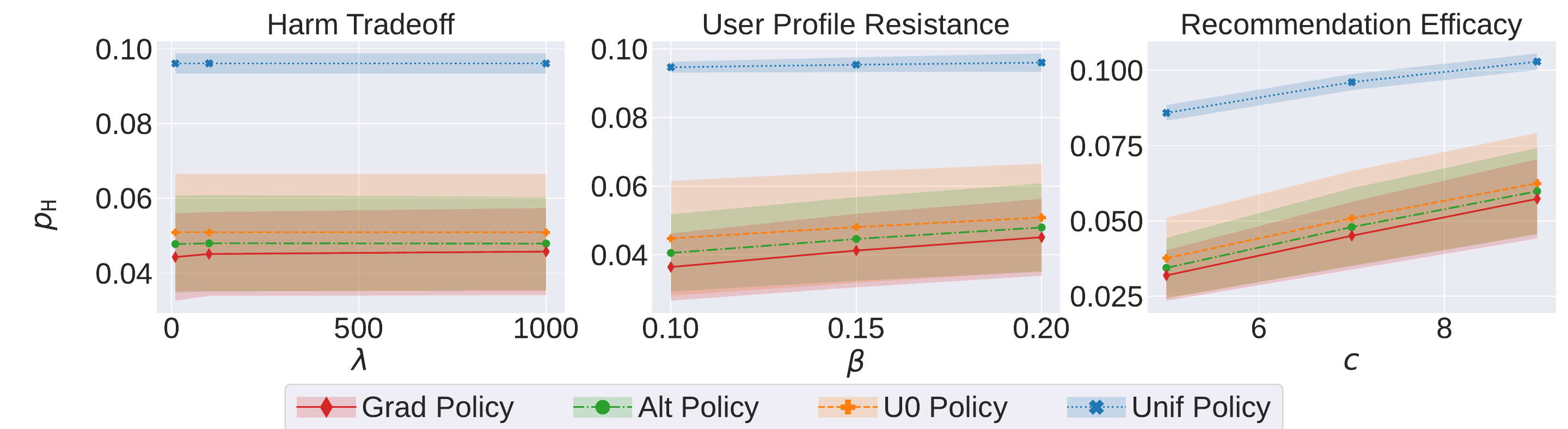}\\
(b) Adventure\\
\includegraphics[width=0.85\textwidth]{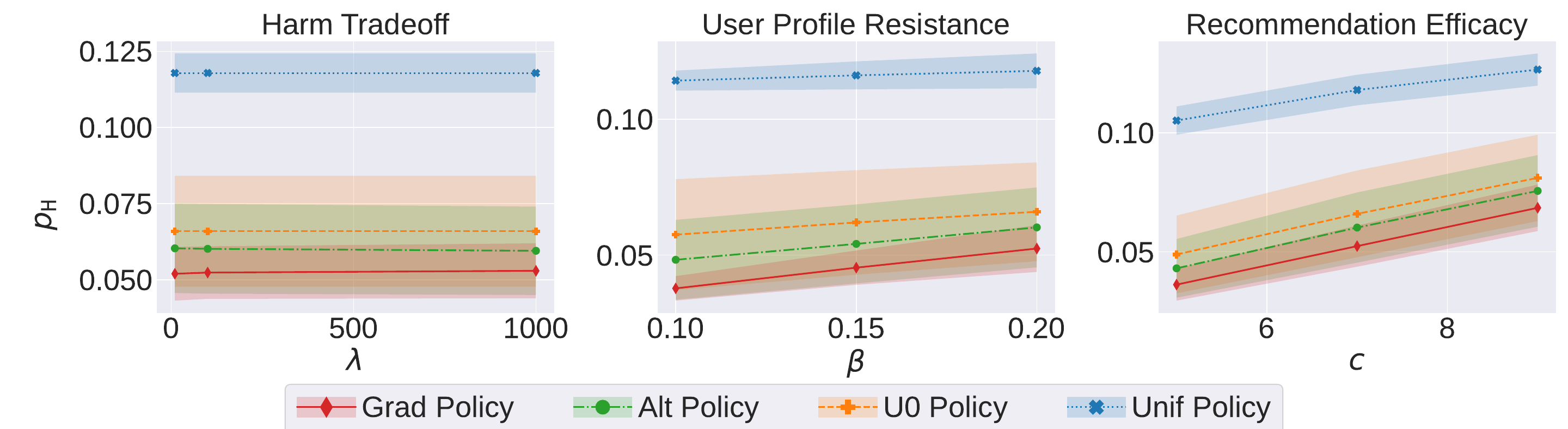}\\
(c) Comedy \\
\includegraphics[width=0.85\textwidth]{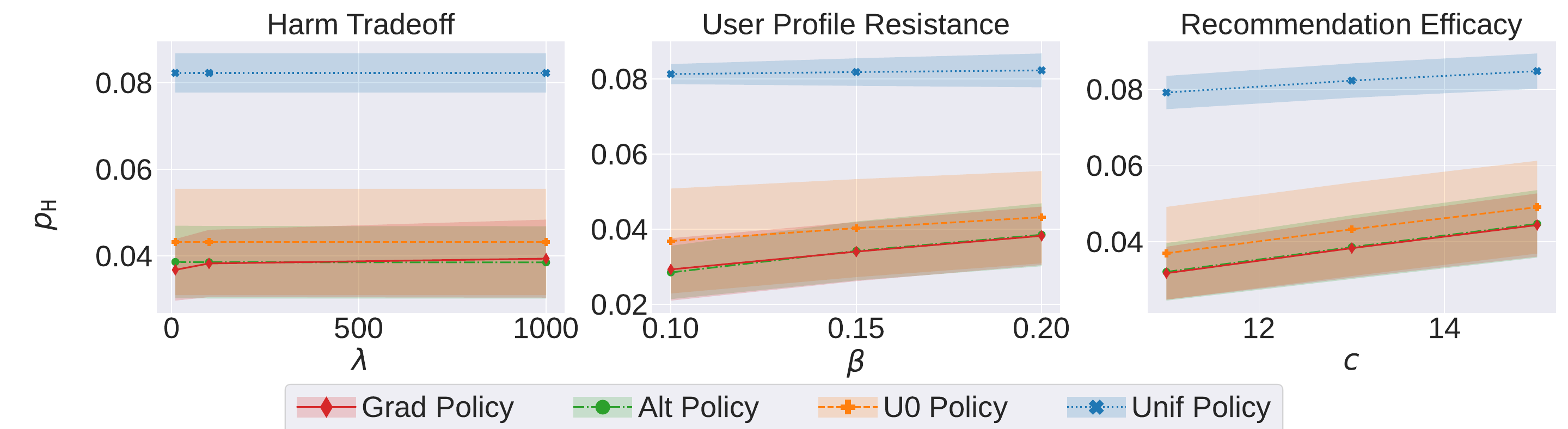}\\
(d) Fantasy \\
\includegraphics[width=0.85\textwidth]{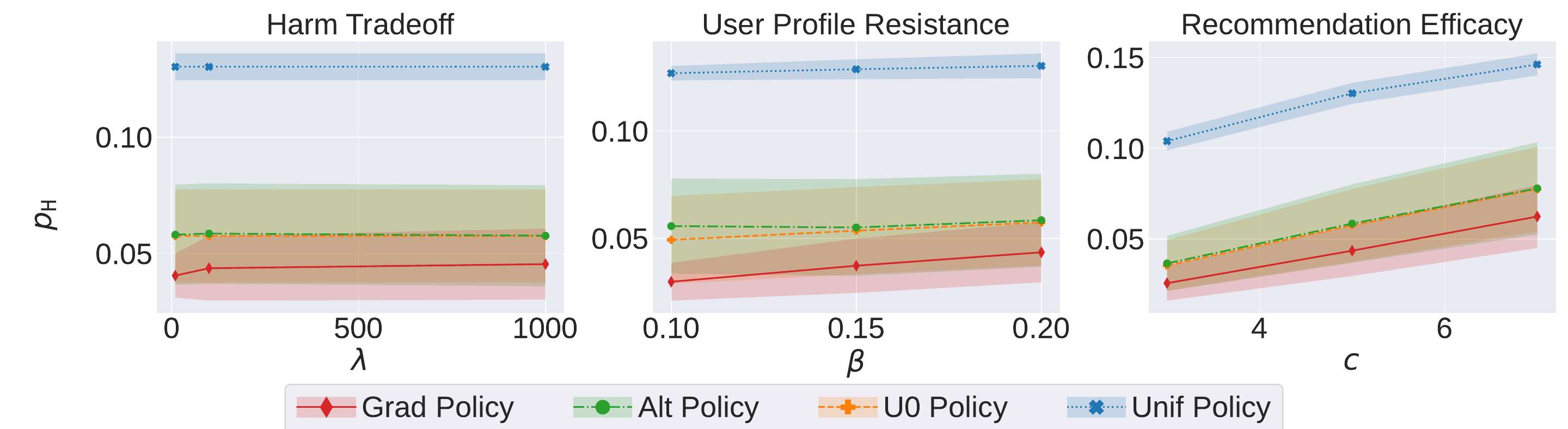}\\
(e) Sci-Fi
\end{center}
\caption{Effect of modifying $\lambda$, $\beta$, and $c$ on $p_\hm$ attained by different policies for different genres. Increasing any parameter decrease the objective attained by every policy.  We observe that increasing $\lambda$ naturally increases the performance gap of the \emph{Grad} policy. Increasing $\beta$ has the opposite effect, as it limits the ability of the policy to impact a user's profile. Parameter $c$ also increases the improvement of \emph{Grad} over other policies as, the larger $c$ is, the less likely the recommendation is to be accepted, and the more important it becomes to minimize harm.  }\label{fig:mod_phm_all}
\end{figure*}
 \noindent\textbf{Understanding the Effect of Model Parameters.}
We plot the change in the objective as we modify parameters for different genres in Figure~\ref{fig:mod_obj_all}. We observe the same patterns we saw in Fig.~\ref{fig:mod_obj_small} for the Action genre. As the user coefficient $\beta$ increases, we see a consistent degradation in performance for the policies (gradient, alternating) which account for user dynamics. 
This is expected as higher $\beta$ makes a user more resistant to changing their initial profile.
As the recommendation efficacy decreases (higher $c$), we see that $p_\clk$ decreases and $p_\hm$ increases as users go to more $\org$ interactions.

We also plot the corresponding $p_\clk$ and $p_\hm$ values in Figures~\ref{fig:mod_pclk_all} and Figure~\ref{fig:mod_phm_all}, respectively.
As we increase the harm penalty $\lambda$ we see that $p_\clk$ and $p_\hm$ remain stable. We see an improved performance on the grad policy when $\lambda=0$ w.r.t. $p_\hm$, which is counterintuitive. This could be either because the problem is highly non-convex, and none of the above solutions are guaranteed to be optimal, or because \emph{Grad} finds a better tradeoff at improving $p_\clk$ rather than directly minimizing $p_\hm$, even when $\lambda>0$, precisely because of the relatively small value of $p_\hm$ relative to $p_\clk$. In all cases, however, \emph{Grad} acheives a higher $p_\clk$ \emph{and} lower $p_\hm$ than other policies.

\begin{figure}
\centering
\begin{minipage}[t]{0.32\linewidth}
\centering
\includegraphics[width=1\linewidth]{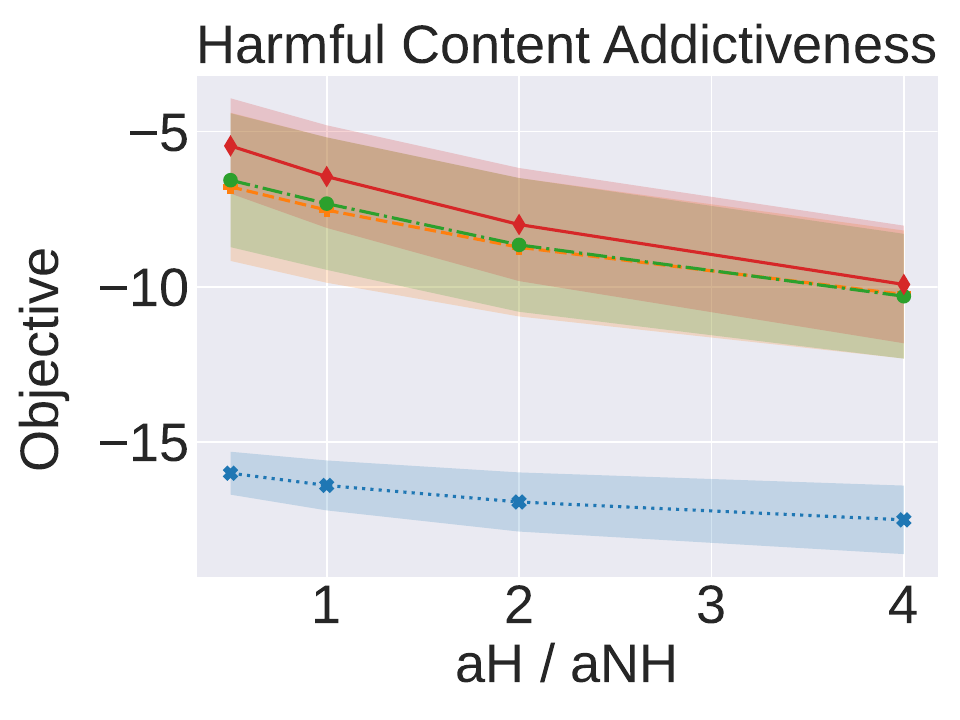}
\vspace{-2em}
\caption*{(a) Action}
\end{minipage}
\begin{minipage}[t]{0.32\linewidth}
\centering
\includegraphics[width=1\linewidth]{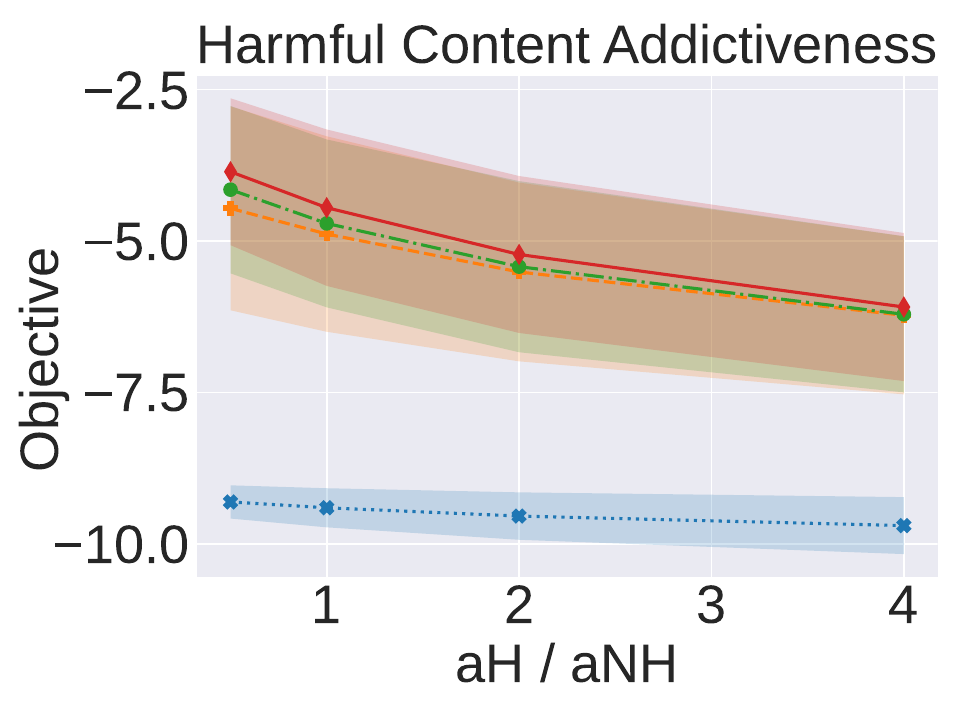}
\vspace{-2em}
\caption*{(b) Adventure}
\end{minipage}
\begin{minipage}[t]{0.32\linewidth}
\centering
\includegraphics[width=1\linewidth]{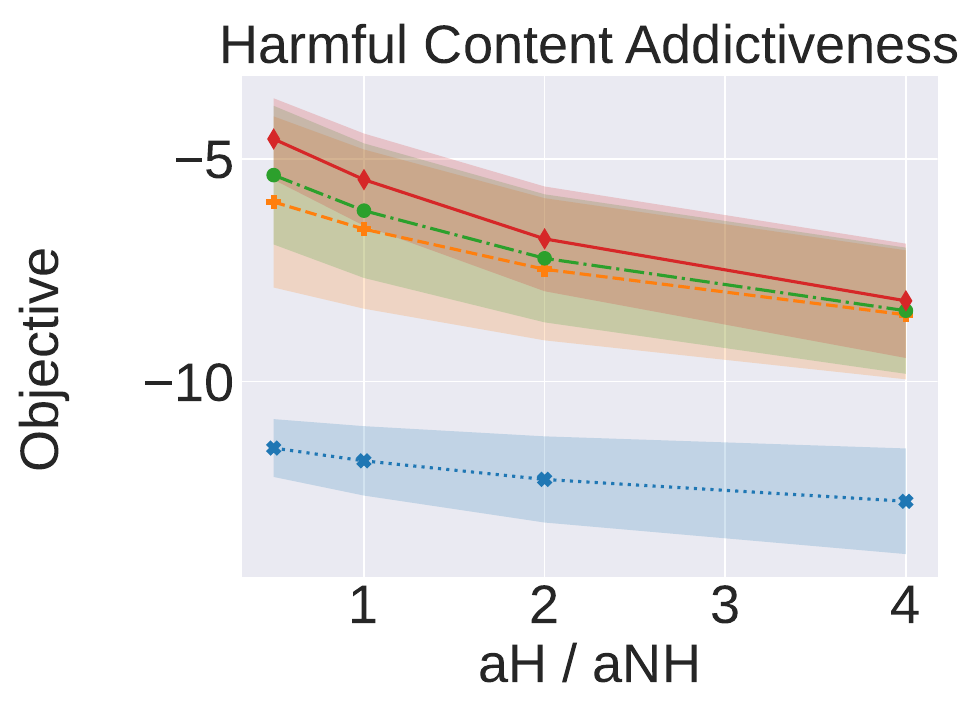}
\vspace{-2em}
\caption*{(c) Comedy}
\end{minipage}
\begin{minipage}[t]{0.32\linewidth}
\centering
\includegraphics[width=1\linewidth]{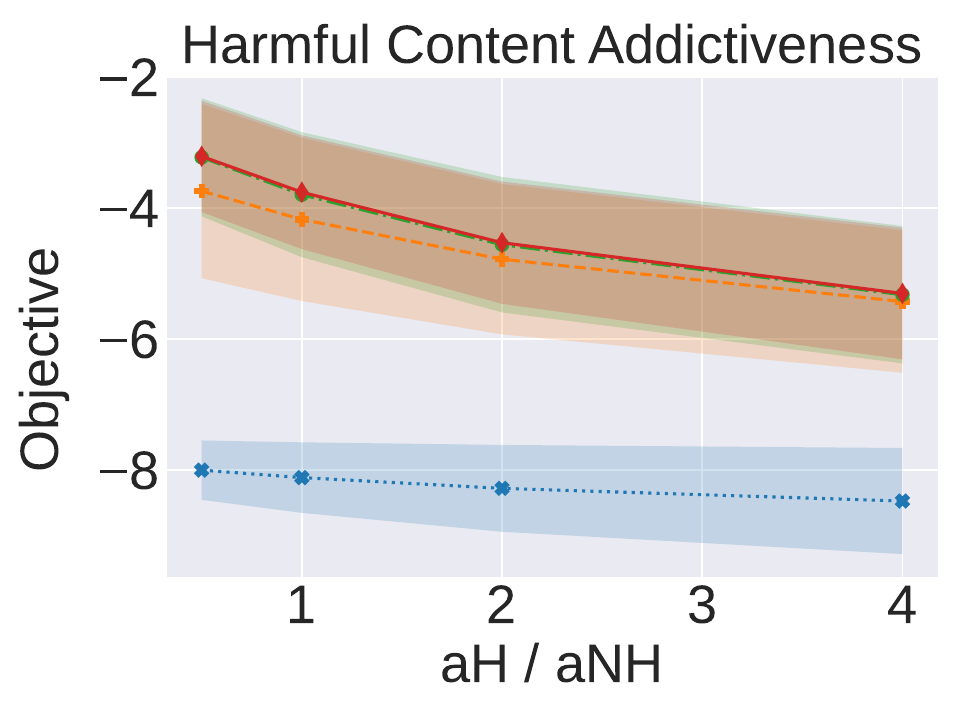}
\vspace{-2em}
\caption*{(d) Fantasy}
\end{minipage}
\begin{minipage}[t]{0.32\linewidth}
\centering
\includegraphics[width=1\linewidth]{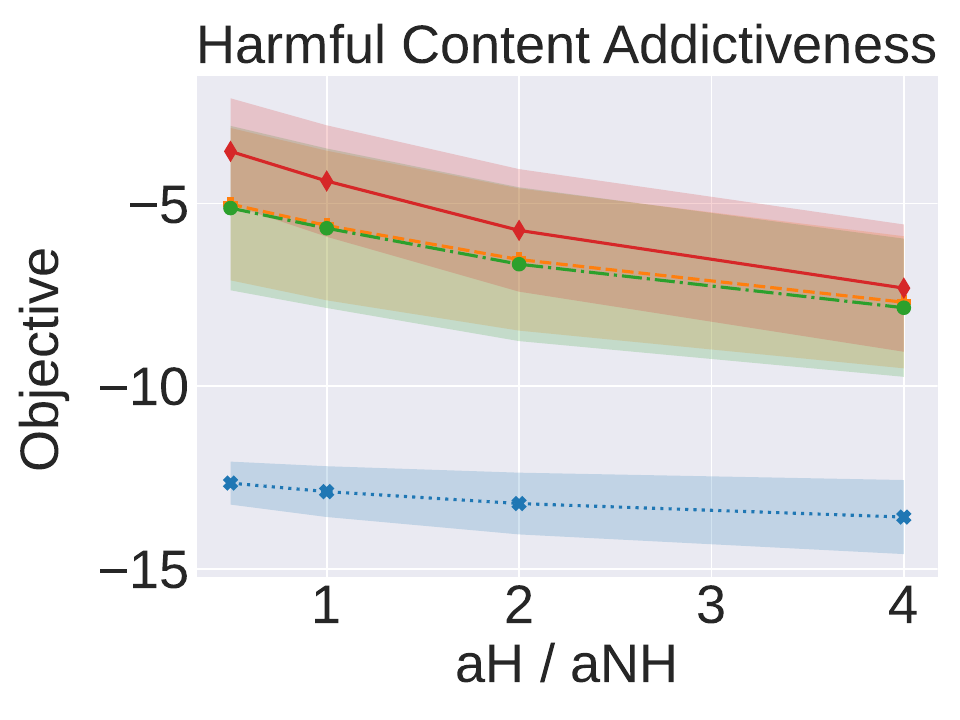}
\vspace{-2em}
\caption*{(e) Sci-Fi}
\end{minipage}
\caption{
Effect of modifying ratio $\alpha_\hm/\alpha_\nh$ on the objective attained by different policies for different genres. Increasing the ratio decreases the objective attained by every policy as well as the gap from \emph{Grad}, as policies have less leeway in minimizing harm. 
}
\label{fig:ahanh_all}
\end{figure}

\begin{figure}
\centering
\begin{minipage}[t]{0.32\linewidth}
\centering
\includegraphics[width=1\linewidth]{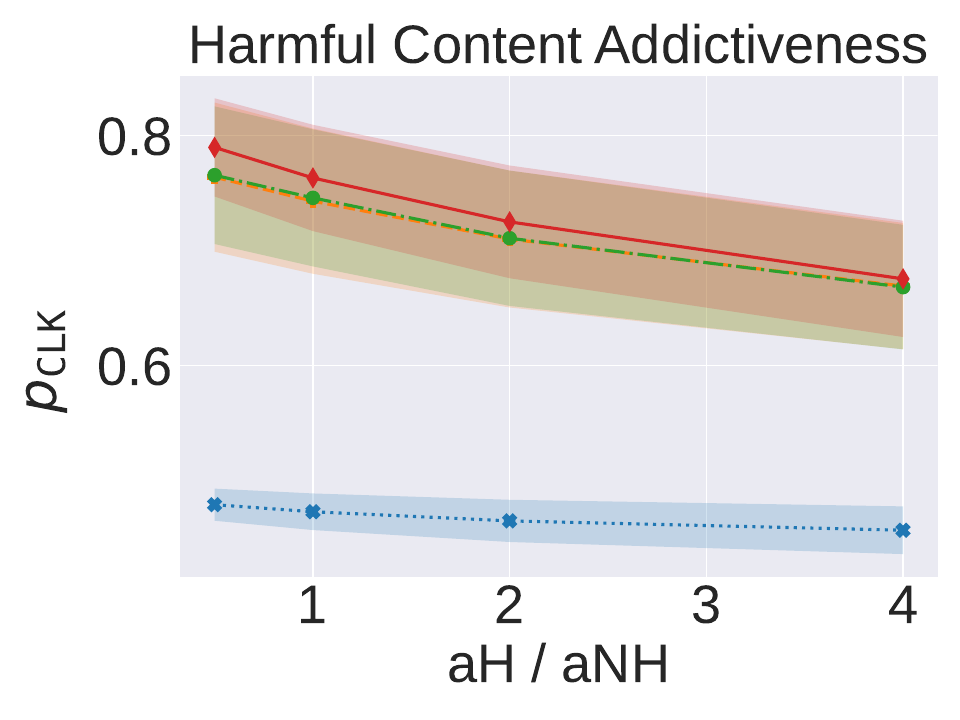}
\vspace{-2em}
\caption*{(a) Action}
\end{minipage}
\begin{minipage}[t]{0.32\linewidth}
\centering
\includegraphics[width=1\linewidth]{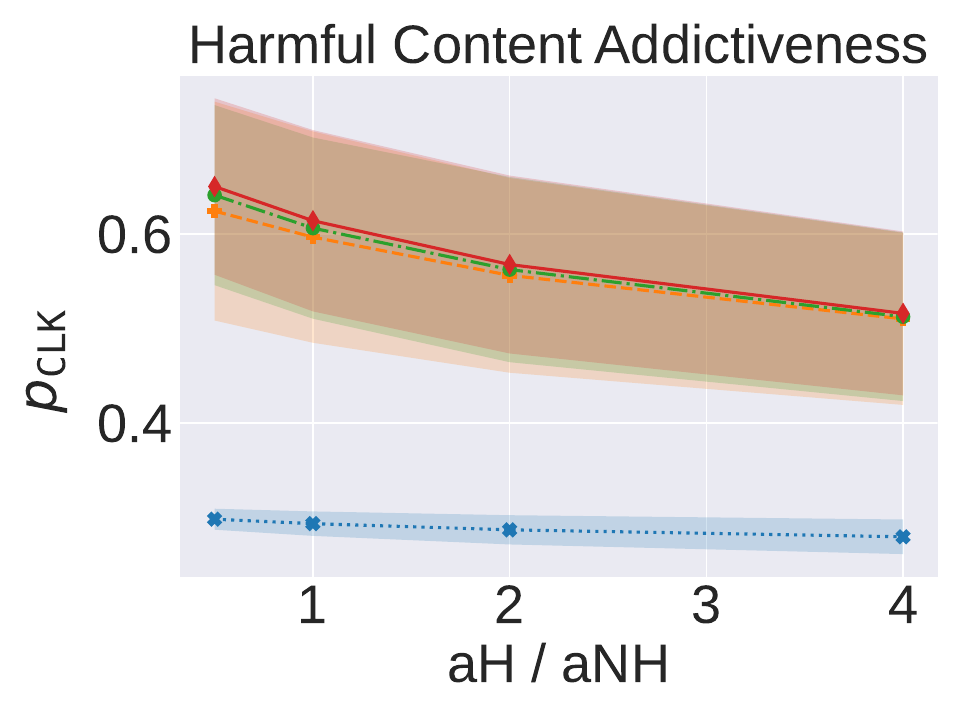}
\vspace{-2em}
\caption*{(b) Adventure}
\end{minipage}
\begin{minipage}[t]{0.32\linewidth}
\centering
\includegraphics[width=1\linewidth]{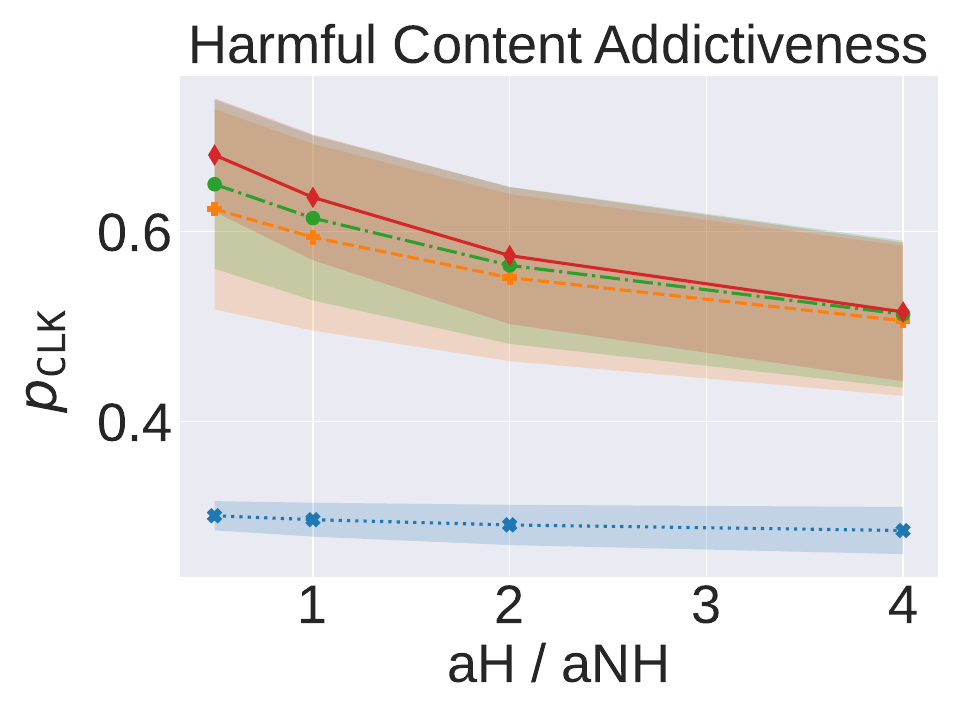}
\vspace{-2em}
\caption*{(c) Comedy}
\end{minipage}
\begin{minipage}[t]{0.32\linewidth}
\centering
\includegraphics[width=1\linewidth]{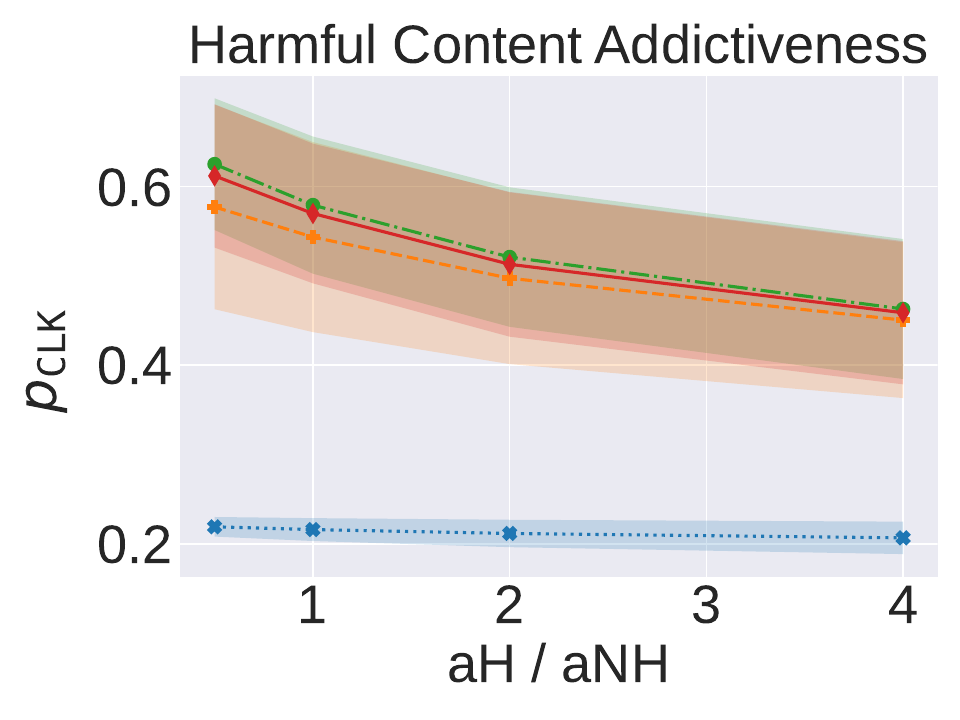}
\vspace{-2em}
\caption*{(d) Fantasy}
\end{minipage}
\begin{minipage}[t]{0.32\linewidth}
\centering
\includegraphics[width=1\linewidth]{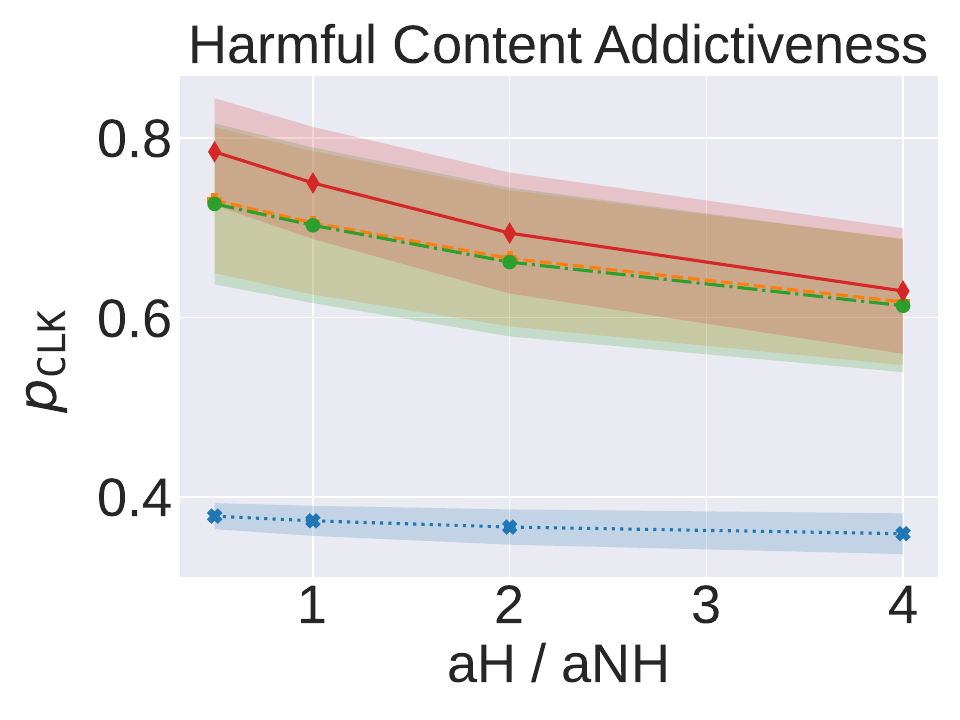}
\vspace{-2em}
\caption*{(e) Sci-Fi}
\end{minipage}
\caption{
Effect of modifying ratio $\alpha_\hm/\alpha_\nh$ on $p_\clk$ attained by different policies for the Action and Adventure genres.  Increasing the ratio decreases the $p_\clk$ attained by every policy as well as the gap from \emph{Grad}, as policies have less leeway in minimizing harm. 
}
\label{fig:ahanh_pclick_all}
\end{figure}

\begin{figure}
\centering
\begin{minipage}[t]{0.32\linewidth}
\centering
\includegraphics[width=1\linewidth]{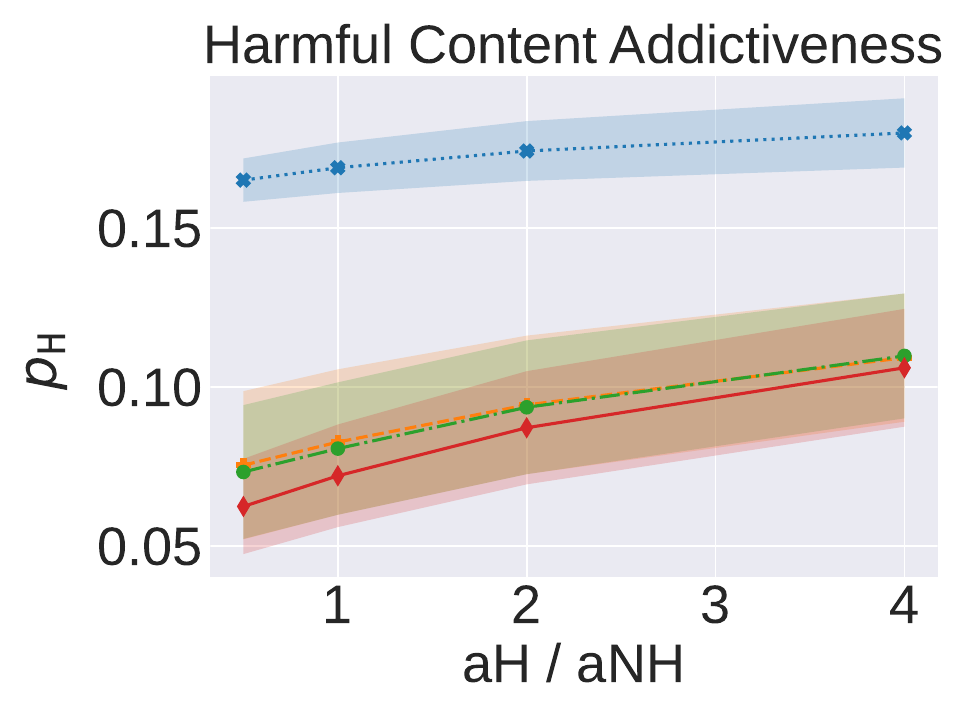}
\vspace{-2em}
\caption*{(a) Action}
\end{minipage}
\begin{minipage}[t]{0.32\linewidth}
\centering
\includegraphics[width=1\linewidth]{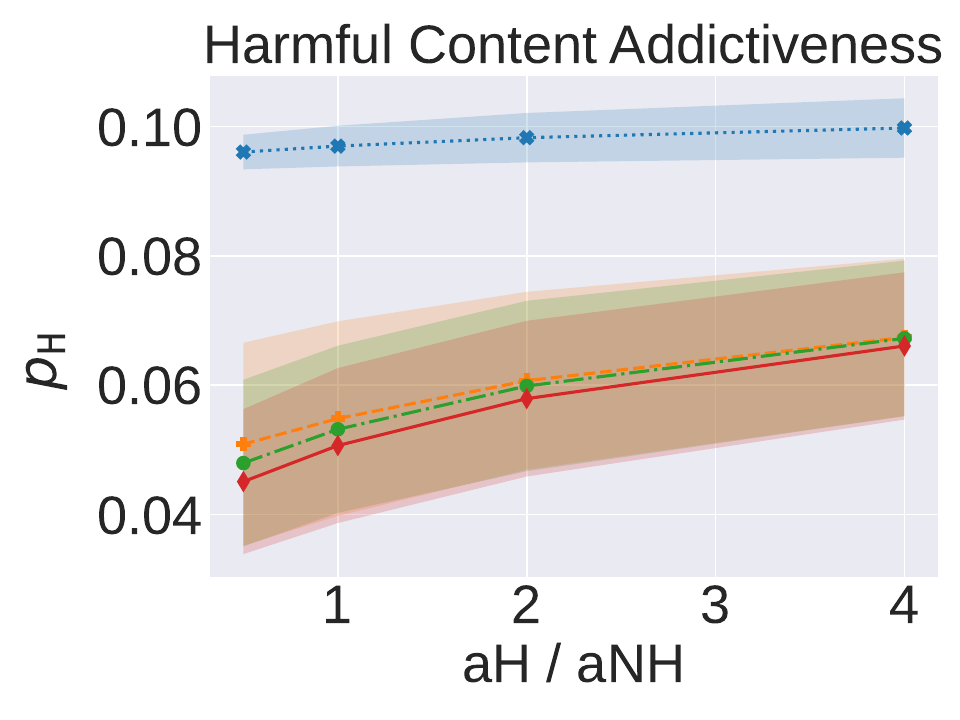}
\vspace{-2em}
\caption*{(b) Adventure}
\end{minipage}
\begin{minipage}[t]{0.32\linewidth}
\centering
\includegraphics[width=1\linewidth]{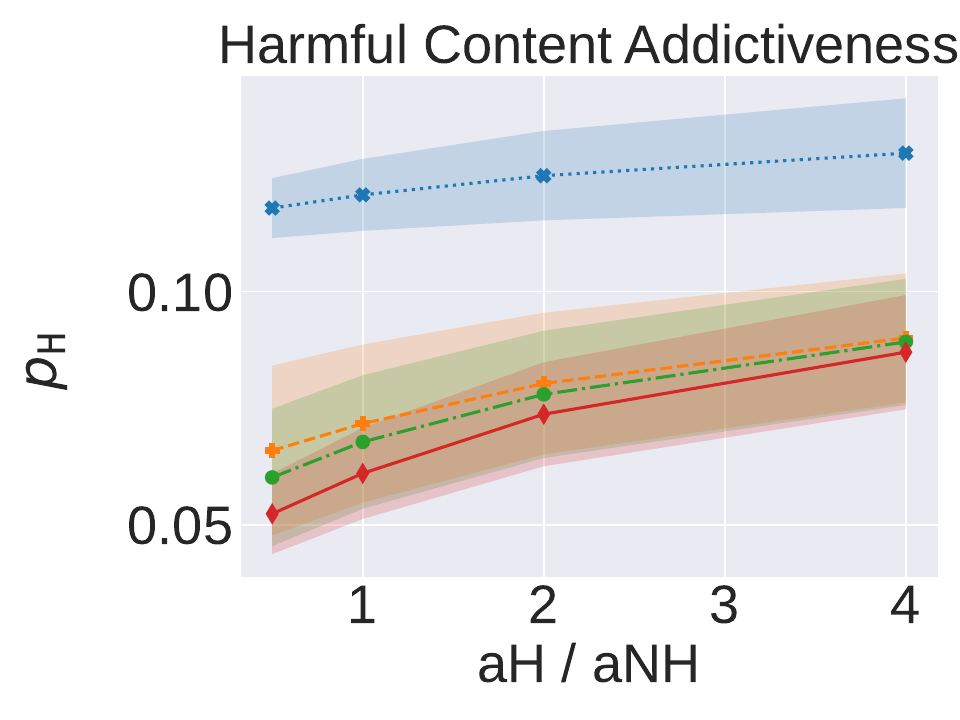}
\vspace{-2em}
\caption*{(c) Comedy}
\end{minipage}
\begin{minipage}[t]{0.32\linewidth}
\centering
\includegraphics[width=1\linewidth]{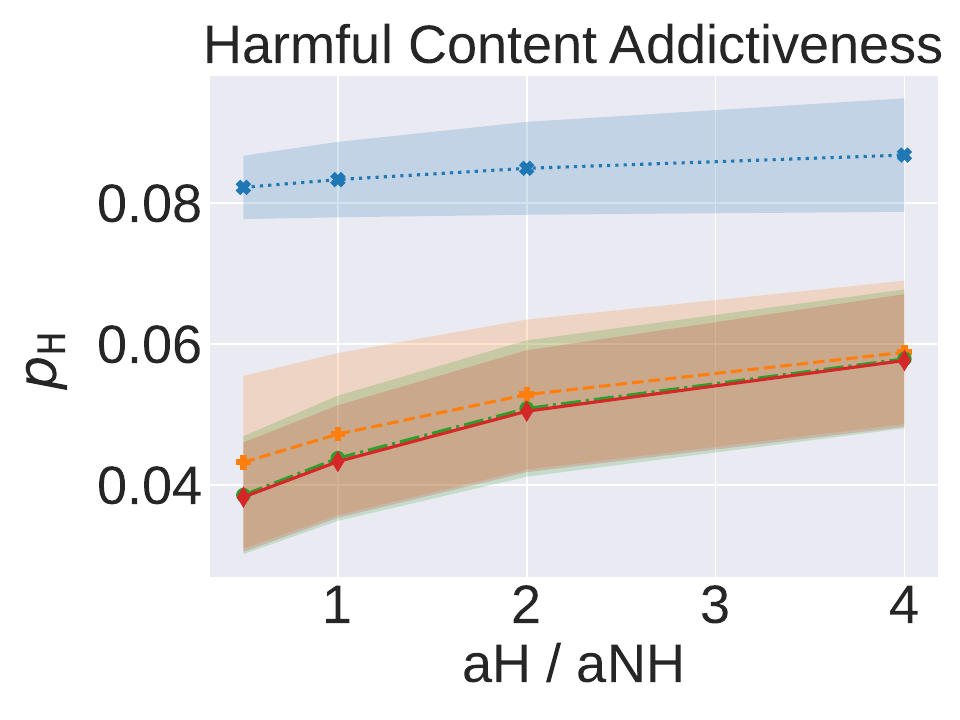}
\vspace{-2em}
\caption*{(d) Fantasy}
\end{minipage}
\begin{minipage}[t]{0.32\linewidth}
\centering
\includegraphics[width=1\linewidth]{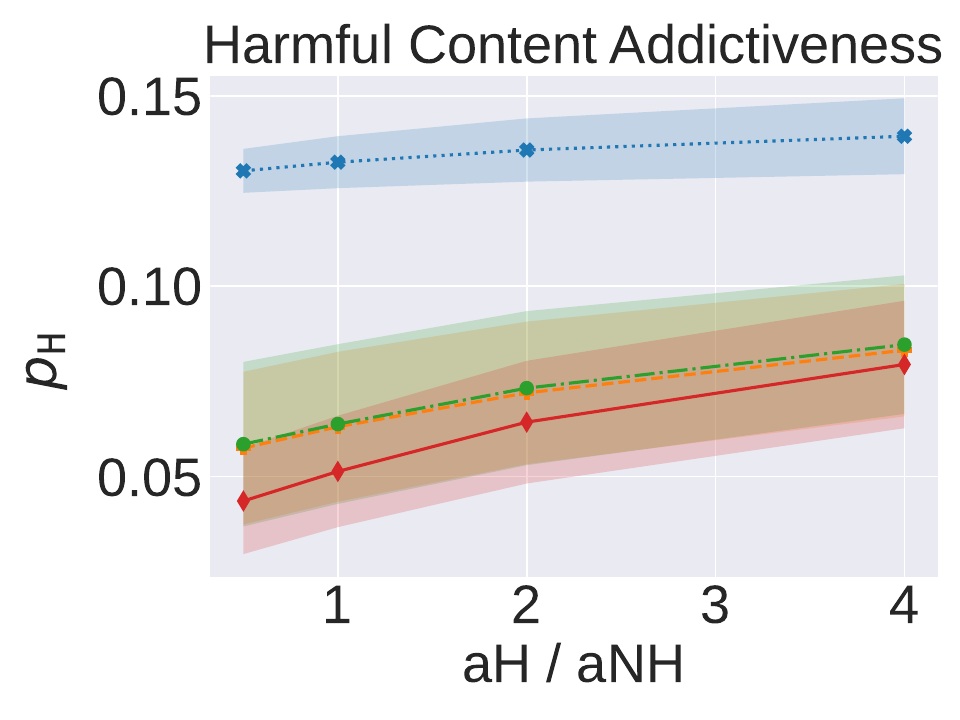}
\vspace{-2em}
\caption*{(e) Sci-Fi}
\end{minipage}
\caption{
Effect of modifying ratio $\alpha_\hm/\alpha_\nh$ on $p_\clk$ attained by different policies for the Action and Adventure genres.  Increasing the ratio increases the $p_\hm$ attained by every policy as well as the gap from \emph{Grad}, as policies have less leeway in minimizing harm. 
}
\label{fig:ahanh_phm_all}
\end{figure} 
We also plot the effect of increasing the ratio between $\alpha_\hm/\alpha_\nh$ in Fig.~\ref{fig:ahanh_all}. As harmful content becomes more attractive, we see that the gradient collapses into  alternating maximization.
All policies also degrade in terms of  $p_\clk$ and $p_\hm$ as harmful content becomes more attractive, as shown in Figs.~\ref{fig:ahanh_pclick_all} and~\ref{fig:ahanh_phm_all}, respectively.

\noindent\textbf{Increasing the Recommendation Set Size $k$.}
Table~\ref{tab:k_c_mod} gives the parameter values for $k$ and $c$ for Figures~\ref{fig:mod_k_obj},~\ref{fig:mod_k_pCLK}, and~\ref{fig:mod_k_pH}, when either $c$ is fixed, or $k/c$ is fixed.
We see in Figures~\ref{fig:mod_k_pCLK} and ~\ref{fig:mod_k_pH} similar trends as was observed in Figure~\ref{fig:mod_k_obj} in the main paper.
Increasing $k$ while $c$ is fixed results in an increase in $p_\clk$, and a resulting decrease in $p_\hm$.

\begin{table}[]
\centering
\begin{tabular}{cc|cc}
\multicolumn{2}{c}{\underline{Fixed $c$}} & \multicolumn{2}{c}{\underline{Fixed $k/c$}}\\
$k$ & $c$ & $k$ & $c$ \\
\midrule
1 & 18 & 1 & 3.6 \\
2 & 18 & 2 & 7.2 \\
4 & 18 & 4 & 14.4 \\
5 & 18 & 5 & 18.0 \\
6 & 18 & 6 & 21.6 \\
8 & 18 & 8 & 28.8 \\
10& 18 & 10& 36.0 \\
\end{tabular}
\caption{
Values for $k$, $c$ for Figures~\ref{fig:mod_k_obj},~\ref{fig:mod_k_pCLK}, and~\ref{fig:mod_k_pH}.
}
\label{tab:k_c_mod}
\end{table}

\begin{figure}
\centering
\begin{minipage}[t]{0.49\linewidth}
\centering
\includegraphics[width=1.05\linewidth]{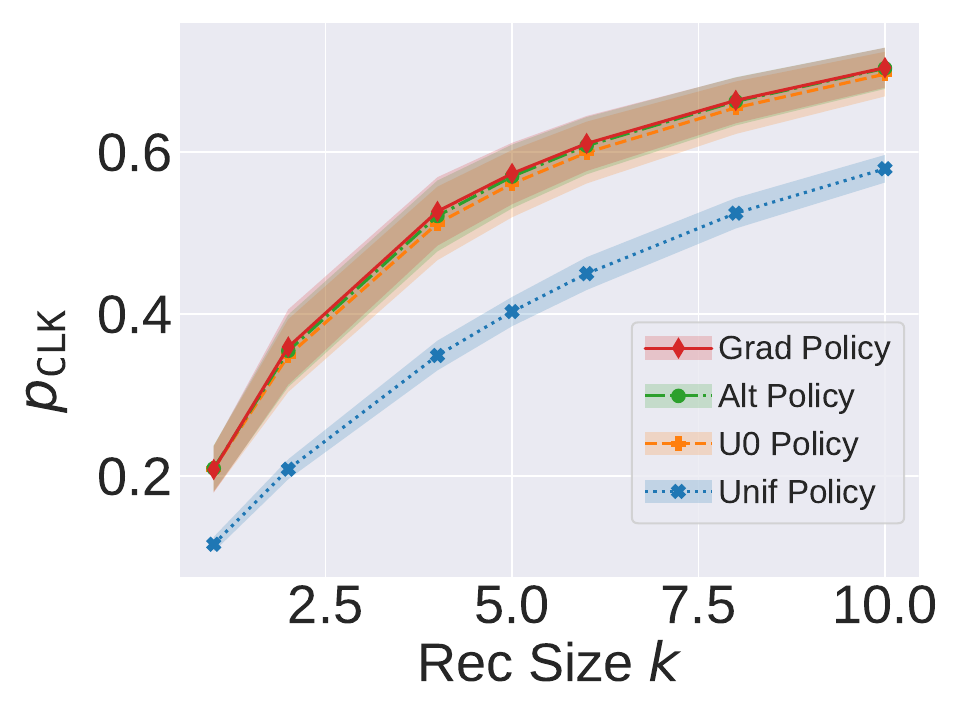}
\vspace{-2em}
\caption*{(a) Fixed $c \approx$ Varying $p_\clk$}
\end{minipage}
\begin{minipage}[t]{0.49\linewidth}
\centering
\includegraphics[width=1.05\linewidth]{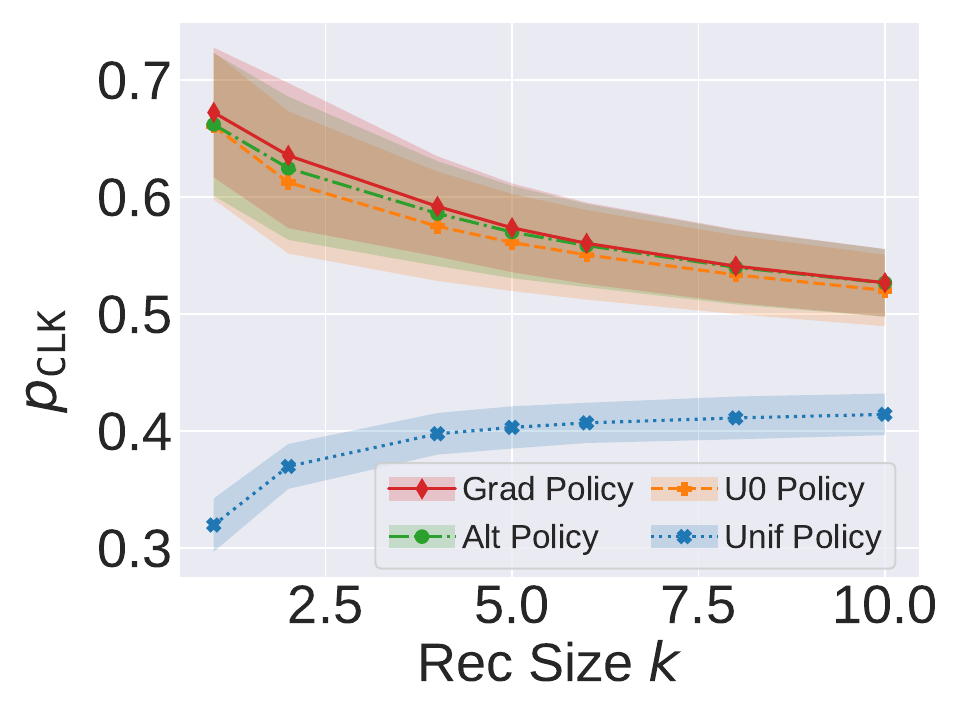}
\vspace{-2em}
\caption*{(a) Fixed $c \approx$ Varying $p_\clk$}
\end{minipage}
\caption{
Effect of increasing the recommendation set size $k$ on $p_\clk$ attained by different policies for the Action genre.
(a) We first increase the number of recommended items $k$. 
From the MNL model in Eq.~\eqref{eq:clk} we see that $S_E$ can increase with larger $k$, effectively increasing $p_\clk$ for all policies. 
(b): we keep the ratio $k/c$ fixed in order to counteract the effect of rising $p_\clk$ upon increasing $k$. 
The gradient-computed policy is superior over a variety of $k$.
} \label{fig:mod_k_pCLK}
\end{figure}

\begin{figure}
\centering
\begin{minipage}[t]{0.49\linewidth}
\centering
\includegraphics[width=1.05\linewidth]{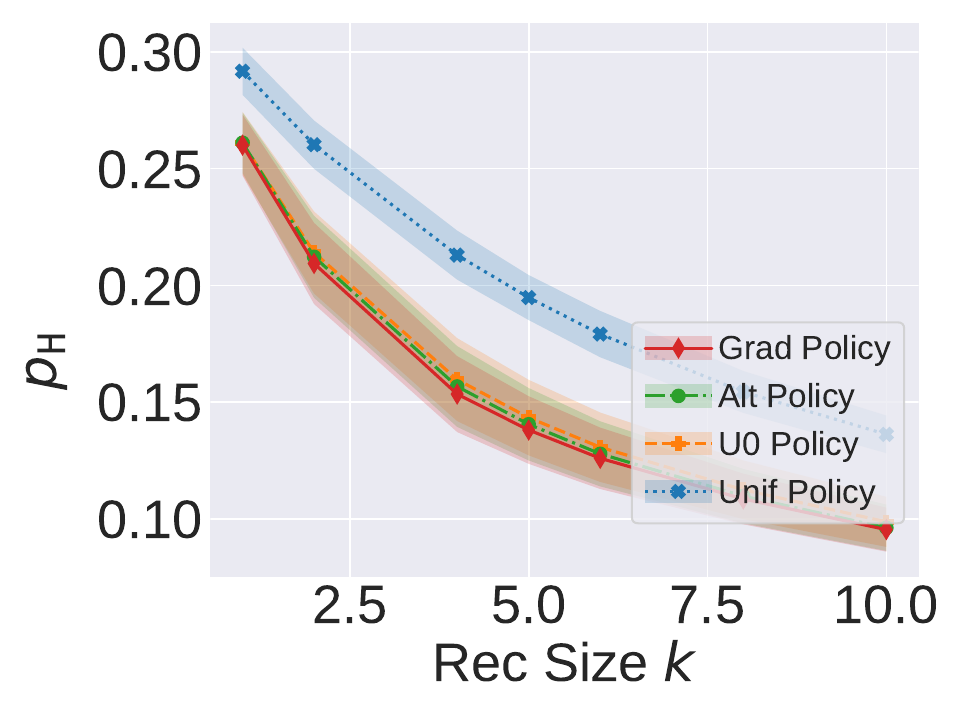}
\vspace{-2em}
\caption*{(b) Fixed $k/c \approx$ Fixed $p_\clk$}
\end{minipage}
\begin{minipage}[t]{0.49\linewidth}
\centering
\includegraphics[width=1.05\linewidth]{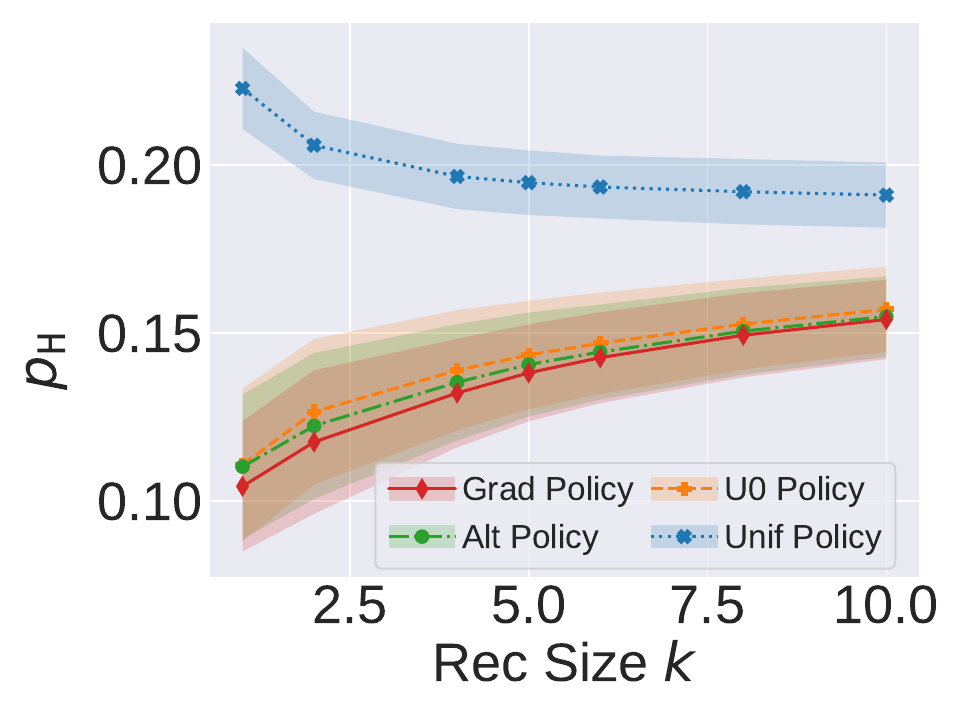}
\vspace{-2em}
\caption*{(b) Fixed $k/c \approx$ Fixed $p_\clk$}
\end{minipage}
\caption{
Effect of increasing the recommendation set size $k$ on $p_\hm$ attained by different policies for the Action genre.
(a) We first increase the number of recommended items $k$. 
From the MNL model in Eq.~\eqref{eq:clk} we see that $S_E$ can increase with larger $k$, effectively increasing $p_\clk$-and thus decrease $p_\hm$-for all policies. 
(b): we keep the ratio $k/c$ fixed in order to counteract the effect of rising $p_\clk$ upon increasing $k$. 
The gradient-computed policy is superior over a variety of $k$.
} \label{fig:mod_k_pH}
\end{figure}  }

\end{document}